\documentclass[10pt,a4paper,journal, onecolumn]{IEEEtran}

\usepackage[utf8]{inputenc}
\usepackage[T1]{fontenc}
\usepackage{url}
\usepackage{ifthen}
\usepackage{cite}
\usepackage[cmex10]{amsmath} % Use the [cmex10] option to ensure complicance
\RequirePackage{bm}

\RequirePackage{amssymb}
\RequirePackage{amsfonts}
\RequirePackage{amsthm}
\RequirePackage{mathtools}
\RequirePackage{dsfont}
\RequirePackage{cases}
\RequirePackage{float}

\RequirePackage[boxed,ruled,vlined]{algorithm2e}
\RequirePackage{algorithmic}

\PassOptionsToPackage{bookmarks=false}{hyperref}
\RequirePackage[colorlinks=true]{hyperref}
\hypersetup{
	linkcolor=blue,          % color of internal links (change box color with linkbordercolor)
	citecolor=blue,        % color of links to bibliography
	filecolor=magenta,      % color of file links
	urlcolor=blue           % color of external links
}

\RequirePackage{enumitem}

\RequirePackage{subfigure}

\RequirePackage{epsfig}
\RequirePackage{epstopdf}

\makeatletter\def\CT{\def\@captype{figure}}\makeatother

\RequirePackage{booktabs}
\RequirePackage{rotating}
\RequirePackage{lscape}
\RequirePackage{pdflscape}
\RequirePackage{longtable}

\makeatletter
\def\th@plain{%
	\thm@notefont{}% same as heading font
	\itshape % body font
}
\def\th@definition{%
	\thm@notefont{}% same as heading font
	\normalfont % body font
}
\makeatother

\theoremstyle{plain}% default
\newtheorem{theorem}{Theorem}

\newtheorem{lemma}{Lemma}

\newtheorem{corollary}{Corollary}

\newtheorem{assumption}{Assumption}
\theoremstyle{definition}

\newtheorem{definition}{Definition}

\newtheorem{remark}{Remark}

\usepackage{mysty}
\begin{document}
\title{A Sharp Algorithmic Analysis of Covariate Adjusted Precision Matrix Estimation with General Structural Priors}

% %%% Single author, or several authors with same affiliation:

\author{%
	\IEEEauthorblockN{Xiao~Lv\IEEEauthorrefmark{1}, Wei~Cui\IEEEauthorrefmark{1} and Yulong~Liu\IEEEauthorrefmark{2}}
	
	\IEEEauthorblockA{\IEEEauthorrefmark{1}%
		School of Information and Electronics, Beijing Institute of Technology\\
		Email: \ba{xiaolv, cuiwei}@bit.edu.cn}
%	\and
%	\IEEEauthorblockN{Yulong~Liu}

	\IEEEauthorblockA{\IEEEauthorrefmark{2}%
		School of Physics, Beijing Institute of Technology\\
		Email: yulongliu@bit.edu.cn}
}

%%% Many authors with many affiliations:
% \author{%
%   \IEEEauthorblockN{Albus Dumbledore\IEEEauthorrefmark{1},
%                     Olympe Maxime\IEEEauthorrefmark{2},
%                     Stefan M.~Moser\IEEEauthorrefmark{3}\IEEEauthorrefmark{4},
%                     and Harry Potter\IEEEauthorrefmark{1}}
%   \IEEEauthorblockA{\IEEEauthorrefmark{1}%
%                     Hogwarts School of Witchcraft and Wizardry,
%                     1714 Hogsmeade, Scotland,
%                     \{dumbledore, potter\}@hogwarts.edu}
%   \IEEEauthorblockA{\IEEEauthorrefmark{2}%
%                     Beauxbatons Academy of Magic,
%                     1290 Pyrénées, France,
%                     maxime@beauxbatons.edu}
%   \IEEEauthorblockA{\IEEEauthorrefmark{3}%
%                     ETH Zürich, ISI (D-ITET), ETH Zentrum,
%                     CH-8092 Zürich, Switzerland,
%                     moser@isi.ee.ethz.ch}
%   \IEEEauthorblockA{\IEEEauthorrefmark{4}%
%                     National Chiao Tung University (NCTU),
%                     Hsinchu, Taiwan,
%                     moser@isi.ee.ethz.ch}
% }

\maketitle

%%%%%%
%% Abstract:
%% If your paper is eligible for the student paper award, please add
%% the comment "THIS PAPER IS ELIGIBLE FOR THE STUDENT PAPER
%% AWARD." as a first line in the abstract.
%% For the final version of the accepted paper, please do not forget
%% to remove this comment!
%%
\begin{abstract}
In this paper, we present a sharp analysis for a class of alternating projected gradient descent algorithms which are used to solve the covariate adjusted precision matrix estimation problem in the high-dimensional setting. We demonstrate that these algorithms not only enjoy a linear rate of convergence in the absence of convexity, but also attain the optimal statistical rate (i.e., minimax rate). By introducing the generic chaining, our analysis removes the impractical resampling assumption used in the previous work. Moreover, our results also reveal a time-data tradeoff in this covariate adjusted precision matrix estimation problem. Numerical experiments are provided to verify our theoretical results.
\end{abstract}

%\begin{IEEEkeywords}
%	Time-data tradeoffs, proximal-gradient homotopy method, linear convergence, structured signals recovery.
%\end{IEEEkeywords}

%% The paper must be self-contained. However, if you are referring to
%% a full version for checking certain proofs, please provide the
%% publically accessible location below.  If the paper is completely
%% self-contained, you can remove the following line from your
%% submission.
%\textit{A full version of this paper is accessible at:}
%\url{https://arxiv.org/pdf/21xx.xxxx.pdf}

\section{Introduction}
Multivariate linear regression problems \cite{Yuan2007DimensionReduction} and their variants have received a lot of attention for their diverse applications such as genomics, econometrics, etc. In this paper, we consider one of their variants, the covariate adjusted precision matrix estimation problem \cite{Cai2012CovariateAdj,Chen18CovariateAdj}.

In general multivariate linear regression models, there are $n$ observations $\vy_{i} \in \RR^{m}$ and predictor vectors $\vx_{i} \in \RR^{d}$, and
\begin{equation}
	\yin = \Gammastar^T \xin + \bepi,
\end{equation}
for $i = 1, \cdots, n$, where $\Gammastar \in \RR^{d \times m}$ is the unknown regression coefficient matrix and $\ba{\bepi}_{i = 1}^{n}$ are independent vectors following $\calN(\bmzero, \Sigmastar)$. We could also write this model in the matrix form
\begin{equation} \label{LinearModel_General}
	\bmY = \bmX \Gammastar + \bmE,
\end{equation}
where $\bmX = [\bmx_1, \cdots, \bmx_n]^T \in \RR^{n \times d}$ is the predictor matrix, $\bmY = [\bmy_1, \cdots, \bmy_n]^T \in \RR^{n \times m}$ is the data matrix, and $\bmE = [\bmepsilon_1, \cdots, \bmepsilon_n]^T \in \RR^{n \times m}$ is the noise matrix.

The objective of the covariate adjusted precision matrix estimation problem is to estimate the regression parameter $\Gammastar$ and the precision matrix $\Omegastar  = \Sigmastar^{-1}$ simultaneously. Both of the two parameters provide insights for exploring the interaction among data, especially in the high-dimensional setting. For instance, in graph theory, $\Gammastar$ and $\Omegastar$ represent the directed graph and the undirected graph respectively. The edges of directed graphs indicate casual relationships and those of undirected graphs reveal conditional dependency relationships \cite{Bhlmann2011StatisticForHigh, Lin2016PenalizedM}.

The estimation of precision matrices and regression coefficient matrices has been widely explored in a separate way. For example, estimating precision matrices is the objective of graphical models. Gaussian graphical models \cite{Lauritzen1996Graphical} are routinely applied to infer the precision matrix. They have achieved a great success in practical applications, such as interpreting the conditional independence between genes at the transcriptional level \cite{Segal2005FromSignatures}. In the high-dimensional setting, the ambient dimension might be much larger than the number of observations and additional structural assumptions are required to guarantee the consistent estimation. With the sparsity prior, a neighborhood selection procedure is proposed in \cite{Meinshausen2006HighDimensionalGraphs} and penalized maximum likelihood approaches are also used in \cite{Yuan2007ModelS,Friedman2007SparseInverse,Rothman2008SparseP,Ravikumar2011HighDimensionalC}. On the other hand, in the high-dimensional regime, regression coefficient matrices could be estimated through least squares combined with structural information such as the reduced rank \cite{Yuan2007DimensionReduction} and the group sparsity \cite{Obozinski2011SupportU}.

Besides the respective success, considering regression parameters and precision matrices jointly could even lead to a better result in many application scenarios. When applying the Gaussian graphical model to gene expression data, the introduction of genetic variants as the regression parameter would benefit the interpretation of gene regulation relationships \cite{Cheung2002TheGenetics, Brem2005TheLandscape}. In \cite{Lin2016PenalizedM}, the influence from the key macroeconomic indicators to the returns of financial assets is modeled as regression parameters and the co-dependency relationships between the economic variables and the returns could be viewed as undirected edges in the layered network structures.

Compared with the diverse applications, the theoretical guarantee for the covariate adjusted precision matrix estimation is still being studied. Rothman et~al. \cite{Rothman2010SparseMultivariate} use the multivariate regression with covariance estimation (MRCE) method to estimate the regression parameters with the incorporation of the covariance information. In \cite{Yin2011ASparseConditional}, Yin and Li introduce a sparse conditional Gaussian graphical model (cGGM) to estimate the sparse gene expression network and provide the asymptotic convergence result for the penalized likelihood estimation. Lee and Liu also consider the penalized maximum likelihood estimator for the joint estimation and explore its asymptotic convergence property in \cite{Lee2012Simultaneous}. Both \cite{Yin2011ASparseConditional} and \cite{Lee2012Simultaneous} only consider the asymptotic properties of the estimators, and neither of them explores the optimization performance guarantee for the algorithms. Compared with the mentioned asymptotic analysis, Cai et~al. provide the non-asymptotic analysis for the statistical error of a two-stage procedure to jointly estimate the regression coefficients and the precision matrix in \cite{Cai2012CovariateAdj}, while there is no algorithmic analysis about the algorithm. At the same time, the two-stage approach might lose the interdependency between the two parameters, as stated in \cite{Chen18CovariateAdj}. To the best of our knowledge, \cite{Chen18CovariateAdj} is the only work providing the non-asymptotic optimization performance guarantee for the algorithm to solve the covariate-adjusted precision matrix estimation problem. However, their analysis is based on an impractical resampling assumption, which requires a fresh batch of samples for each iteration. Moreover, their theoretical results are not sharp, since there is an additional logarithmic factor in the finial estimation error compared with the minimax rate and there is also an additional logarithmic factor in the requirement of measurements compared with the minimal requirement.

In this paper, we first improve the analysis of the alternating gradient descent with hard thresholding applied to the covariate adjusted precision matrix estimation problem in \cite{Chen18CovariateAdj} in the following three aspects:  (1): By introducing the generic chaining, our analysis removes the impractical resampling assumption used in \cite{Chen18CovariateAdj}, which leads to a sharper analysis for this algorithm. More precisely, our analysis illustrates that this algorithm not only converges linearly in the absence of convexity, but also attains the minimax rate. At the same time, the requirement of samples to guarantee the successful recovery also matches the order of the minimal requirement. (2): We theoretically demonstrate that the increase of samples will accelerate the convergence rate of this algorithm, which reveals that a time-data tradeoff exists for this problem. (3): Considering the non-convex property of this model, we also suggest a simplified initialization procedure with less input parameters, which could make the whole algorithm achieve a better performance.  We then generalize our analysis framework to the alternating projected gradient descent with general convex structural constraints. Our analysis shows that the class of algorithms enjoys a similar performance with alternating projected gradient descent with non-convex structural constraints.

%In this paper, we improve the analysis for the alternating gradient descent with hard thresholding applied to the covariate adjusted precision matrix estimation problem in \cite{Chen18CovariateAdj}. First, our non-asymptotic optimization performance guarantee for this model does not rely on the resampling assumption. Secondly, through the generic chaining, we illustrate this algorithm not only converges linearly, but also attains the minimax rate. At the same time, the requirement of samples to guarantee the successful recovery also matches the order of the minimal requirement. Thirdly, we demonstrate that the increase of samples could accelerate the convergence rate of this algorithm. This phenomenon illustrates that the time-data tradeoffs exist for this problem. Considering the non-convex property of this model, we also propose a simplified initialization algorithm with less input parameters, which could make the whole algorithm converge faster. Furthermore, we extend our analysis framework to the alternating projected gradient descent for the models with general structural priors.

%\input{tex/Algorithm}
\section{Model and Algorithm}
To estimate the regression coefficient matrix $\Gammastar$ and the precision matrix $\Omegastar$ in \eqref{LinearModel_General} jointly, we consider the maximum likelihood estimator according to the Gaussian mapping. Based on \cite{Yin2011ASparseConditional,Lee2012Simultaneous,Chen18CovariateAdj}, the corresponding conditional negative log-likelihood function for model \eqref{LinearModel_General} could be represented as (neglect the constants)
\begin{equation}
	\begin{split}
		&f_n(\bmGamma, \bmOmega) \\
		%		&= - \log\abs{\bmOmega} + \frac{1}{n} \fnorm{(\bmY - \bmX \bmGamma) \bmOmega^{\frac{1}{2}}}^2 \\
		&= - \log\abs{\bmOmega} + \frac{1}{n} \tr \left\{ (\bmY - \bmX \bmGamma) \bmOmega (\bmY - \bmX \bmGamma)^T \right\}.
	\end{split}
\end{equation}

In the high-dimensional and underdetermined case, we need to refer to the structural information of parameters to guarantee the performance of estimation. The sparsity priors of $\Gammastar$ and $\Omegastar$ have been considered in \cite{Yin2011ASparseConditional, Cai2012CovariateAdj, Lin2016PenalizedM, Chen18CovariateAdj}. In this paper, we follow the line of \cite{Chen18CovariateAdj} and consider the following optimization problems
\begin{equation}
	\begin{split}
		\umin{\bmGamma, \mOmega} & - \log\abs{\bmOmega} + \frac{1}{n} \tr \left\{(\bmY - \bmX \bmGamma) \bmOmega (\bmY - \bmX \bmGamma)^T \right\} \\
		\st &\quad \lzeronorm{\Vect{\bmGamma^T}} \leq \lzeronorm{\Vect{\Gammastar^T}}, \\
		& \quad \lzeronorm{\Vect{\bmOmega^T}} \leq \lzeronorm{\Vect{\Omegastar^T}}.
	\end{split} \label{Sparse_model}
\end{equation}

The key challenge to analyze the model (\ref{Sparse_model}) is that the function $f_n(\bmGamma, \bmOmega)$ is not jointly convex about $\bmGamma$ and $\bmOmega$. There is another line of research \cite{Sohn2012JointEstimation,Yuan2014PartialGaussian,McCarter2014OnSparse, McCarter2016LargeScale} adopting a different parameterization which makes the objective function convex. The difference and comparison between these two models are provided in \cite{Lin2016PenalizedM} and \cite{Chen18CovariateAdj}.

Despite the absence of the joint convexity, the loss function $f_n(\bmGamma, \bmOmega)$ is still bi-convex. The bi-convexity guarantees the loss function is convex with respect to $\mGamma$ $(\mOmega)$ when $\mOmega$ $(\mGamma)$ is fixed. In this way, the alternating method is a natural choice. Alternating methods have been widely used to solve joint estimation problems, latent variable models and matrix factorization problems, such as \cite{Banerjee2017Alt,Banerjee2018ImprovedAlt,Balakrishnan2017StatisticalGuar,Klusowski2019Estimation,Sun2016GuaranteedMatrix}. However, the sharp analysis of the optimization performance guarantee for the model \eqref{Sparse_model} is still absent.

Based on the bi-convex property of (\ref{Sparse_model}), \cite{Chen18CovariateAdj} applies the alternating gradient descent with hard thresholding (Algorithm \ref{Alg_AltIHT}) to jointly estimate $\Gammastar$ and $\Omegastar$. Here $\HT(\mGamma, s)$ represents the hard thresholding operator, which only remains the top $s$ entries of $\mGamma$ in terms of magnitude \cite{Jain2014OnIterativeHard}.

%two sets $\calK_{\Gamma}$ and $\calK_{\Omega}$, where
%\begin{align}
%	\calK_{\Gamma} &\coloneqq \ba{\mGamma \mid \lonenorm{\Vect{\bmGamma^T}} \leq \lonenorm{\Vect{\Gammastar^T}}}, \label{Constraint_Gamma} \\
%	\calK_{\Omega} &\coloneqq \ba{\mOmega \mid \lonenorm{\Vect{\mOmega^T}} \leq \lonenorm{\Vect{\Omegastar^T}}}. \label{Constraint_Omega}
%\end{align}

\begin{algorithm}[htb]
	\caption{Alternating Gradient Descent with Hard Thresholding \cite{Chen18CovariateAdj}}
	\label{Alg_AltIHT}
	\begin{algorithmic}
		\STATE {\bfseries Input:} Iteration number $T$, step size $\eta_{\Gamma}$, $\eta_{\Omega}$, sparsity $s_{\Gamma}$, $s_{\Omega}$.
		\FOR{$t = 0$ {\bfseries to} $T - 1$}
		\STATE $\Gammatp = \HT (\Gammat - \eta_{\Gamma} \nabla_{\Gamma} f_n(\Gammat, \Omegat), s_{\Gamma})$
		\STATE $\Omegatp = \HT(\Omegat - \eta_{\Omega} \nabla_{\Omega} f_n(\Gammat, \Omegat), s_{\Omega})$
		\ENDFOR
		\STATE {\bfseries Output:} $\bmGamma_T$, $\bmOmega_T$
	\end{algorithmic}
\end{algorithm}

Considering the non-convexity of the objective function in (\ref{Sparse_model}), a good initialization (Algorithm \ref{Alg_AltIHT_Initialization}) is required to guarantee the estimation performance. We suggest the following initialization procedure. This procedure can be viewed as a simplification of the one in \cite{Chen18CovariateAdj} by avoiding the use of two unknown parameters $\lambda_{\Gamma}$ and $\lambda_{\Omega}$, which have complicated upper bounds in the supplementary of \cite{Chen18CovariateAdj}.

\begin{algorithm}[htb]
	\caption{Initialization}
	\label{Alg_AltIHT_Initialization}
	\begin{algorithmic}
		\STATE {\bfseries Input:} Sparsity $s_{\Gamma}$, $s_{\Omega}$.
		\STATE $\bmGamma_{\text{ini}} = \argmin_{\lzeronorm{\Vect{\mGamma^T}} \leq s_{\Gamma}} \frac{1}{2} \fnorm{\bmY - \bmX \bmGamma}^2$
		\STATE $\bmS = \frac{1}{n} (\bmY - \bmX \bmGamma_{\text{ini}})^T(\bmY - \bmX \bmGamma_{\text{ini}})$
		\STATE $\bmOmega_{\text{ini}} = \HT(\bmS^{-1}, s_{\Omega})$
		\STATE {\bfseries Output:} $\bmGamma_{\text{ini}}$, $\bmOmega_{\text{ini}}$
	\end{algorithmic}
\end{algorithm}

It is worth noting that the traditional optimization theory predicts that the alternating minimization method could only reach a sublinear rate even for jointly convex loss functions (without strongly convexity) \cite[Theorem 4.1]{Jain2017NonConvexO}. We will show in the next section that if we promote structural priors by projection operations (the hard thresholding operator $\HT(\cdot, s)$ could be viewed as the projection onto the set $\ba{\mGamma \mid \lzeronorm{\Vect{\mGamma^T}} \leq s}$. ), Algorithm \ref{Alg_AltIHT} would enjoy a linear rate even though the loss function in \eqref{Sparse_model} is not jointly convex.

\section{Main Theory}

\subsection{Improved analysis of the alternating gradient descent with hard thresholding}
In this section, we first present an improved analysis of the alternating gradient descent with hard thresholding in \cite{Chen18CovariateAdj}. We begin by introducing two assumptions which are required by our analysis.

%Before diving into the main theoretical results, we first present two assumptions required by our analysis.
\begin{assumption} \label{SingularValue_Omega}
	The rows of $\bmE$ are independent with the distribution $\calN(\bmzero, \Omegastar^{-1})$. We suppose the eigenvalues of $\Omegastar$ satisfy
	\begin{equation}
		\nu_{\min} \leq \lambda_{\min}(\Omegastar) \leq \lambda_{\max}(\Omegastar) \leq \nu_{\max},
	\end{equation}
	where $\nu_{\min} > 0$.
\end{assumption}
%This assumption is also declared in \cite{Cai2012CovariateAdj,Lee2012Simultaneous,Chen18CovariateAdj}.

\begin{assumption} \label{SingularValue_Sigma}
	Suppose $\bmX$ is independent with $\bmE$ and the rows of $\bmX$ are independent following the distribution $\calN(\bmzero, \bmSigma_{\bmX})$. Further, the eigenvalues of $\bmSigma_{\bmX}$ satisfy
	\begin{equation}
		\tau_{\min} \leq \lambda_{\min}(\bmSigma_{\bmX}) \leq \lambda_{\max}(\bmSigma_{\bmX}) \leq \tau_{\max},
	\end{equation}
	where $\tau_{\min} > 0$.
\end{assumption}

The Gaussian assumption about $\bmX$ is required by the Hanson-Wright inequality \cite{Rudelson2013HansonWri} used in the proofs of Lemma \ref{Quadratic_form_XX} and \ref{Quadratic_form_Independent} (in supplementary material). This assumption could be extended to the case where $\Vect{\bmX^T}$ satisfies the convex concentration property \cite{Adamczak2015ANoteOnTheHanson}.

\begin{remark}[Comparison with assumptions in \cite{Chen18CovariateAdj}]
	In \cite{Chen18CovariateAdj}, the eigenvalues of $\Gammastar$ and $\Omegastar$ are required to satisfy $1 / \nu \leq \lambda_{\min}(\Omegastar) \leq \lambda_{\max}(\Omegastar) \leq \nu$ and $1 / \tau \leq \lambda_{\min}(\bmSigma_{\bmX}) \leq \lambda_{\max}(\bmSigma_{\bmX}) \leq \tau$, where $\nu \geq 1$, $\tau \geq 1$. Their assumptions only adapt to the case where the eigenvalues of $\Gammastar$ and $\Omegastar$ are centered around $1$. If all the eigenvalues deviate from $1$, large $\nu$ and $\tau$ are required, which would lead to pessimistic steps in the algorithm. Our Assumption \ref{SingularValue_Omega} and \ref{SingularValue_Sigma} are not only weaker than the ones in \cite{Chen18CovariateAdj}, but also adapt to more general $\Gammastar$ and $\Omegastar$. Moreover, the analysis in \cite{Cai2012CovariateAdj} and \cite{Chen18CovariateAdj} requires $\linfnorm{\Omegastar} \leq M$, where $\linfnorm{\Omegastar} = \max_{1 \leq i \leq m} \sum_{j = 1}^{m} (\Omegastar)_{ij}$. Our analysis does not rely on this condition.
\end{remark}

%In \cite{Chen18CovariateAdj}, the authors also require $\linfnorm{\Omegastar} \leq M$, where $\linfnorm{\Omegastar} = \max_{1 \leq i \leq m} \sum_{j = 1}^{m} (\Omegastar)_{ij}$. Our analysis does not rely on this condition.

Then we introduce some notations that are useful for our analysis.
\begin{definition}[Gaussian width]
	The Gaussian width is a simple way to quantify the size of a set $\calC$
	\begin{equation*}
		\gwidth{\calC} \coloneqq \EE \usup{\bmx \in \calC} \iprod{\bmg}{\bmx}, \text{ where } \bmg \sim \calN(\bmzero, \bmI).
	\end{equation*}
\end{definition}

In our analysis, we would frequently use the Gaussian widths of two sets, $\calC_{2s_{\Gamma}} \cap \calS^{dm - 1}$ and $\calC_{2s_{\Omega}} \cap \calS^{m^2 - 1}$. Here $\calS^{dm - 1}$ and $\calS^{m^2 - 1}$ represent the spheres with unit Frobenius norm in $\RR^{d \times m}$ and $\RR^{m \times m}$ respectively. $\calC_{2s_{\Gamma}}$ and $\calC_{2s_{\Omega}}$ are two sets defined as
\begin{align}
	\calC_{2s_{\Gamma}} &\coloneqq \ba{\mGamma \in \RR^{d \times m} \mid \lzeronorm{\Vect{\mGamma^T}} \leq 2 s_{\Gamma}}, \label{Sparsity_Gamma} \\
	\calC_{2s_{\Omega}} &\coloneqq \ba{\mOmega \in \RR^{m \times m} \mid \lzeronorm{\Vect{\mOmega^T}} \leq 2 s_{\Omega}}. \label{Sparsity_Omega}
\end{align}
%where $\cone(\calC)$ represents the conic hull of the set $\calC$, $\calK_{\Gamma}$ and $\calK_{\Omega}$ are defined in \eqref{Constraint_Gamma} and \eqref{Constraint_Omega}.

For simplicity, we write $\omega_{\Gamma} = \gwidth{\calC_{2s_{\Gamma}} \cap \calS^{dm - 1}}$ and $\omega_{\Omega} = \gwidth{\calC_{2s_{\Omega}} \cap \calS^{m^2 - 1}}$ in the remained part.

%We define $\calC_{2s_{\Gamma}} = \ba{\bmGamma \in \RR^{d \times m} \mid \lzeronorm{\Vect{\bmGamma^T}} \leq 2 s_{\Gamma}}$ which is the set composing of matrices with at most $2 s_{\Gamma}$ nonzero entries. Similarly, we set $\calC_{2s_{\Omega}} = \ba{\bmOmega \in \RR^{m \times m} \mid \lzeronorm{\Vect{\bmOmega^T}} \leq 2 s_{\Omega}}$. For simplicity, we write $\omega_{\Gamma} = \gwidth{\calC_{2s_{\Gamma}} \cap \calS^{dm - 1}}$ and $\omega_{\Omega} = \gwidth{\calC_{2s_{\Omega}} \cap \calS^{m^2 - 1}}$.

We are now ready to exhibit the non-asymptotic optimization performance guarantee of the alternating gradient descent with hard thresholding (Algorithm \ref{Alg_AltIHT}) for the problem \eqref{Sparse_model}.

\begin{theorem}[Linear convergence] \label{ConvergenceAltIHT}
	Suppose the numbers of non-zero entries of $\Gammastar$ and $\Omegastar$ are $s_{\Gamma}^{\star}$ and $s_{\Omega}^{\star}$ respectively. Under Assumption \ref{SingularValue_Omega} and \ref{SingularValue_Sigma}, let $R = \min(\tau_{\min} \nu_{\min} / (2 \tau_{\max}), 1 / (8 \tau_{\max} \nu_{\max}^2), 1)$. Algorithm \ref{Alg_AltIHT} starts from $\bmGamma_0$ and $\bmOmega_0$ satisfying $\max(\fnorm{\bmGamma_0 - \Gammastar}, \fnorm{\bmOmega_0 - \Omegastar}) \leq R$. We set $s_{\Gamma} \geq (1 + 4(1 / \rho_{\text{pop}} - 1)^2) s_{\Gamma}^{\star}$, $s_{\Omega} \geq (1 + 4(1 / \rho_{\text{pop}} - 1)^2) s_{\Omega}^{\star}$, and set the step sizes as
	\begin{align}
		\eta_{\Gamma} &= \frac{1}{\nu_{\max} \tau_{\max} + \nu_{\min} \tau_{\min}}, \\
		\eta_{\Omega} &= \frac{8 \nu_{\max}^2 \nu_{\min}^2 }{16 \nu_{\max}^2 + \nu_{\min}^2 }.
	\end{align}
	If the number of measurements satisfies
	\begin{equation}
		n \geq C_1 \frac{(\omega_{\Gamma} + \omega_{\Omega} + u)^2}{\rho_{\text{pop}}(1 - \sqrt{\rho_{\text{pop}}})^2 R^2} ,
	\end{equation}
	the alternating gradient descent with hard thresholding (Algorithm \ref{Alg_AltIHT}) would converge linearly and each iteration obeys
	\begin{equation}
		\Deltatp \leq  \rho^{t + 1} \Delta_0 + \frac{\epsilon}{1 - \rho},
	\end{equation}
	with probability $1 - 14 \exp(- u^2)$, where $\Deltat = \max(\fnorm{\Gammat - \Gammastar}, \fnorm{\Omegat - \Omegastar})$, $\rho = \sqrt{\rho_{\text{pop}}} + \rho_{\text{sam}}$,
	\begin{align*}
		\rho_{\text{pop}} &\leq \max  \bBa{1 - \frac{\tau_{\min} \nu_{\min}}{\tau_{\max} \nu_{\max} + \tau_{\min} \nu_{\min}} , \\
			&~~~~~~~~~~~~ 1 - \frac{\nu_{\min}^2}{16 \nu_{\max}^2 + \nu_{\min}^2}} , \\
		\rho_{\text{sam}} &\leq \frac{C_2}{\sqrt{\rho_{\text{pop}}}} \frac{\omega_{\Gamma} + \omega_{\Omega} + u}{\sqrt{n}},
	\end{align*}
	%	\begin{equation}
	%		\begin{split}
	%			\rho_{\text{pop}} &\leq \max(1 - \frac{\tau_{\min} \nu_{\min}}{\tau_{\max} \nu_{\max} + \tau_{\min} \nu_{\min}}, \\
	%			&\qquad 1 - \frac{\nu_{\min}^2}{16 \nu_{\max}^2 + \nu_{\min}^2}),
	%		\end{split}
	%	\end{equation}
	%	\begin{equation}
	%		\rho_{\text{sam}} \leq C_2 \frac{\omega_{\Gamma} + \omega_{\Omega} + u}{\sqrt{n}},
	%	\end{equation}
	and
	\begin{equation*}
		\epsilon = \frac{C_3}{\sqrt{\rho_{\text{pop}}}} \max\bBa{\frac{1}{\sqrt{\nu_{\max} \tau_{\max}}} \frac{\omega_{\Gamma} + u}{\sqrt{n}}, \nu_{\min} \frac{\omega_{\Omega} + u}{\sqrt{n}}}.
	\end{equation*}
	Here, $C_1$, $C_2$ and $C_3$ are positive constants without relationships with $\omega_{\Gamma}$, $\omega_{\Omega}$, $n$.
\end{theorem}

\begin{remark}[Comparison with the results in \cite{Chen18CovariateAdj}]
	Compared with \cite{Chen18CovariateAdj}, Theorem 1 improves theirs in the following three aspects. First, our proof does not rely on the impractical resampling assumption, which is used in \cite{Chen18CovariateAdj} to simplify the analysis. Secondly, our estimation error attains the minimax rate and the requirement of samples is also rate-optimal, while there is an additional logarithmic factor in the estimation error and the requirement of samples in \cite{Chen18CovariateAdj} caused by the resampling procedure. Thirdly, our result clearly reveals a time-data tradeoff in this problem.
\end{remark}

\begin{remark}[Time-data tradeoffs]
	It is not hard to find that the component $\rho_{\text{sam}}$ in the convergence rate will decrease as the increase of samples. This implies that with the increase of the number of samples, Algorithm \ref{Alg_AltIHT} will achieve a faster convergence rate, which theoretical demonstrates that a time-data tradeoff exists for the model \eqref{Sparse_model}. It is worth noting that the appearance of $\rho_{\text{sam}}$ is a special product of our analysis. In \cite{Chen18CovariateAdj}, the components of $\rho_{\text{sam}}$ are included in the noise part and they only consider the influence of the population loss function on the convergence rate.
	%	the total convergence rate $\rho$ is composed of two parts, the population part $\rho_{\text{pop}}$ and the sample part $\rho_{\text{sam}}$. Particularly, the sample part $\rho_{\text{sam}}$ indicates the time-data tradeoffs that more data would accelerate the convergence rate.
\end{remark}

\begin{remark}[Sharpness]
	%	Suppose the numbers of non-zero entries of $\Gammastar$ and $\Omegastar$ are $s_{\Gamma}^{\star}$ and $s_{\Omega}^{\star}$ respectively.
	When the Gaussian width $\omega_{\Gamma}$ is dominant, our estimation error about $\Gammastar$ is in the order of \\$\calO(\sqrt{s_{\Gamma}^{\star} \log(edm / s_{\Gamma}^{\star})} / \sqrt{n})$ \cite[Exercise 10.3.8]{vershynin2018HDP}, which is in similar flavor with the results of linear inverse problems \cite{Raskutti2011MinimaxR, Oymak2015SharpMSE, oymak2018sharp}. Additionally, our requirement of measurements is in the order of $s_{\Gamma}^{\star} \log(edm / s_{\Gamma}^{\star})$, which also matches the minimal number of measurements to guarantee the successful recovery in \cite{chandrasekaran2013computational,Amelunxen2013Living}. When the Gaussian width $\omega_{\Omega}$ is dominant, our estimation error about $\Omegastar$ is in the order of $\calO(\sqrt{s_{\Omega}^{\star} \log(em^2 / s_{\Omega}^{\star})} / \sqrt{n})$, which coincides with the minimax lower bound for sparse precision matrix estimation in \cite{Cai2016EstimatingSparse}. However, the estimation error of $\Omegastar$ in \cite{Chen18CovariateAdj} is in the order of $\calO(\sqrt{\log n}\sqrt{s_{\Omega}^{\star} \log(m^2)} / \sqrt{n})$ and there is an additional logarithmic factor compared with the minimax rate. Furthermore, the requirement of measurements in \cite{Chen18CovariateAdj} also has an additional logarithmic factor caused by the resampling step.
\end{remark}

\begin{remark}[Technique to remove resampling]
	To remove the resampling assumption in \cite{Chen18CovariateAdj}, we have introduced the technique of the generic chaining \cite{Talagrand2005TheGenericCh} into our analysis. Actually, similar idea is also used in \cite{Banerjee2018ImprovedAlt}. However, compared with \cite{Banerjee2018ImprovedAlt}, we have considered different observation model with different recovery algorithms. More importantly, we need to develop new mathematical tools to perform our theoretical analysis (e.g., two deviation inequalities: Lemma \ref{Quadratic_form_XX} and \ref{Quadratic_form_Independent} in supplementary material).
	
	%there is non-trivial effort in our analysis. First, the measurement models considered in two works are different.
	%%	The model in \cite{Banerjee2018ImprovedAlt} could be viewed as a vanilla linear model $\vy = \mX \thetastar + \veta$, where $\mX$ has i.i.d. rows and the entries of $\veta$ are correlated. In contrast, both the rows of the data matrix  and the entries of the noise vector would be coupled, when we reformulate the model \eqref{LinearModel_General} to a vanilla linear model.
	%	Secondly, the alternating minimization in \cite{Banerjee2018ImprovedAlt} requires the solution of a constrained least squares problem at each iteration for the regression parameter. Instead, Algorithm \ref{Alg_AltIHT} in our work is composed of two simple hard thresholding operations. Thirdly, considering the different model and algorithm, we develop new mathematical tools to handle the tail bounds of the random processes appeared in our analysis specially (e.g., two deviation inequalities: Lemma \ref{Quadratic_form_XX} and \ref{Quadratic_form_Independent} in supplementary material).
	%	When analyzing this model, Chen et~al. \cite{Chen18CovariateAdj} impose the resampling assumption, that a fresh piece of data is used for every iteration. Under this assumption, $\Gammat$ and $\Omegat$ are viewed as fixed matrices at each iteration, which simplifies the analysis. However, this assumption neither coincides with the practical algorithm nor produces a tight result (an additional logarithmic factor).
\end{remark}

%\begin{remark}[Extension]
%	Combined with Lemma 4.1 in \cite{Li2016StochasticVar}, our analysis also adapt to the covariate adjusted precision matrix estimation problem with sparsity constraint. In this way, our analysis framework could remove the resampling assumption used in the analysis of \cite{Chen18CovariateAdj} and produce the result with the minimax rate. On the other hand, our results could be extended to the case with general structural prior promoted by any convex regularizer as the result in \cite{oymak2018sharp}, for Lemma \ref{Difference_Projection} (in supplementary material) is valid for general convex regularizers.
%\end{remark}

Then, we present the convergence result for the initialization (Algorithm \ref{Alg_AltIHT_Initialization}).

\begin{theorem}[Initialization] \label{InitializationIHT}
	Under Assumption \ref{SingularValue_Omega} and \ref{SingularValue_Sigma}, if the number of measurements satisfies
	\begin{equation}
		\begin{split}
			n &\geq C_4 \frac{(m + \omega_{\Gamma} + u)^2}{R^2} ,
		\end{split}
	\end{equation}
	then the output of Algorithm \ref{Alg_AltIHT_Initialization} satisfies
	\begin{equation}
		\max(\fnorm{\Gammaini - \Gammastar}, \fnorm{\Omegaini - \Omegastar}) \leq R,
	\end{equation}
	with probability at least $1 - 18 \exp(- u^2)$. Here $C_4$ is a positive constant without relationship with $\omega_{\Gamma}$, $m$, $n$.
\end{theorem}

\begin{remark}
	We adopt a different initialization algorithm from \cite{Chen18CovariateAdj} to avoid the selection of two unknown parameters $\lambda_{\Gamma}$ and $\lambda_{\Omega}$. The simulation results illustrate that this initialization could make the whole algorithm achieve a better performance.
\end{remark}

\begin{remark}
	In Theorem 4.7 of \cite{Chen18CovariateAdj}, the requirement of measurements contains the coefficient $d^2$, which is of the same order as $m^2$ in most situations.
	%	We also note that if we adopt the Graphical Lasso estimator to estimate $\bmOmega_{\text{ini}}$, we could expect tighter sample complexity with the price of higher computation complexity.
\end{remark}

\subsection{Extension to the model with general convex constraints}
In many practical applications of machine learning, convex constraints are widely utilized to promote the structures. 
This fact motivates us to extend the above theoretical analysis to the model with general convex constraints.

For the regression parameter $\Gammastar$ and the precision matrix $\Omegastar$ with general structural priors, we promote their structures by two convex functions $\calR_{\Gamma}(\cdot)$ and $\calR_{\Omega}(\cdot)$ respectively, and consider the following optimization problems
\begin{equation}
	\begin{split}
		\umin{\bmGamma, \mOmega} & - \log\abs{\bmOmega} + \frac{1}{n} \tr \left\{(\bmY - \bmX \bmGamma) \bmOmega (\bmY - \bmX \bmGamma)^T \right\} \\
		\st &\quad \calR_{\Gamma}(\mGamma) \leq \calR_{\Gamma}(\Gammastar), \\
		& \quad \calR_{\Omega}(\mOmega) \leq \calR_{\Omega}(\Omegastar).
	\end{split} \label{General_Structure_model}
\end{equation}

Similarly, based on the bi-convex property of (\ref{General_Structure_model}), we apply the alternating projected gradient descent (Algorithm \ref{General_AltPGD}) to jointly estimate $\Gammastar$ and $\Omegastar$. Here the two operators $\calP_{\calK_{\Gamma}}$ and $\calP_{\calK_{\Omega}}$ represent the orthogonal projection onto two sets $\calK_{\Gamma}$ and $\calK_{\Omega}$, where
\begin{align}
	\calK_{\Gamma} &\coloneqq \ba{\mGamma \in \RR^{d \times m} \mid \calR_{\Gamma}(\mGamma) \leq \calR_{\Gamma}(\Gammastar)}, \label{Constraint_Gamma} \\
	\calK_{\Omega} &\coloneqq \ba{\mOmega \in \RR^{m \times m} \mid \calR_{\Omega}(\mOmega) \leq \calR_{\Omega}(\Omegastar)}. \label{Constraint_Omega}
\end{align}

\begin{algorithm}[htb]
	\caption{Alternating Projected Gradient Descent}
	\label{General_AltPGD}
	\begin{algorithmic}
		\STATE {\bfseries Input:} Iteration number $T$, step size $\eta_{\Gamma}$, $\eta_{\Omega}$, constraint set $\calK_{\Gamma}$, $\calK_{\Omega}$.
		\FOR{$t = 0$ {\bfseries to} $T - 1$}
		\STATE $\Gammatp = \calP_{\calK_{\Gamma}}(\Gammat - \eta_{\Gamma} \nabla_{\Gamma} f_n(\Gammat, \Omegat))$
		\STATE $\Omegatp = \calP_{\calK_{\Omega}}(\Omegat - \eta_{\Omega} \nabla_{\Omega} f_n(\Gammat, \Omegat))$
		\ENDFOR
		\STATE {\bfseries Output:} $\bmGamma_T$, $\bmOmega_T$
	\end{algorithmic}
\end{algorithm}

Likewise, considering the non-convexity of the objective function of (\ref{General_Structure_model}), we also refer to an initialization (Algorithm \ref{General_Initialization}) for general structural priors to guarantee the estimation performance.

\begin{algorithm}[htb]
	\caption{Initialization}
	\label{General_Initialization}
	\begin{algorithmic}
		\STATE {\bfseries Input:} Constraint set $\calK_{\Gamma}$, $\calK_{\Omega}$.
		\STATE $\bmGamma_{\text{ini}} = \argmin_{\mGamma \in \calK_{\Gamma}} \frac{1}{2} \fnorm{\bmY - \bmX \bmGamma}^2$
		\STATE $\bmS = \frac{1}{n} (\bmY - \bmX \bmGamma_{\text{ini}})^T(\bmY - \bmX \bmGamma_{\text{ini}})$
		\STATE $\bmOmega_{\text{ini}} = \calP_{\calK_{\Omega}}(\bmS^{-1})$
		%		\STATE {\bfseries Input:} Iteration number $T$.
		%		\FOR{$t = 0$ {\bfseries to} $T - 1$}
		%		\STATE $\Gammatp = \calP_{\calK_{\Gamma}}(\Gammat - \frac{1}{n} \bmX^T(\bmX \Gammat - \bmY))$
		%		\STATE $\bmS = \frac{1}{n} (\bmY - \bmX \Gammatp)^T(\bmY - \bmX \Gammatp)$
		%		\STATE $\Omegatp = \calP_{\calK_{\Omega}}(\bmS^{-1})$
		%		\ENDFOR
		\STATE {\bfseries Output:} $\bmGamma_{\text{ini}}$, $\bmOmega_{\text{ini}}$
	\end{algorithmic}
\end{algorithm}

In the remained analysis, we would frequently use the Gaussian widths of two sets, $\calC_{\Gamma} \cap \calS^{dm - 1}$ and $\calC_{\Omega} \cap \calS^{m^2 - 1}$. Here, $\calC_{\Gamma}$ and $\calC_{\Omega}$ are two descent cones defined as
\begin{align}
	\calC_{\Gamma} &\coloneqq \cone(\calK_{\Gamma} - \Gammastar), \\
	\calC_{\Omega} &\coloneqq \cone(\calK_{\Omega} - \Omegastar),
\end{align}
where $\cone(\calC)$ represents the conic hull of the set $\calC$, $\calK_{\Gamma}$ and $\calK_{\Omega}$ are defined in \eqref{Constraint_Gamma} and \eqref{Constraint_Omega}. For simplicity, we write $\bar{\omega}_{\Gamma} = \gwidth{\calC_{\Gamma} \cap \calS^{dm - 1}}$ and $\bar{\omega}_{\Omega} = \gwidth{\calC_{\Omega} \cap \calS^{m^2 - 1}}$ in the remained part.

We are now ready to exhibit the linear convergence of the alternating projected gradient descent (Algorithm \ref{General_AltPGD}) for the problem \eqref{General_Structure_model}.

\begin{theorem}[Linear convergence] \label{ConvergenceAltPGD}
	Under Assumption \ref{SingularValue_Omega} and \ref{SingularValue_Sigma}, suppose $R = \min(\tau_{\min} \nu_{\min} / (2 \tau_{\max}), 1 / (8 \tau_{\max} \nu_{\max}^2), 1)$. We start from $\bmGamma_0$ and $\bmOmega_0$ satisfying $\max(\fnorm{\bmGamma_0 - \Gammastar}, \fnorm{\bmOmega_0 - \Omegastar}) \leq R$ and set the step sizes as
	\begin{align*}
		\eta_{\Gamma} &= \frac{1}{\nu_{\max} \tau_{\max} + \nu_{\min} \tau_{\min}},~~~ \eta_{\Omega} = \frac{8 \nu_{\max}^2 \nu_{\min}^2 }{16 \nu_{\max}^2 + \nu_{\min}^2 }.
	\end{align*}
	If the number of measurements satisfies
	\begin{equation}
		n \geq C_5 \frac{(\bar{\omega}_{\Gamma} + \bar{\omega}_{\Omega} + u)^2}{(1 - \rho_{\text{pop}})^2 R^2} ,
	\end{equation}
	the alternating projected gradient descent (Algorithm \ref{General_AltPGD}) would converge linearly and each iteration obeys
	\begin{equation}
		\Deltatp \leq  \rho^{t + 1} \Delta_0 + \frac{\epsilon}{1 - \rho},
	\end{equation}
	with probability $1 - 14 \exp(- u^2)$, where $\Deltat = \max(\fnorm{\Gammat - \Gammastar}, \fnorm{\Omegat - \Omegastar})$, $\rho = \rho_{\text{pop}} + \rho_{\text{sam}}$,
	\begin{align*}
		\rho_{\text{pop}} &\leq \max\bBa{1 - \frac{\tau_{\min} \nu_{\min}}{\tau_{\max} \nu_{\max} + \tau_{\min} \nu_{\min}}, \\
			&~~~~~~~~~~~~ 1 - \frac{\nu_{\min}^2}{16 \nu_{\max}^2 + \nu_{\min}^2}}, \\
		\rho_{\text{sam}} &\leq C_6 \frac{\bar{\omega}_{\Gamma} + \bar{\omega}_{\Omega} + u}{\sqrt{n}},
	\end{align*}
	%	\begin{equation}
	%		\begin{split}
	%			\rho_{\text{pop}} &\leq \max(1 - \frac{\tau_{\min} \nu_{\min}}{\tau_{\max} \nu_{\max} + \tau_{\min} \nu_{\min}}, \\
	%			&\qquad 1 - \frac{\nu_{\min}^2}{16 \nu_{\max}^2 + \nu_{\min}^2}),
	%		\end{split}
	%	\end{equation}
	%	\begin{equation}
	%		\rho_{\text{sam}} \leq C_6 \frac{\bar{\omega}_{\Gamma} + \bar{\omega}_{\Omega} + u}{\sqrt{n}},
	%	\end{equation}
	and
	\begin{equation}
		\epsilon = C_7 \max\bBa{\frac{1}{\sqrt{\nu_{\max} \tau_{\max}}} \frac{\bar{\omega}_{\Gamma} + u}{\sqrt{n}}, \nu_{\min} \frac{\bar{\omega}_{\Omega} + u}{\sqrt{n}}}.
	\end{equation}
	Here, $C_5$, $C_6$ and $C_7$ are positive constants without relationships with $\bar{\omega}_{\Gamma}$, $\bar{\omega}_{\Omega}$, $n$.
\end{theorem}

Then, we present the corresponding result for the initialization (Algorithm \ref{General_Initialization}).

\begin{corollary}[Initialization] \label{InitializationPGD}
	Under Assumption \ref{SingularValue_Omega} and \ref{SingularValue_Sigma}, if the number of measurements satisfies
	\begin{equation}
		\begin{split}
			n &\geq C_8 \frac{(m + \bar{\omega}_{\Gamma} + u)^2}{R^2} ,
		\end{split}
	\end{equation}
	then the output of Algorithm \ref{General_Initialization} satisfies
	\begin{equation}
		\max(\fnorm{\Gammaini - \Gammastar}, \fnorm{\Omegaini - \Omegastar}) \leq R,
	\end{equation}
	with probability at least $1 - 18 \exp(- u^2)$. Here $C_8$ is a positive constant without relationship with $\bar{\omega}_{\Gamma}$, $m$, $n$.
\end{corollary}

When $\Omegastar$ is known, the model \eqref{General_Structure_model} degrades to the vanilla multivariate regression problem \eqref{Sparse_model_PGD} and the alternating method reduces to the projected gradient descent (PGD). The details of PGD is provided in Algorithm \ref{General_PGD}, where the constraint set $\calK_{\Gamma}$ is defined as \eqref{Constraint_Gamma}.
\begin{equation}
	\begin{split}
		\umin{\bmGamma} &\quad f_n(\mGamma) = \frac{1}{2n} \tr((\bmY - \bmX \bmGamma) \Omegastar (\bmY - \bmX \bmGamma)^T) \\
		\st &\quad \calR_{\Gamma}(\mGamma) \leq \calR_{\Gamma}(\Gammastar).
	\end{split} \label{Sparse_model_PGD}
\end{equation}

\begin{algorithm}[htb]
	\caption{Projected Gradient Descent}
	\label{General_PGD}
	\begin{algorithmic}
		\STATE {\bfseries Input:} Iteration number $T$, step size $\eta_{\Gamma}$, constraint set $\calK_{\Gamma}$.
		\FOR{$t = 0$ {\bfseries to} $T - 1$}
		\STATE $\Gammatp = \calP_{\calK_{\Gamma}}(\Gammat - \eta_{\Gamma} \nabla f_n(\Gammat))$
		\ENDFOR
		\STATE {\bfseries Output:} $\bmGamma_T$
	\end{algorithmic}
\end{algorithm}

Our analysis in Theorem \ref{ConvergenceAltPGD} naturally adapts to this condition. In Corollary \ref{PGD_Sparse}, we present the optimization performance guarantee of PGD, which could be viewed as an extension of the result in \cite{oymak2018sharp} to the multivariate regression problem.

\begin{corollary}[Linear convergence of PGD] \label{PGD_Sparse}
	Under Assumption \ref{SingularValue_Omega} and \ref{SingularValue_Sigma}, we apply PGD starting from $\bmGamma_0 = \bmzero$ with the step size $\eta_{\Gamma} = 2 / (\tau_{\max} \nu_{\max} + \tau_{\min} \nu_{\min})$. When the number of measurements satisfies
	\begin{equation}
		n \geq C_9 \frac{(\bar{\omega}_{\Gamma} + u)^2}{(1 - \rho_{\text{pop}})^2} ,
	\end{equation}
	we have
	\begin{equation}
		\fnorm{\Gammatp - \Gammastar} \leq \rho \fnorm{\Gammat - \Gammastar} + \epsilon
		%		 \leq \rho^{t + 1} \fnorm{\bmGamma_0 - \Gammastar} + \frac{1}{1 - \rho} \epsilon
	\end{equation}
	with probability at least $1 - 4 \exp(- u^2)$. Here $\rho = \rho_{\text{pop}} + \rho_{\Gamma, \text{sam}}$,
	\begin{align*}
		\rho_{\text{pop}} &\leq 1 - \frac{2 \tau_{\min} \nu_{\min}}{\tau_{\max} \nu_{\max} + \tau_{\min} \nu_{\min}}, \\
		\rho_{\Gamma, \text{sam}} &\leq C_{10} \frac{\bar{\omega}_{\Gamma} + u}{\sqrt{n}},
	\end{align*}
	%\begin{equation}
	%	\rho_{\Gamma, \text{sam}} \leq C_{10} \frac{\bar{\omega}_{\Gamma} + u}{\sqrt{n}}
	%\end{equation}
	and
	\begin{equation}
		\epsilon \leq C_{11} \frac{1}{\sqrt{\tau_{\max} \nu_{\max}}} \frac{\bar{\omega}_{\Gamma} + u}{\sqrt{n}}.
	\end{equation}
	Here, $C_9$, $C_{10}$, and $C_{11}$ are positive constants without relationship with $\bar{\omega}_{\Gamma}$, $n$.
\end{corollary}

%\begin{remark}
%	The result in Corollary \ref{PGD_Sparse} could be extended to the case with general structural prior promoted by any convex regularizer and it could be viewed as an extension of the result in \cite{oymak2018sharp} to the multivariate regression problem.
%%	Compared with the traditional analysis of IHT in \cite{Jain2014OnIterativeHard}, our analysis reveals the time-data tradeoffs directly. The result in \cite{oymak2018sharp} could be applied to the analysis of IHT. However, its analysis is based on the linear regression problem, which is a special case of the multivariate regression problem we consider in this paper.
%\end{remark}

\section{Experiments}
%\subsection{Synthetic data}
In this section, we verify our theoretical results with numerical simulations. Through the experiments, the support of $\Gammastar$ is selected at random and its entries have i.i.d $\calN(0, 1)$ values. In our initialization algorithm, we perform 2 projected gradient descent iterations. All simulations are run on a PC with Intel i5-6500 and 16GB memory. 
\subsection{Comparison of estimation error and running time}
In this part, we compare the estimation error and the running time of three methods. The first is the method in \cite{Chen18CovariateAdj}. The second is Algorithm \ref{Alg_AltIHT} and our initialization Algorithm \ref{Alg_AltIHT_Initialization}. The third is Algorithm \ref{General_AltPGD} and \ref{General_Initialization} with the $l_1$-norm as the regularizers.

We consider three scenarios. The rows of the predictor matrix $\bmX$ are generated independently from the distribution $\calN(\vzero, \mSigma_{\mX})$. The covariance matrix $\mSigma_{\mX}$ follows a band graph, where $\mSigma_{ii}^{\mX} = 0.5$, $\mSigma_{i, i + 1}^{\mX} = \mSigma_{i + 1, i}^{\mX} = 0.15$ and $\mSigma_{ij}^{\mX} = 0$, for $\abs{i - j} > 1$. The precision matrix also follows a band graph, where $\bmOmega_{ii}^{\star} = 0.6$, $\bmOmega_{i, i + 1}^{\star} = \bmOmega_{i + 1, i}^{\star} = 0.18$ and $\bmOmega_{ij}^{\star} = 0$, for $\abs{i - j} > 1$. We set $s_{\Gamma}^{\star} = 200$ and record the average running time and the average relative estimation errors of 50 experiments.
\begin{table}[ht]
	\caption{Comparison between three methods.}
	\label{Initialization_Comparision}
	\vskip -0.15in
	\begin{center}
		\begin{small}
			\begin{tabular}{lccr}
				\toprule
				&\multicolumn{3}{c}{$n = 6000, m = 100, d = 100$}\\
				Methods& $\frac{\fnorm{\hat{\mGamma} - \Gammastar}}{\fnorm{\Gammastar}}$ & $\frac{\fnorm{\hat{\mOmega} - \Omegastar}}{\fnorm{\Omegastar}}$ & Time \\
				\midrule
				\cite{Chen18CovariateAdj}& 0.034 & 0.024 & 55.98\\
				Algorithm \ref{Alg_AltIHT} and \ref{Alg_AltIHT_Initialization}& \textbf{0.033} & \textbf{0.023} & \textbf{4.02}\\
				Algorithm \ref{General_AltPGD} and \ref{General_Initialization}& 0.055 & 0.062 & \textbf{3.67}\\
				\midrule
				&\multicolumn{3}{c}{$n = 18000, m = 150, d = 150$}\\
				Methods& $\frac{\fnorm{\hat{\mGamma} - \Gammastar}}{\fnorm{\Gammastar}}$ & $\frac{\fnorm{\hat{\mOmega} - \Omegastar}}{\fnorm{\Omegastar}}$ & Time \\
				\midrule
				\cite{Chen18CovariateAdj}& 0.102 & 0.017 & 165.77\\
				Algorithm \ref{Alg_AltIHT} and \ref{Alg_AltIHT_Initialization}& \textbf{0.018} & \textbf{0.014} & \textbf{12.96}\\
				Algorithm \ref{General_AltPGD} and \ref{General_Initialization}& 0.035 & 0.041 & \textbf{12.18}\\
				\midrule
				&\multicolumn{3}{c}{$n = 20000, m = 200, d = 200$} \\
				Methods& $\frac{\fnorm{\hat{\mGamma} - \Gammastar}}{\fnorm{\Gammastar}}$ & $\frac{\fnorm{\hat{\mOmega} - \Omegastar}}{\fnorm{\Omegastar}}$ & Time \\
				\midrule
				\cite{Chen18CovariateAdj}& 0.104 & 0.016 & 235.18\\
				Algorithm \ref{Alg_AltIHT} and \ref{Alg_AltIHT_Initialization}& \textbf{0.017} & \textbf{0.013} & \textbf{21.04}\\
				Algorithm \ref{General_AltPGD} and \ref{General_Initialization}& 0.035 & 0.041 & \textbf{19.97}\\
				\bottomrule
			\end{tabular}
			%			\end{sc}
		\end{small}
	\end{center}
	\vskip -0.1in
\end{table}

In Table \ref{Initialization_Comparision}, the smaller estimation error and less running time of Algorithm \ref{Alg_AltIHT} and \ref{Alg_AltIHT_Initialization} (compared with the method in \cite{Chen18CovariateAdj}) come from the different initialization procedures. The larger estimation error of Algorithm \ref{General_AltPGD} and \ref{General_Initialization} (compared with Algorithm \ref{Alg_AltIHT} and \ref{Alg_AltIHT_Initialization}) is because we use the convex $l_1$-norm as a surrogate of the nonconvex $l_0$-norm.

\subsection{Comparison of requirement for samples to guarantee successful recovery}
In this part, we illustrate how many samples are required to guarantee the successful recovery by three methods. The first is the method in \cite{Chen18CovariateAdj} labeled as AltIHT. The second is Algorithm \ref{Alg_AltIHT} and our initialization Algorithm \ref{Alg_AltIHT_Initialization}. The third is Algorithm \ref{General_AltPGD} and \ref{General_Initialization} with the $l_1$-norm as the regularizers.

We set $d = m = 50$, $s_{\Gamma}^{\star} = 200$. The rows of the predictor matrix $\bmX$ are generated independently from the distribution $\calN(\vzero, \mI_{d})$. The precision matrix follows a block diagonal graph. Every block has the format $(\begin{smallmatrix}
	1 & 0.2 \\ 0.2 & 1
\end{smallmatrix})$. We record the empirical success rate averaged over 100 replications. Here a replication is successful if the relative estimation errors of $\Gammastar$ and $\Omegastar$ satisfy $\fnorm{\hat{\mGamma} - \Gammastar} / \fnorm{\Gammastar} < 10^{-1}$ and $\fnorm{\hat{\mOmega} - \Omegastar} / \fnorm{\Omegastar} < 10^{-1}$.
\begin{figure}[ht]
	%	\centering
	\vskip -0.1in
	%\captionsetup{aboveskip=0pt}
	\centerline{\includegraphics[width=0.5\linewidth]{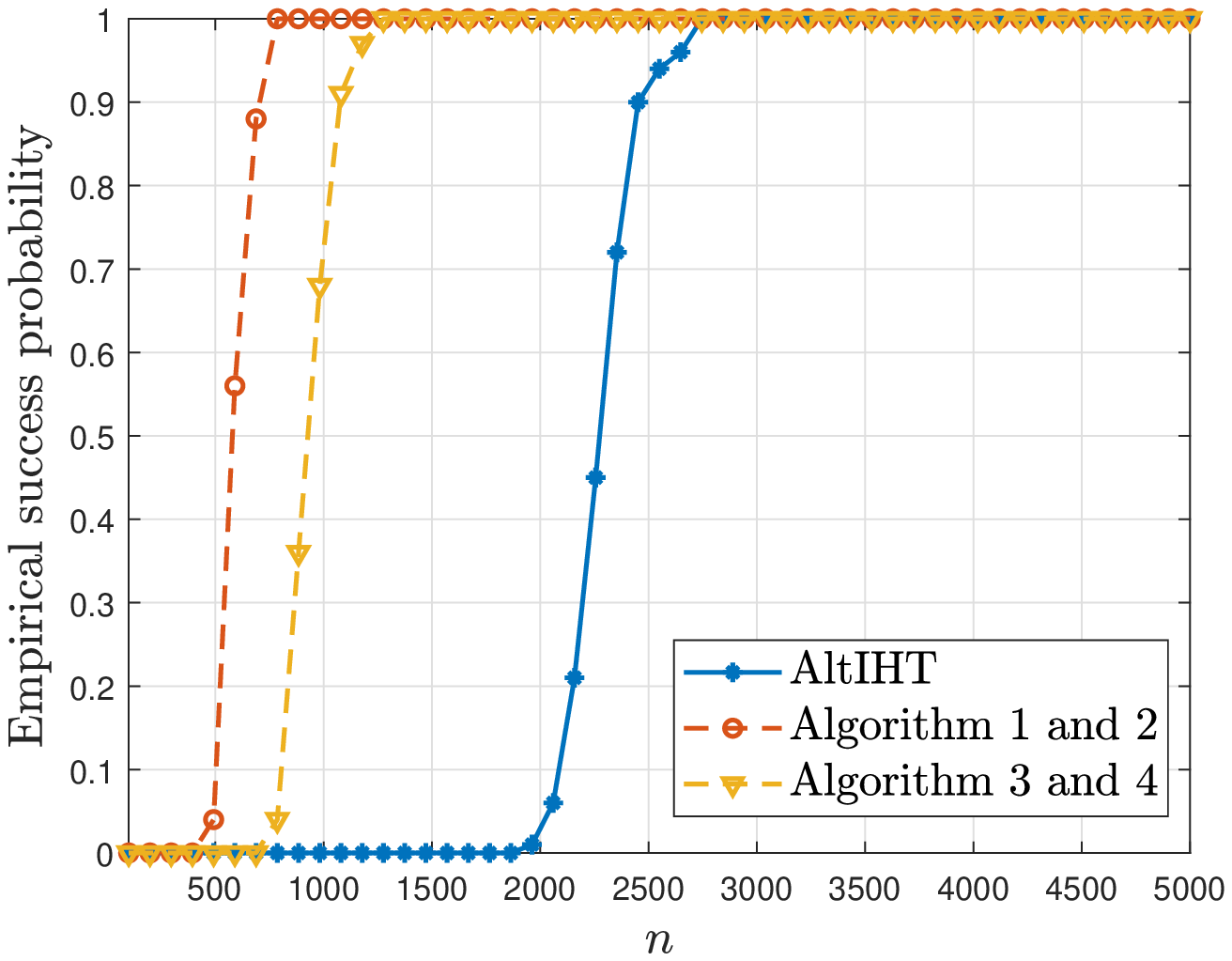}}
	\vskip -0.1in
	\caption{Empirical success rates of three methods under different number of samples.}
	\label{FigEmpiricalSuccessfulRate}
\end{figure}

In Figure \ref{FigEmpiricalSuccessfulRate}, the method of Algorithm \ref{Alg_AltIHT} and \ref{Alg_AltIHT_Initialization} benefits from our initialization and requires the least samples. Though the method of Algorithm \ref{General_AltPGD} and \ref{General_Initialization} also adopts our initialization, it requires more samples because of using the $l_1$-norm instead of the nonconvex $l_0$-norm. This point also matches the phenomenon that the $l_0$-norm would lead to a sharper phase transition curve for linear inverse problems in \cite{oymak2018sharp}. The benefit of our initialization could also be verified from the fact that the original AltIHT in \cite{Chen18CovariateAdj} requires the most samples.
%we present the empirical success rates of two algorithms under different number of samples. Here we label the algorithm in \cite{Chen18CovariateAdj} as AltIHT. Figure \ref{FigEmpiricalSuccessfulRate} displays that our method requires less samples to guarantee the successful recovery compared with the algorithm in \cite{Chen18CovariateAdj}.

\subsection{Time-data tradeoffs}
To verify the time-data tradeoffs phenomenon, we perform Algorithm \ref{Alg_AltIHT} and our initialization (Algorithm \ref{Alg_AltIHT_Initialization}) under different numbers of measurements $n_1 = 3000$, $n_2 = 4000$, $n_3 = 5000$. We set $d = m = 100$, $s_{\Gamma}^{\star} = 400$. The rows of the predictor matrix $\bmX$ are generated independently from the distribution $\calN(\vzero, \mI_{d})$. The precision matrix follows a band graph, where $\bmOmega_{ii}^{\star} = 1$, $\bmOmega_{i, i + 1}^{\star} = \bmOmega_{i + 1, i}^{\star} = 0.4$ and $\bmOmega_{ij}^{\star} = 0$, for $\abs{i - j} > 1$. Each scenario is repeated for 50 trials.
%In the initialization algorithm, two projected gradient descent operations are conducted.
\begin{figure}[ht]
	%	\centering
	\vskip -0.1in
	%\captionsetup{aboveskip=0pt}
	\centerline{\subfigure[]{\includegraphics[width=0.5\linewidth]{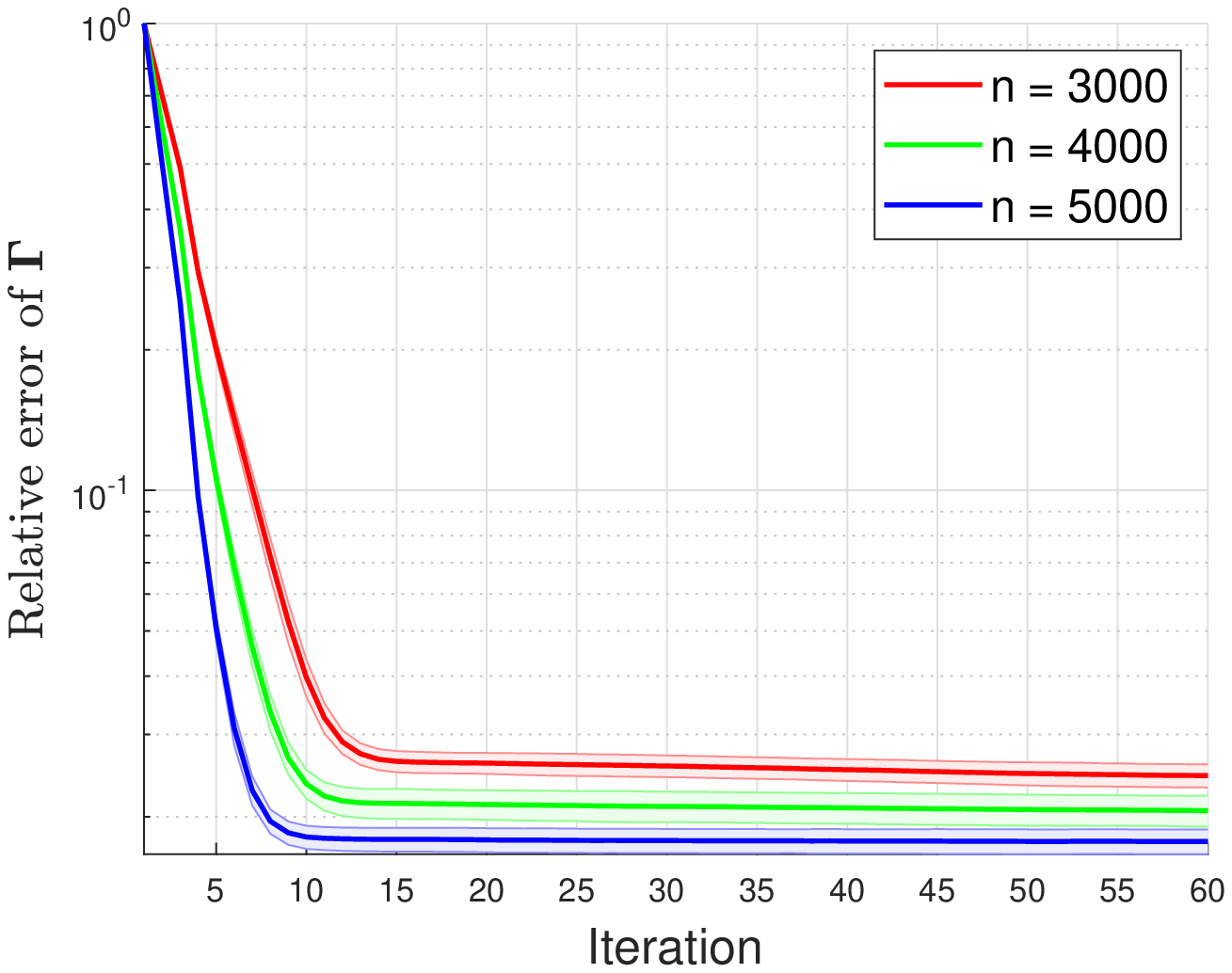}
			% where an .eps filename suffix will be assumed under latex,
			% and a .pdf suffix will be assumed for pdflatex
			\label{Tradeoff_Gamma}}
		%		\hspace{0.1em}%
		\hfill
		\subfigure[]{\includegraphics[width=0.5\linewidth]{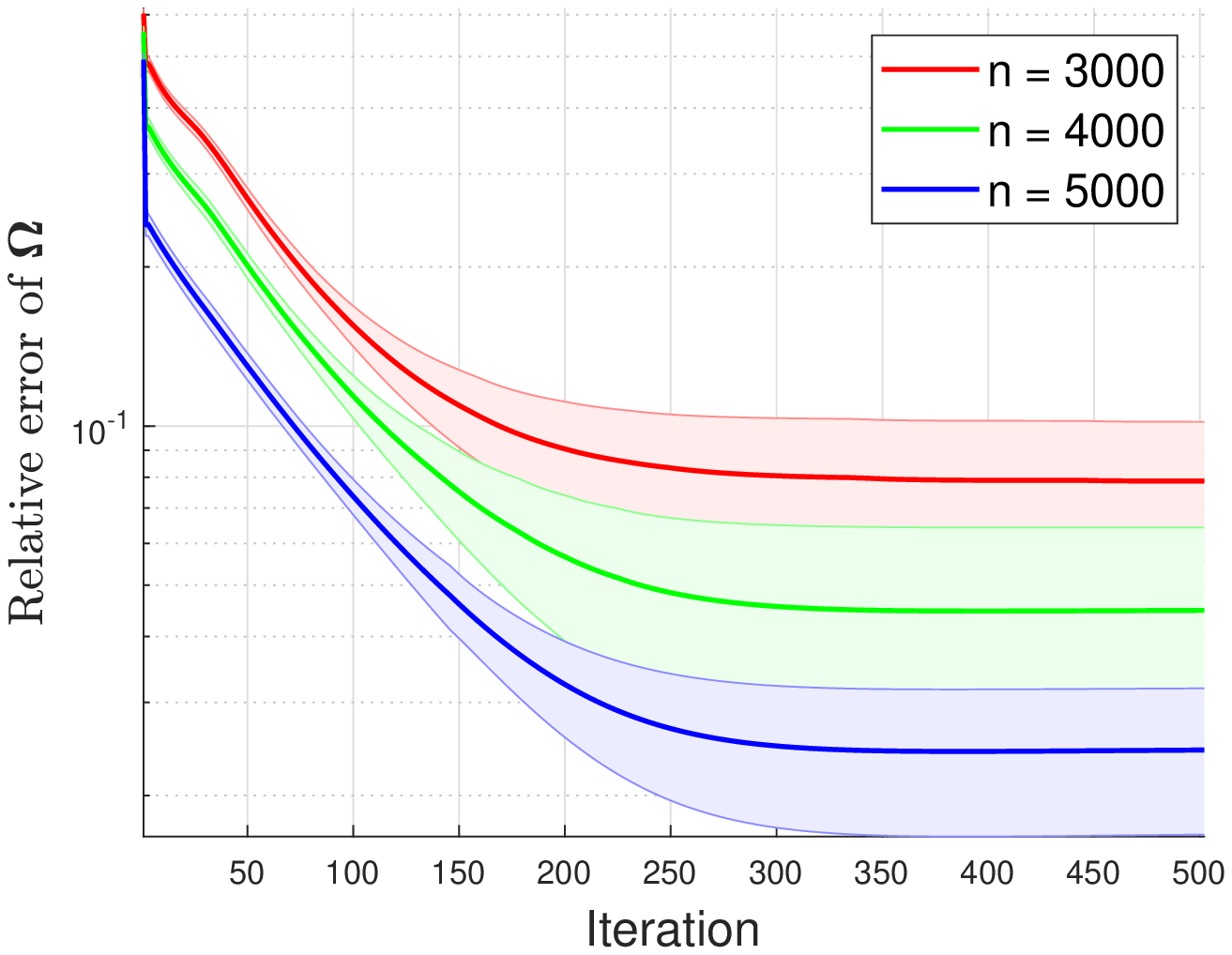}
			% where an .eps filename suffix will be assumed under latex,
			% and a .pdf suffix will be assumed for pdflatex
			\label{Tradeoff_Omega}}}
	\vskip -0.1in
	\caption{(a) Convergence of $\fnorm{\Gammat - \Gammastar} / \fnorm{\Gammastar}$. (b) Convergence of $\fnorm{\Omegat - \Omegastar} / \fnorm{\Omegastar}$.}
	\label{FigTradeoff}
\end{figure}

In Figure \ref{Tradeoff_Gamma} and \ref{Tradeoff_Omega}, we present the convergence results for $\fnorm{\Gammat - \Gammastar} / \fnorm{\Gammastar}$ and $\fnorm{\Omegat - \Omegastar} / \fnorm{\Omegastar}$. From the figures we could illustrate more data would lead to faster convergence rates and smaller estimation errors, which support the theoretical result in Theorem \ref{ConvergenceAltIHT}. For Algorithm \ref{General_AltPGD} and \ref{General_Initialization} with the $l_1$-norm, the results are similar and we do not include them in this manuscript.
\subsection{Statistical estimation error}
In this part, we verify the scaling of the statistical estimation error of Algorithm \ref{Alg_AltIHT} and our initialization (Algorithm \ref{Alg_AltIHT_Initialization}).

We consider two different scenarios, the $\Gammastar$-sparsity dominated case and the $\Omegastar$-sparsity dominated case. For the $\Gammastar$-sparsity dominated case, we set $d = m = 50$ and consider $s_{\Gamma}^{\star} = 200, 250, 300$ three conditions.
%The number of measurements varies from $1200$ to $3400$.
For the $\Omegastar$-sparsity dominated case, we set $d = 50$ and consider $m = 56, 66, 76$ three conditions corresponding to $s_{\Omega}^{\star} = 112, 132, 152$.
%The number of measurements varies from $1500$ to $4100$.
The rows of the predictor matrix $\bmX$ are generated independently from the distribution $\calN(\vzero, \mI_{d})$. The precision matrix follows a block diagonal graph. Every block has the format $(\begin{smallmatrix}
	1 & 0.3 \\ 0.3 & 1
\end{smallmatrix})$. Each scenario is repeated for 400 trials.

\begin{figure}[ht]
	%	\centering
	\vskip -0.1in
	%\captionsetup{aboveskip=0pt}
	\centerline{\subfigure[Original numbers of measurements]{\includegraphics[width=0.5\linewidth]{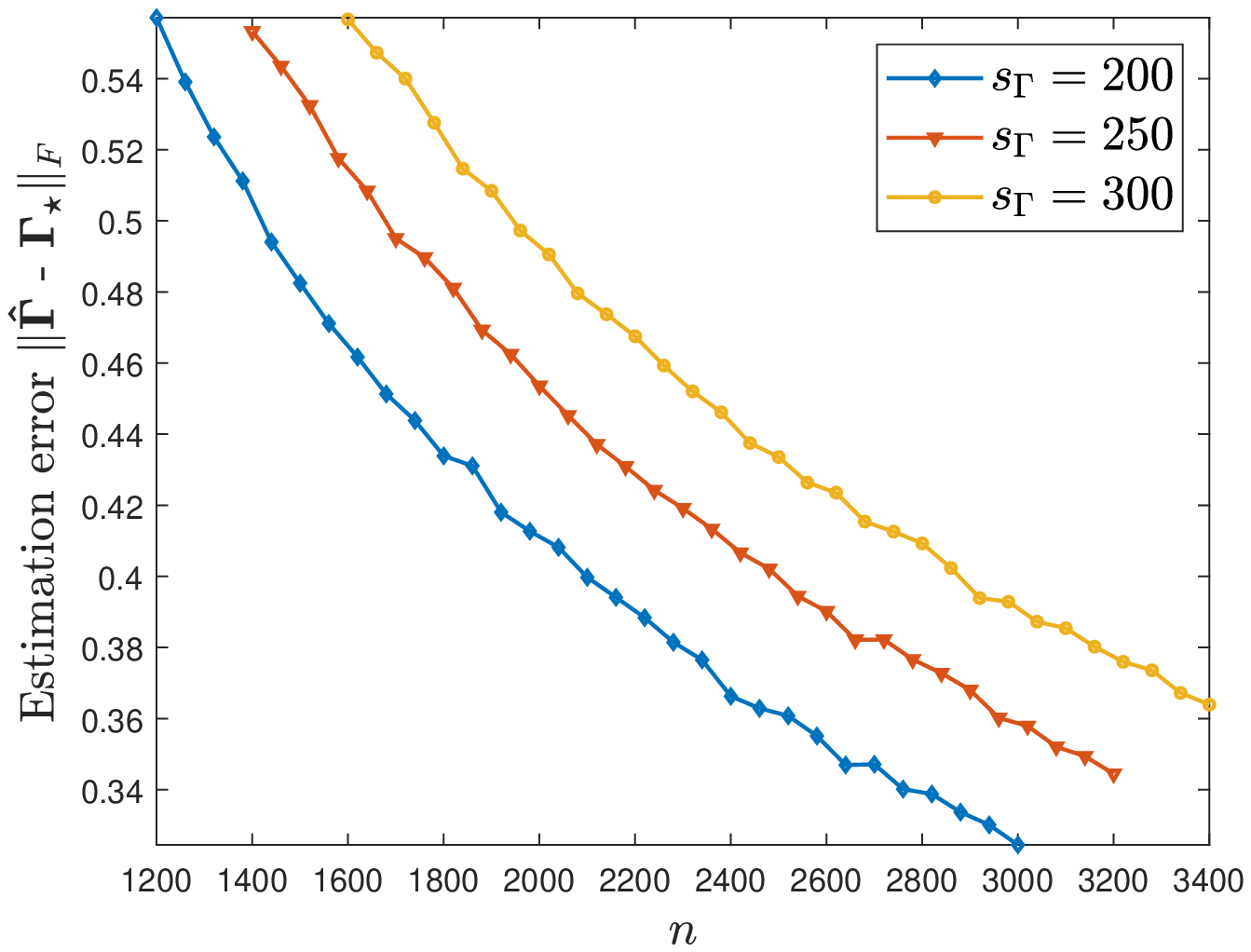}
			% where an .eps filename suffix will be assumed under latex,
			% and a .pdf suffix will be assumed for pdflatex
			\label{ErrorScaleGammaOriginal}}
		%		\hspace{0.1em}%
		\hfill
		\subfigure[Rescaled numbers of measurements]{\includegraphics[width=0.5\linewidth]{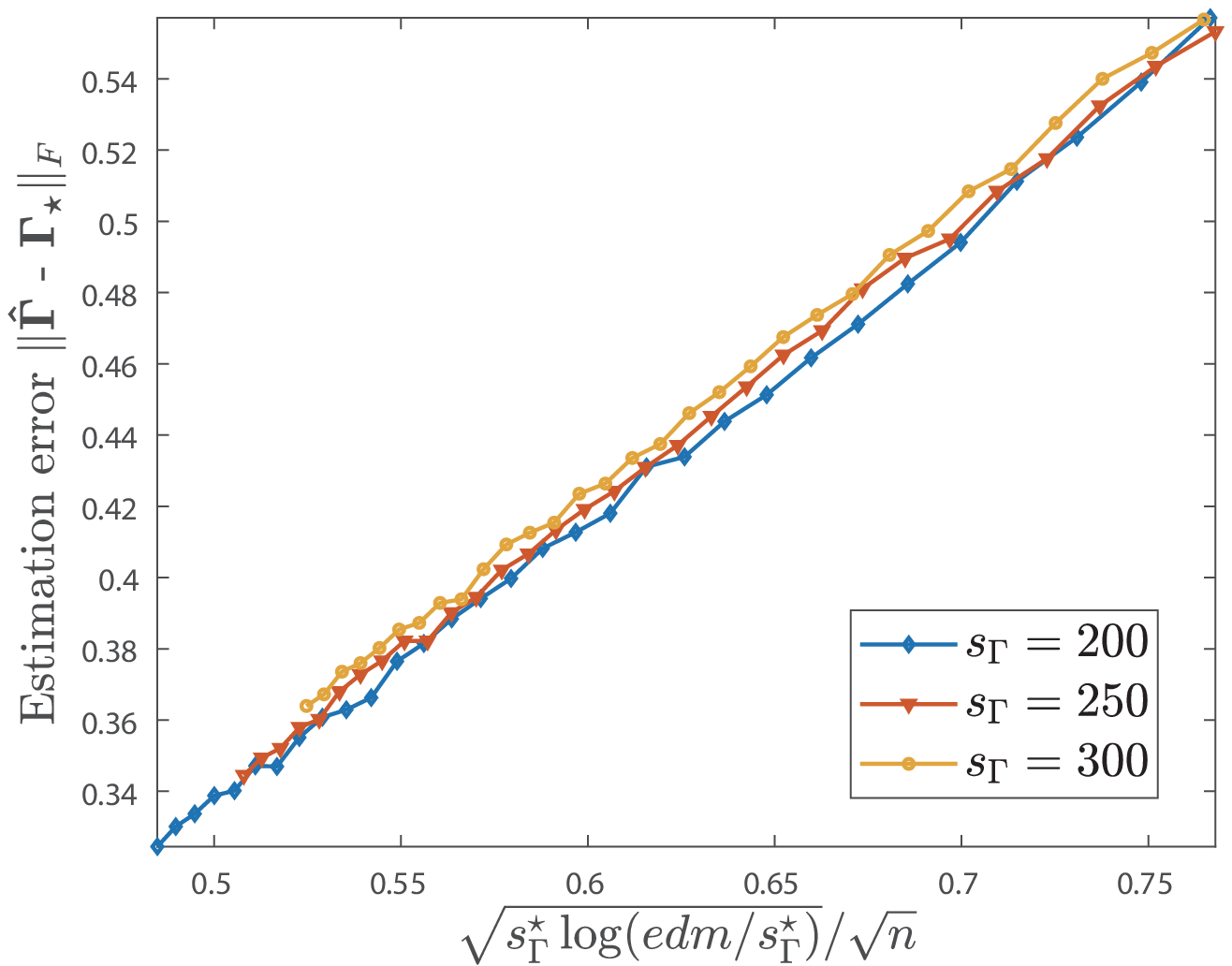}
			% where an .eps filename suffix will be assumed under latex,
			% and a .pdf suffix will be assumed for pdflatex
			\label{ErrorScaleGammaRescaled}}}
	\vskip -0.1in
	\caption{Estimation error $\fnorm{\hat{\mGamma} - \Gammastar}$ under different numbers of measurements $n$ and different sparsity levels.}
	\label{Fig_ErrorScale_Gamma}
\end{figure}

\begin{figure}[ht]
	%	\centering
	\vskip -0.1in
	%\captionsetup{aboveskip=0pt}
	\centerline{\subfigure[Original numbers of measurements]{\includegraphics[width=0.5\linewidth]{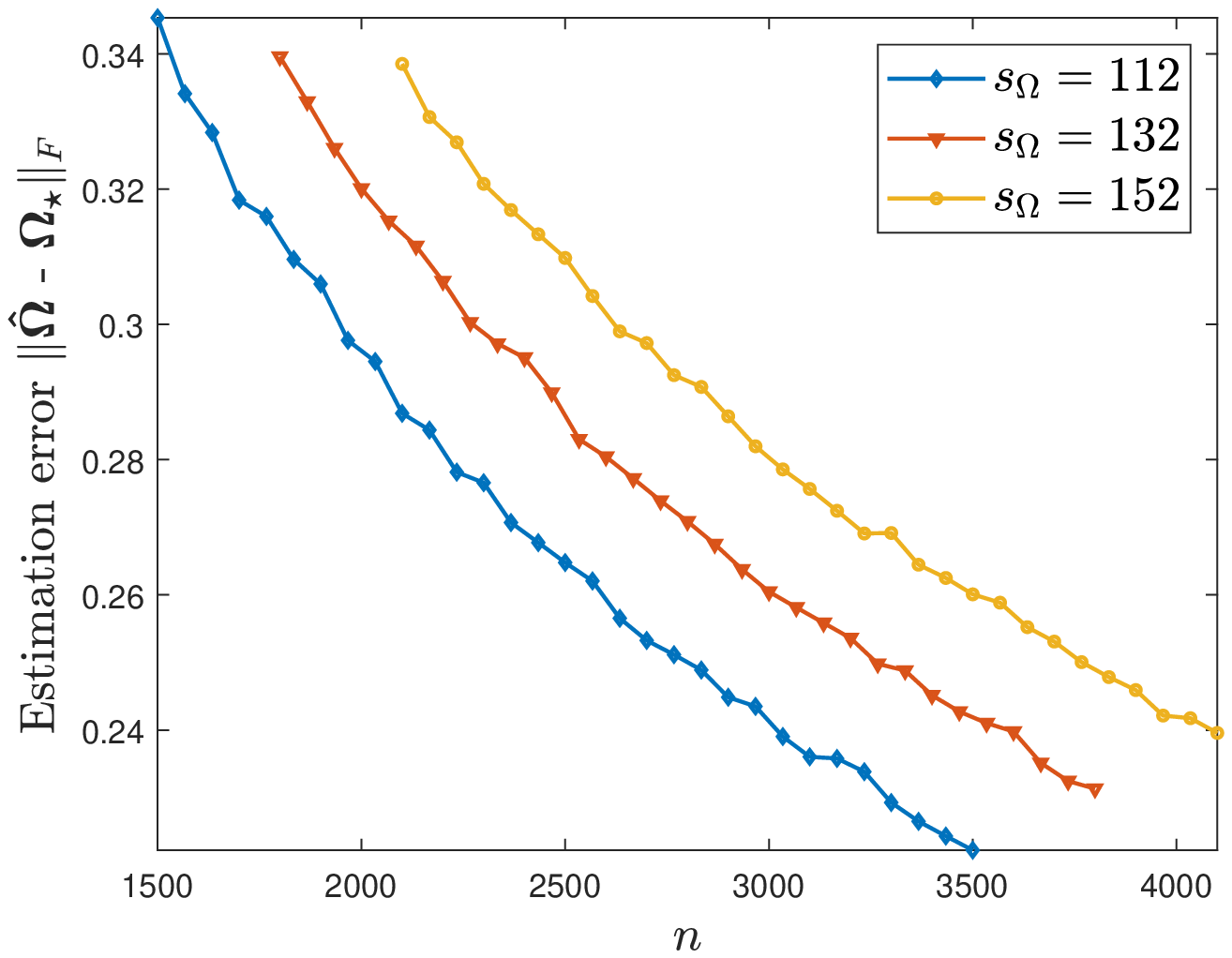}
			% where an .eps filename suffix will be assumed under latex,
			% and a .pdf suffix will be assumed for pdflatex
			\label{ErrorScaleOmegaOriginal}}
		%		\hspace{0.1em}%
		\hfill
		\subfigure[Rescaled numbers of measurements]{\includegraphics[width=0.5\linewidth]{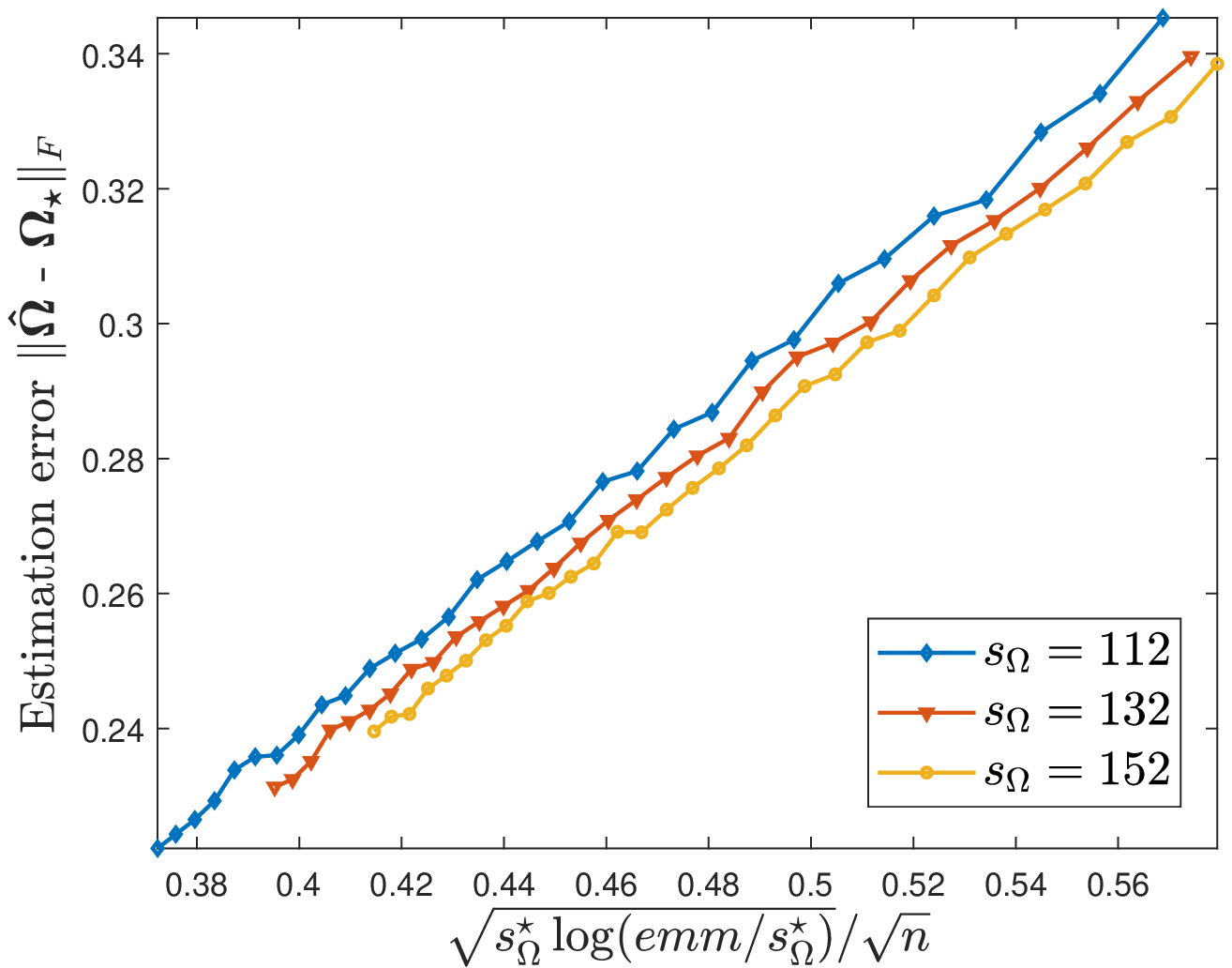}
			% where an .eps filename suffix will be assumed under latex,
			% and a .pdf suffix will be assumed for pdflatex
			\label{ErrorScaleOmegaRescaled}}}
	\vskip -0.1in
	\caption{Estimation error $\fnorm{\hat{\mOmega} - \Omegastar}$ under different numbers of measurements $n$ and different sparsity levels.}
	\label{Fig_ErrorScale_Omega}
\end{figure}

The scalings of estimation errors about $\Gammastar$ and $\Omegastar$ are presented in Figure \ref{Fig_ErrorScale_Gamma} and \ref{Fig_ErrorScale_Omega}. The diagrams illustrate the estimation errors of $\Gammastar$ and $\Omegastar$ are proportion to $\omega_{\Gamma} / \sqrt{n}$ and $\omega_{\Omega} / \sqrt{n}$ respectively without any logarithmic factor, which verifies our theoretical result in Theorem \ref{ConvergenceAltIHT}. For Algorithm \ref{General_AltPGD} and \ref{General_Initialization} with the $l_1$-norm, the results are similar and we do not include them in this manuscript.

\subsection{Network structure learning on S\&P 500 stock data}
In this part, we apply Algorithm \ref{Alg_AltIHT} and our initialization (Algorithm \ref{Alg_AltIHT_Initialization}) to analyze the network structure of the stocks in the S\&P 500 index. The stock data consists of 1259 daily closing prices for 434 companies in the S\&P 500 index between February 8, 2013 and February 7, 2018 \cite{SPStock}. In this way, we get 1259 data vectors, each of which contains the closing prices of all stocks on a trading day. To make the data stationary, we calculate the log-returns $\ba{\vr_{t}}_{t = 1}^{T - 1}$ of stocks by
\begin{equation}
	r_{t, i} = \log(\frac{p_{t + 1, i}}{p_{t, i}}), ~~~ t = 1 , \cdots, T - 1,
\end{equation}
where $p_{t, i}$ represents the closing price of stock $i$ at day $t$. Then we construct the predictor matrix $\mX = [\vr_{1}, \cdots, \vr_{T-2}]^T$ and the data matrix $\mY = [\vr_{2}, \cdots, \vr_{T-1}]^T$. In the simulation, the step sizes and the constraint parameters are selected through 5-fold cross validation. 

\begin{figure}[ht]
	%	\centering
	\vskip -0.1in
	%\captionsetup{aboveskip=0pt}
	\centerline{\subfigure[Precision matrix estimated by Algorithm \ref{Alg_AltIHT} and  \ref{Alg_AltIHT_Initialization}]{\includegraphics[width=0.5\linewidth]{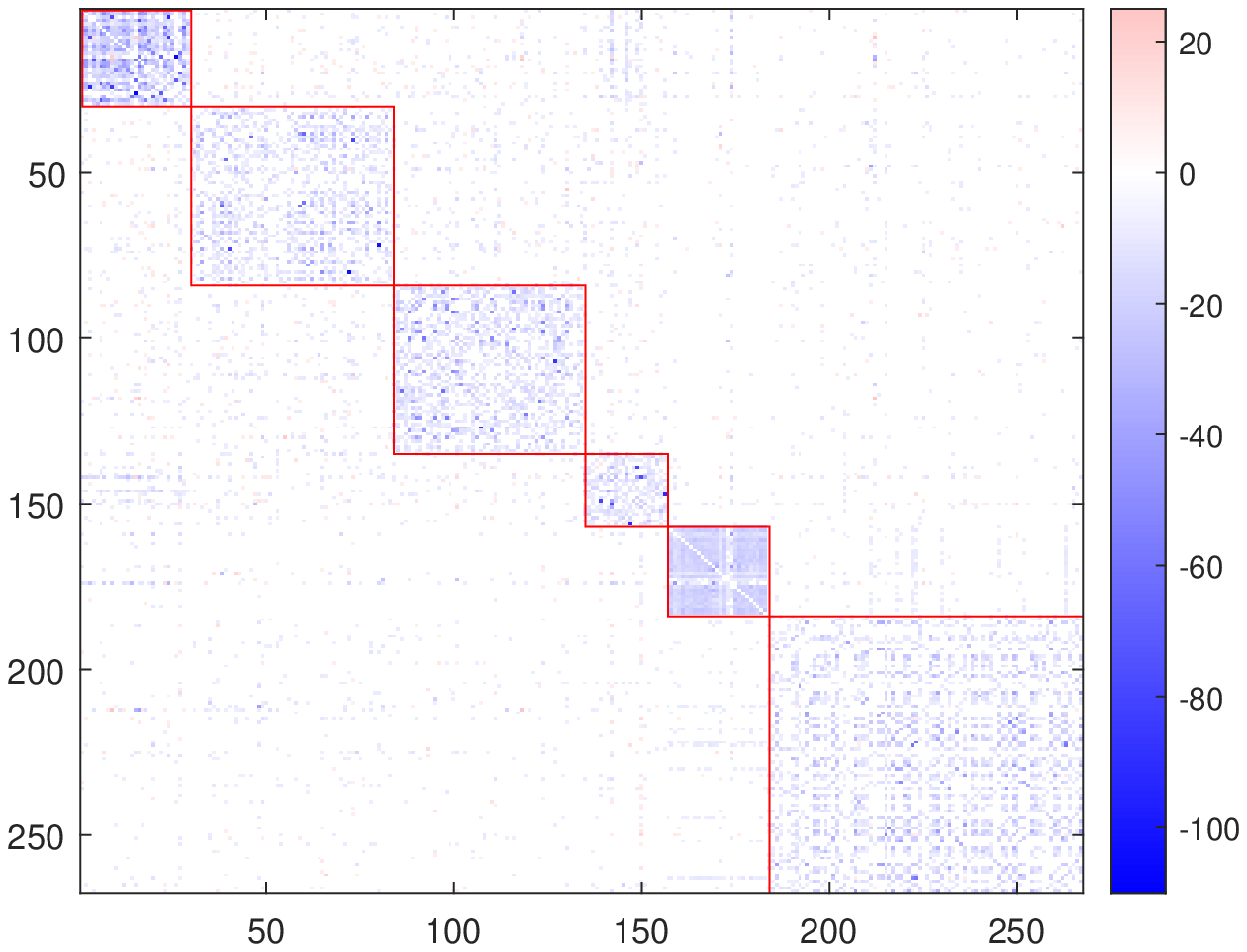}
			% where an .eps filename suffix will be assumed under latex,
			% and a .pdf suffix will be assumed for pdflatex
			\label{PrecisionMatrixOurs}}
		%		\hspace{0.1em}%
		\hfill
		\subfigure[Precision matrix estimated by the method in \cite{Chen18CovariateAdj}]{\includegraphics[width=0.5\linewidth]{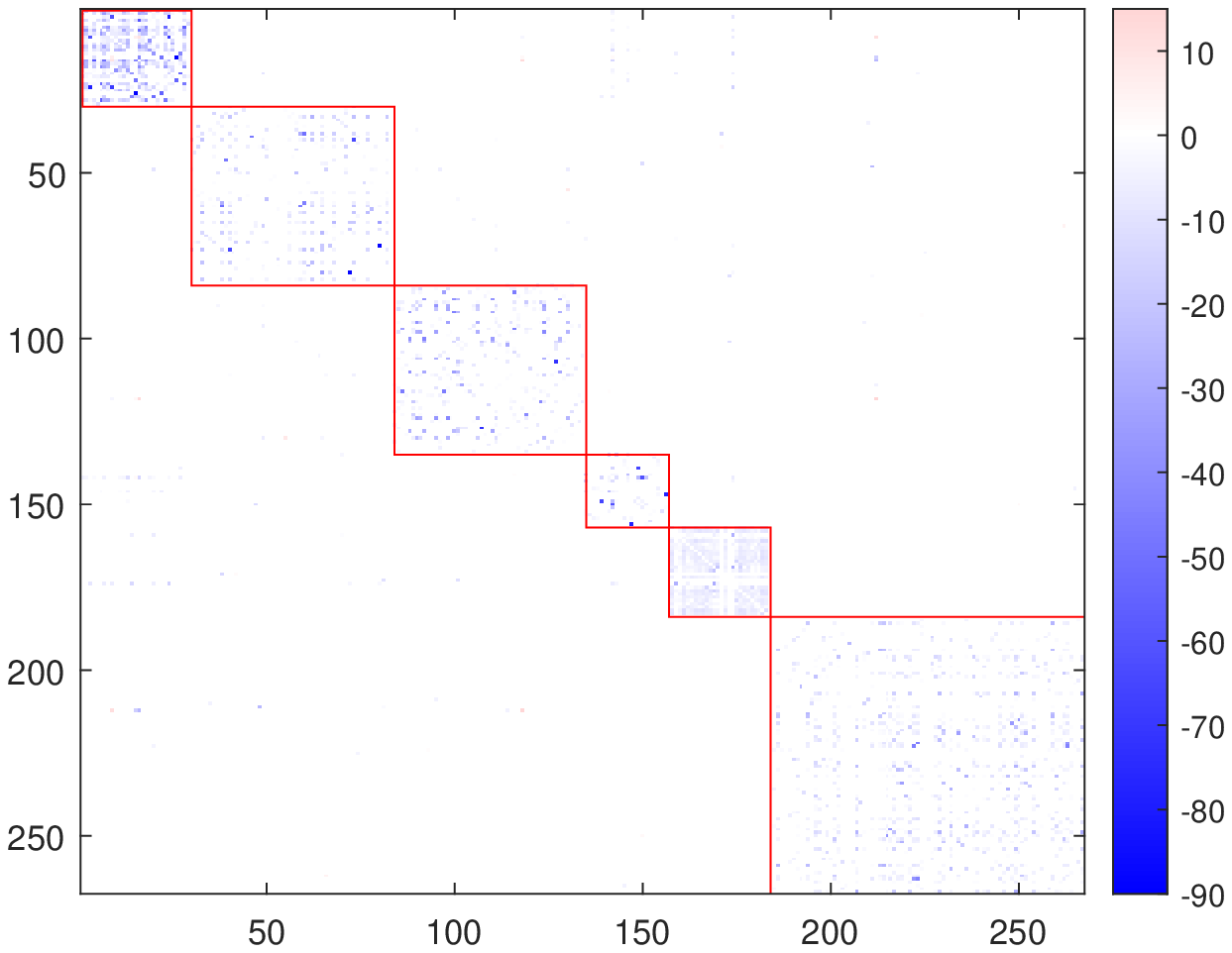}
			% where an .eps filename suffix will be assumed under latex,
			% and a .pdf suffix will be assumed for pdflatex
			\label{PrecisionMatrixAltIHT}}}
	\vskip -0.1in
	\caption{Sparsity patterns of precision matrices estimated by two methods. From top left to bottom right, the selected sectors are Energy, Information Technology, Health Care, Materials, Utilities and Financials. }
	\label{Fig_StockData}
\end{figure}

In Figure \ref{PrecisionMatrixOurs}, the sparsity pattern of the precision matrix estimated by Algorithm \ref{Alg_AltIHT} and \ref{Alg_AltIHT_Initialization} illustrates that there are strong conditional dependency relationships among the stocks in the same sector. This phenomenon is also recorded in \cite{Fan2016IncorporatingG}. In Figure \ref{PrecisionMatrixAltIHT}, we also present the sparsity pattern of the precision matrix estimated by the method in \cite{Chen18CovariateAdj} for comparison, which indicates similar relationships among the stocks in the same sector. 

%We adopt the VAR model \eqref{VAR_Model} with the regularizer $\calR(\cdot) = \lonenorm{\cdot}$ to study the evolution of stock log-prices over the 2013-2018 period and then solve the model with PGD. By selecting the constrained parameter through 5-fold cross validation, we derive that the final relative prediction error is $0.19$ and $12.8 \%$ entries of $\hat{\mGamma}$ are non-zero. In Figure \ref{Fig_Correlation}, we present the sparsity patterns of two parts of $\hat{\mGamma}$, which indicate a meaningful relationship among the stock prices. The 434 companies belong to 10 different sectors, such as energy, health care and information technology (IT). In Figure \ref{Fig_EnergyHealthCare} and \ref{Fig_EnegyIT}, the prices of stocks in the same sector tend to be positively correlated. The prices between the energy sector (29 stocks) and the health care sector (51 stocks) are likely to be negatively correlated and the same phenomenon exists between the energy sector and the IT sector (54 stocks).

\section{Discussion}
In this paper, we provide a sharp analysis of a class of alternating projected gradient descent algorithms for the covariate adjusted precision matrix estimation problem. It would be an interesting direction to combine our analysis with practical applications, such as time series models and low rank matrices estimation in \cite{Skripnikov2019JointE}.

\clearpage

%\onecolumn
\begin{center}
	\textbf{\large Supplementary for A Sharp Analysis of Covariate Adjusted Precision Matrix Estimation via Alternating Projected Gradient Descent}
\end{center}
%\title{Supplementary for A Sharp Analysis of Covariate Adjusted Precision Matrix Estimation via Alternating Gradient Descent with Hard Thresholding}

In this supplementary, we present the complete proof for the theoretical results in the paper. We use $C$ and $c$ to denote positive constants which might change from line to line throughout the paper.

\section{Preliminaries}
The core of our analysis is the sample-based analysis for the iterations. The following two lemmas illustrate the mixed tails of terms like $\iprod{\bmU}{\bmX^T \bmX}$ and $\iprod{\bmU}{\bmX^T \bmE}$, which would appear many times in the remained part.
\begin{lemma} \label{Quadratic_form_XX}
	Suppose $\mU \in \RR^{d \times d}$, $\mX \in \RR^{n \times d}$ and $\Vect{\bmX^T}$ follows the distribution $\calN(\bmzero, \bmUpsilon_{\bmX})$. We have the tail bound
	\begin{equation}
		P(\abs{\tr(\bmX \bmU \bmX^T) - \EE \tr(\bmX \bmU \bmX^T)} > u) \leq 2 \exp(- c \min (\frac{u^2}{n \norm{\bmUpsilon_{\bmX}}^2 \fnorm{\bmU}^2}, \frac{u}{\norm{\bmUpsilon_{\bmX}} \fnorm{\bmU}})),
	\end{equation}
	where $c$ is a positive constant.
\end{lemma}

%Under the Gaussian setting, the above lemma doesn't require the independence between the rows of $\bmX$, for the rotation invariance of Gaussian vectors.
\begin{lemma} \label{Quadratic_form_Independent}
	Consider $\mU \in \RR^{m \times d}$, $\mX \in \RR^{n \times d}$ and $\mE \in \RR^{n \times m}$. Suppose $\bmX$ is independent with $\bmE$ and $\Vect{\bmX^T} \sim \calN(\bmzero, \bmUpsilon_{\bmX})$, $\Vect{\bmE^T} \sim \calN(\bmzero, \bmUpsilon_{\bmE})$. Then
	\begin{equation}
		P(\abs{\tr(\bmE \bmU \bmX^T)} > u) \leq 2 \exp(- c \min (\frac{u^2}{n \norm{\bmUpsilon_{\bmE}^{\frac{1}{2}}}^2 \norm{\bmUpsilon_{\bmX}^{\frac{1}{2}}}^2 \fnorm{\bmU}^2}, \frac{u}{\norm{\bmUpsilon_{\bmE}^{\frac{1}{2}}} \norm{\bmUpsilon_{\bmX}^{\frac{1}{2}}} \fnorm{\bmU}})),
	\end{equation}
	where $c$ is a positive constant.
\end{lemma}

The following lemma is the fundamental tool to analyze the suprema of random processes with a mixed tail, which is based on the generic chaining \cite{Talagrand2005TheGenericCh} itself.

\begin{lemma} \cite[Theorem 3.5]{Dirksen2015TailBounds} \label{Suprema}
	Let $d_1$, $d_2$ be two semi-metrics on $T$. Suppose the random process $(X_t)_{t \in T}$ has a mixed tail
	\begin{equation}
		P(\abs{X_{t} - X_{s}} > u) \leq 2 \exp(- \min (\frac{u^2}{d_2(t, s)^2}, \frac{u}{d_1(t, s)})),
	\end{equation}
	then we could derive
	\begin{equation}
		\begin{split}
			P\big(\usup{t \in T} \abs{X_t - X_{t_0}} &> C(\gamma_2(T, d_2) + \gamma_1(T, d_1) + u \Delta_2(T) + u^2 \Delta_1(T))\big) \leq 2 \exp(- u^2),
		\end{split}
	\end{equation}
	where $C$ is a positive constant and $\Delta_2(T)$ ($\Delta_1(T)$) is the diameter of $T$ with respect to $d_2$ ($d_1$).
\end{lemma}

Here, we introduce the definition of $\gamma_{\alpha}$-functional used in the above lemma.

\begin{definition}[$\gamma_{\alpha}$-functional]
	Let $(T, d)$ be a semi-metric space. For any $0 < \alpha < \infty$, the $\gamma_{\alpha}$-functional of $(T, d)$ is defined as
	\begin{align}
		\gamma_{\alpha}(T, d) = \uinf{\calT} \usup{t \in T} \sum_{n = 0}^{\infty} 2^{\frac{n}{\alpha}} d(t, T_{n}), \label{Gamma_Alpha_Functional}
	\end{align}
	where $d(t, T_{n}) = \uinf{s \in T_{n}} d(t, s)$ and the infimum in \eqref{Gamma_Alpha_Functional} is taken over all admissible sequences.
\end{definition}

\section{Model}
The corresponding negative log-likelihood function is
\begin{equation}
	\begin{split}
		f_n(\bmGamma, \bmOmega) &= - \log\abs{\bmOmega} + \frac{1}{n} \tr \bBa{(\bmY - \bmX \bmGamma) \bmOmega (\bmY - \bmX \bmGamma)^T} \\
		&= - \log\abs{\bmOmega} + \frac{1}{n} \tr \bBa{(\bmGamma - \bmGamma_{\star})^T \bmX^T \bmX (\bmGamma - \bmGamma_{\star}) \bmOmega - 2 \bmE^T \bmX(\bmGamma - \bmGamma_{\star}) \bmOmega + \bmE^T \bmE \bmOmega},
	\end{split}
\end{equation}
where $\bmX \in \RR^{n \times d}$, $\bmY \in \RR^{n \times m}$, $\bmGamma \in \RR^{d \times m}$, $\bmOmega \in \RR^{m \times m}$. Without the generality, we suppose $\bmOmega$ is symmetric.

The population loss function is
\begin{equation}
	f(\bmGamma, \bmOmega) = - \log\abs{\bmOmega} + \tr \bBa{(\bmGamma - \bmGamma_{\star})^T \bmSigma_{\bmX} (\bmGamma - \bmGamma_{\star}) \bmOmega + \bmOmega_{\star}^{-1} \bmOmega}.
\end{equation}
For the convenience of analysis, we collect the corresponding gradients and Hessian matrices here
\begin{align}
	\nabla_{\Gamma} f_n (\bmGamma, \bmOmega) &= \frac{2}{n} \bmX^T \bmX (\bmGamma - \Gammastar) \bmOmega - \frac{2}{n} \bmX^T \bmE \bmOmega, \\
	\nabla_{\Omega} f_n (\bmGamma, \bmOmega) &= -\bmOmega^{-1} + \frac{1}{n} (\bmGamma - \bmGamma_{\star})^T \bmX^T \bmX (\bmGamma - \bmGamma_{\star}) - \frac{2}{n} (\bmGamma - \bmGamma_{\star})^T \bmX^T \bmE + \frac{1}{n} \bmE^T \bmE, \\
	\nabla_{\Gamma} f (\bmGamma, \bmOmega) &= 2\bmSigma_{\bmX} (\bmGamma - \Gammastar) \bmOmega, \\
	\nabla_{\Omega} f (\bmGamma, \bmOmega) &= -\bmOmega^{-1} + (\bmGamma - \bmGamma_{\star})^T \bmSigma_{\bmX} (\bmGamma - \bmGamma_{\star}) + \bmOmega_{\star}^{-1}, \\
	\nabla_{\Gamma}^{2} f (\bmGamma, \bmOmega) &= \bmOmega \otimes 2 \bmSigma_{\bmX}, \\
	\nabla_{\Omega}^{2} f (\bmGamma, \bmOmega) &= \bmOmega^{-1} \otimes \bmOmega^{-1}.
\end{align}
Here $\nabla_{\Gamma}^{2} f (\bmGamma, \bmOmega)$ and $\nabla_{\Omega}^{2} f (\bmGamma, \bmOmega)$ is in the sense of vectorization.

In \cite{Chen18CovariateAdj}, the authors introduce the following local properties of the population function $f(\bmGamma, \bmOmega)$ required by the analysis.
\begin{lemma} \label{SSGamma}
	Under Assumption \ref{SingularValue_Omega} and \ref{SingularValue_Sigma}, for any $\bmGamma, \bmGamma' \in \ball{F}(\Gammastar; R)$, we have
	\begin{equation}
		\nu_{\min} \tau_{\min} \fnorm{\bmGamma' - \bmGamma}^2 \leq f(\bmGamma', \Omegastar) - f(\bmGamma, \Omegastar) - \iprod{\nabla_{\Gamma} f(\bmGamma, \Omegastar)}{\bmGamma' - \bmGamma} \leq \nu_{\max} \tau_{\max} \fnorm{\bmGamma' - \bmGamma}.
	\end{equation}
\end{lemma}
\begin{lemma} \label{SSOmega}
	Under Assumption \ref{SingularValue_Omega} and \ref{SingularValue_Sigma}, for any $\bmOmega, \bmOmega' \in \ball{F}(\Omegastar; R)$ where $R \leq \frac{\nu_{\min}}{2}$, we have
	\begin{equation}
		\frac{1}{8 \nu_{\max}^2} \fnorm{\bmOmega' - \bmOmega}^2 \leq f(\Gammastar, \bmOmega') - f(\Gammastar, \bmOmega) - \iprod{\nabla_{\Omega} f(\Gammastar, \bmOmega)}{\bmOmega' - \bmOmega} \leq  \frac{2}{\nu_{\min}^2} \fnorm{\bmOmega' - \bmOmega}.
	\end{equation}
\end{lemma}

\begin{lemma} \label{LipschitzGradient}
	Under Assumption \ref{SingularValue_Omega} and Assumption \ref{SingularValue_Sigma}, for any $\bmOmega \in \ball{F}(\Omegastar; R)$, we could derive
	\begin{equation}
		\fnorm{\nabla_{\Gamma} f(\bmGamma, \Omegastar) - \nabla_{\Gamma} f(\bmGamma, \bmOmega)} \leq 2 \tau_{\max} R \fnorm{\bmOmega - \Omegastar}.
	\end{equation}
	For any $\bmGamma \in \ball{F}(\Gammastar; R)$, we could derive
	\begin{equation}
		\fnorm{\nabla_{\Omega} f(\Gammastar, \bmOmega) - \nabla_{\Omega} f(\bmGamma, \bmOmega)} \leq \tau_{\max} R \fnorm{\bmGamma - \Gammastar}.
	\end{equation}
\end{lemma}

\section{Analysis of the alternating gradient descent with hard thresholding (Proof of Theorem 1)}
Our analysis is based on the facts $\Gammat \in \ball{F}(\Gammastar, R)$ and $\Omegat \in \ball{F}(\Omegastar, R)$.
\subsection{Analysis of the iteration about $\bmGamma$}
First, we introduce two helpful lemmas for our analysis.

With the following lemma, we could deal with terms with the hard thresholding operator.
\begin{lemma} \cite{Li2016StochasticVar} \label{Contraction_HardTh}
	Suppose $\xreal$ is a sparse vector satisfying $\lzeronorm{\xreal} \leq s_{\star}$. $\HT(\cdot, s)$ is the hard thresholding operator with $s \geq s_{\star}$. Then we could bound the difference $\ltwonorm{\HT(\bmx, s) - \xreal}$ for any $\bmx$ by
	\begin{equation}
		\ltwonorm{\HT(\bmx, s) - \xreal}^2 \leq (1 + \frac{2 \sqrt{s_{\star}}}{\sqrt{s - s_{\star}}}) \ltwonorm{\bmx - \xreal}^2.
	\end{equation}
\end{lemma}

The following lemma lays a foundation for the convergence analysis of gradient descent iterations.
\begin{lemma} \cite{Nesterov:2014:ILC:2670022} \label{StrongConvexContrac}
	Suppose $f(\bmx)$ is $\mu$-strongly convex and $L$-smooth. With the step size $\eta = 2 / (L + \mu)$, the gradient descent iteration would contract as
	\begin{equation}
		\ltwonorm{(\bmx - \eta \nabla f(\bmx) - \xreal)} \leq \frac{L - \mu}{L + \mu} \ltwonorm{\bmx - \xreal},
	\end{equation}
	where $\xreal$ is the optimal point.
\end{lemma}

We set the step sizes as
\begin{equation}
	\eta_{\Gamma} = \frac{1}{\nu_{\max} \tau_{\max} + \nu_{\min} \tau_{\min}} \quad \text{and} \quad \eta_{\Omega} = \frac{8 \nu_{\max}^2 \nu_{\min}^2 }{16 \nu_{\max}^2 + \nu_{\min}^2 }.
\end{equation}

We write $\calI = \calI_{t+1} \cup \calI_{\star}$, where $\calI_{t+1}$ and $\calI_{\star}$ are the support sets of $\Gammatp$ and $\Gammastar$, respectively.

Now, we could rewrite $\fnorm{\Gammatp - \Gammastar}$ as
\begin{align}
	&\fnorm{\Gammatp - \Gammastar} \nonumber \\
	%	&= \fnorm{\calP_{\calK_{\Gamma}}(\Gammat - \eta_{\Gamma} \nabla_{\Gamma} f_n(\Gammat, \Omegat)) - \Gammastar} \\
	%	&\leq \usup{\mV \in \calC_{2s_{\Gamma}} \cap \calS^{dm - 1}} \iprod{\mV}{\Gammat - \Gammastar - \eta_{\Gamma} \nabla_{\Gamma} f_n(\Gammat, \Omegat)} \\
	%	&= \usup{\mV \in \calC_{2s_{\Gamma}} \cap \calS^{dm - 1}} \iprod{\mV}{\Gammat - \Gammastar - \eta_{\Gamma} \nabla_{\Gamma} f(\Gammat, \Omegastar) \\
	%		&\qquad + \eta_{\Gamma} (\nabla_{\Gamma} f(\Gammat, \Omegastar) - \nabla_{\Gamma} f(\Gammat, \Omegat)) + \eta_{\Gamma} (\nabla_{\Gamma} f(\Gammat, \Omegat) - \nabla_{\Gamma} f_n(\Gammat, \Omegat))} \\
	%	&\leq \fnorm{\Gammat - \eta_{\Gamma} \nabla_{\Gamma} f(\Gammat, \Omegastar) - \Gammastar} + \eta_{\Gamma} \fnorm{\nabla_{\Gamma} f(\Gammat, \Omegastar) - \nabla_{\Gamma} f(\Gammat, \Omegat)} \\
	%	&\qquad + \eta_{\Gamma} \usup{\mV \in \calC_{2s_{\Gamma}} \cap \calS^{dm - 1}} \iprod{\mV}{\nabla_{\Gamma} f(\Gammat, \Omegat) - \nabla_{\Gamma} f_n(\Gammat, \Omegat)}, \\
	&= \fnorm{\HT((\Gammat - \eta_{\Gamma} \nabla_{\Gamma} f_n(\Gammat, \Omegat))_{\calI}, s_{\Gamma}) - \Gammastar} \\
	&\leq \sqrt{1 + \frac{2\sqrt{s_{\Gamma}^{\star}}}{\sqrt{s_{\Gamma} - s_{\Gamma}^{\star}}}} \fnorm{(\Gammat - \eta_{\Gamma} \nabla_{\Gamma} f_n(\Gammat, \Omegat))_{\calI} - \Gammastar} \\
	&\leq \sqrt{1 + \frac{2\sqrt{s_{\Gamma}^{\star}}}{\sqrt{s_{\Gamma} - s_{\Gamma}^{\star}}}} \big(\fnorm{(\Gammat - \eta_{\Gamma} \nabla_{\Gamma} f(\Gammat, \Omegastar))_{\calI} - \Gammastar} + \eta_{\Gamma} \fnorm{(\nabla_{\Gamma} f(\Gammat, \Omegastar) - \nabla_{\Gamma} f_n(\Gammat, \Omegat))_{\calI}} \big) \nonumber \\
	&\leq \sqrt{1 + \frac{2\sqrt{s_{\Gamma}^{\star}}}{\sqrt{s_{\Gamma} - s_{\Gamma}^{\star}}}} \big(\fnorm{\Gammat - \eta_{\Gamma} \nabla_{\Gamma} f(\Gammat, \Omegastar) - \Gammastar} + \eta_{\Gamma} \fnorm{(\nabla_{\Gamma} f(\Gammat, \Omegastar) - \nabla_{\Gamma} f_n(\Gammat, \Omegat))_{\calI}} \big), \label{Iteration_Gamma}
\end{align}
where the first inequality is based on Lemma \ref{Contraction_HardTh}.

%where the first inequality is based on Lemma \ref{Difference_Projection} and the last inequality is from the Cauchy–Schwarz inequality.

The first term of (\ref{Iteration_Gamma}) could be bounded by the strong convexity and the smoothness of the population function $f(\bmGamma, \Omegastar)$ about $\bmGamma$ in Lemma \ref{SSGamma} and the corresponding convergence result in Lemma \ref{StrongConvexContrac}
\begin{equation}
	\fnorm{\Gammat - \eta_{\Gamma} \nabla_{\Gamma} f(\Gammat, \Omegastar) - \Gammastar} \leq \frac{\tau_{\max} \nu_{\max} - \tau_{\min} \nu_{\min}}{\tau_{\max} \nu_{\max} + \tau_{\min} \nu_{\min}} \fnorm{\Gammat - \Gammastar}. \label{StrongConvexityGamma}
\end{equation}

The second term of (\ref{Iteration_Gamma}) could be rewritten as
\begin{equation}
	\fnorm{(\nabla_{\Gamma} f(\Gammat, \Omegastar) - \nabla_{\Gamma} f_n(\Gammat, \Omegat))_{\calI}} \leq \fnorm{\nabla_{\Gamma} f(\Gammat, \Omegastar) - \nabla_{\Gamma} f(\Gammat, \Omegat)} + \fnorm{(\nabla_{\Gamma} f(\Gammat, \Omegat) - \nabla_{\Gamma} f_n(\Gammat, \Omegat))_{\calI}}.
\end{equation}
%The first item could be bounded by the Lipschitz property of $\nabla_{\Gamma} f(\bmGamma, \cdot)$ about $\bmOmega$ around $\Omegastar$ in Lemma \ref{LipschitzGradient}
%\begin{equation}
%	\fnorm{\nabla_{\Gamma} f(\Gammat, \Omegastar) - \nabla_{\Gamma} f(\Gammat, \Omegat)} \leq 2 \tau R \fnorm{\Omegat - \Omegastar}. \label{LipschitzGradientGamma}
%\end{equation}

The first part could be bounded by the Lipschitz property of $\nabla_{\Gamma} f(\bmGamma, \cdot)$ about $\bmOmega$ around $\Omegastar$ in Lemma \ref{LipschitzGradient}
\begin{equation}
	\fnorm{\nabla_{\Gamma} f(\Gammat, \Omegastar) - \nabla_{\Gamma} f(\Gammat, \Omegat)} \leq 2 \tau_{\max} R \fnorm{\Omegat - \Omegastar}. \label{LipschitzGradientGamma}
\end{equation}

The second part is associated with the sample loss function $f_n(\bmGamma, \bmOmega)$ and needs the sample-based analysis in the following lemma.
\begin{lemma} \label{Sample_Gamma}
	Under Assumption \ref{SingularValue_Omega} and \ref{SingularValue_Sigma}, we set $\eta_{\Gamma} = \frac{1}{\nu_{\max} \tau_{\max} + \nu_{\min} \tau_{\min}}$. For any $\Gammat \in \ball{F}(\Gammastar, R)$ and $\Omegat \in \ball{F}(\Omegastar, R)$, the difference $(\nabla_{\Gamma} f(\Gammat, \Omegat) - \nabla_{\Gamma} f_n(\Gammat, \Omegat))_{\calI}$ could be bounded by
	\begin{equation}
		\begin{split}
			%			&\eta_{\Gamma} \usup{\mV \in \calC_{2s_{\Gamma}} \cap \calS^{dm - 1}} \iprod{\mV}{\nabla_{\Gamma} f(\Gammat, \Omegat) - \nabla_{\Gamma} f_n(\Gammat, \Omegat)}  \\
			&\eta_{\Gamma} \fnorm{(\nabla_{\Gamma} f(\Gammat, \Omegat) - \nabla_{\Gamma} f_n(\Gammat, \Omegat))_{\calI}}  \\
			& \leq C_{\Gamma, 1} \frac{\nu_{\max} \tau_{\max}}{\nu_{\max} \tau_{\max} + \nu_{\min} \tau_{\min}} (\frac{R}{\nu_{\max}} \frac{\omega_{\Gamma} + \omega_{\Omega} + u}{\sqrt{n}} \fnorm{\Omegat - \Omegastar} + \frac{\omega_{\Gamma} + u}{\sqrt{n}} \fnorm{\Gammat - \Gammastar} \\
			&  + \frac{1}{\sqrt{\tau_{\max} \nu_{\min}} \nu_{\max}} \frac{\omega_{\Gamma} + \omega_{\Omega} + u}{\sqrt{n}} \fnorm{\Omegat - \Omegastar} + \frac{1}{\sqrt{\tau_{\max} \nu_{\max}}} \frac{\omega_{\Gamma} + u}{\sqrt{n}}),
		\end{split}
	\end{equation}
	with probability at least $1 - 8 \exp(- u^2)$, when $n \geq  (\omega_{\Gamma} + \omega_{\Omega} + u)^2$.
\end{lemma}

\subsection{Analysis of the iteration about $\bmOmega$}
Similarly, we write $\calT = \calT_{t+1} \cup \calT_{\star}$, where $\calT_{t+1}$ and $\calT_{\star}$ are the support sets of $\Omegatp$ and $\Omegastar$, respectively.

For $\calT$ contains $\calT_{t+1}$ and $\calT_{\star}$, we could rearrange $\fnorm{\Omegatp - \Omegastar}$ as
\begin{align}
	&\fnorm{\Omegatp - \Omegastar} \nonumber \\
	%	&= \fnorm{\calP_{\calK_{\Omega}}(\Omegat - \eta_{\Omega} \nabla_{\Omega} f_n(\Gammat, \Omegat)) - \Omegastar} \\
	%	&\leq \usup{\mV \in \calC_{2s_{\Omega}} \cap \calS^{m^2 - 1}} \iprod{\mV}{\Omegat - \Omegastar - \eta_{\Omega} \nabla_{\Omega} f_n(\Gammat, \Omegat)} \\
	%	&= \usup{\mV \in \calC_{2s_{\Omega}} \cap \calS^{m^2 - 1}} \iprod{\mV}{\Omegat - \Omegastar - \eta_{\Omega} \nabla_{\Omega} f(\Gammastar, \Omegat) \\
	%		&\qquad + \eta_{\Omega} (\nabla_{\Omega} f(\Gammastar, \Omegat) - \nabla_{\Omega} f(\Gammat, \Omegat)) + \eta_{\Omega} (\nabla_{\Omega} f(\Gammat, \Omegat) - \nabla_{\Omega} f_n(\Gammat, \Omegat))} \\
	%	&\leq \fnorm{\Omegat - \eta_{\Omega} \nabla_{\Omega} f(\Gammastar, \Omegat) - \Omegastar} + \eta_{\Omega} \fnorm{\nabla_{\Omega} f(\Gammastar, \Omegat) - \nabla_{\Omega} f(\Gammat, \Omegat)} \\
	%	&\qquad + \eta_{\Omega} \usup{\mV \in \calC_{2s_{\Omega}} \cap \calS^{m^2 - 1}} \iprod{\mV}{\nabla_{\Omega} f(\Gammat, \Omegat) - \nabla_{\Omega} f_n(\Gammat, \Omegat)}, \\
	&= \fnorm{\HT((\Omegat - \eta_{\Omega} \nabla_{\Omega} f_n(\Gammat, \Omegat))_{\calT}, s_{\Omega}) - \Omegastar} \\
	&\leq \sqrt{1 + \frac{2\sqrt{s_{\Omega}^{\star}}}{\sqrt{s_{\Omega} - s_{\Omega}^{\star}}}} \fnorm{(\Omegat - \eta_{\Omega} \nabla_{\Omega} f_n(\Gammat, \Omegat))_{\calT} - \Omegastar} \\
	&\leq \sqrt{1 + \frac{2\sqrt{s_{\Omega}^{\star}}}{\sqrt{s_{\Omega} - s_{\Omega}^{\star}}}} \big(\fnorm{(\Omegat - \eta_{\Omega} \nabla_{\Omega} f(\Gammastar, \Omegat))_{\calT} - \Omegastar} + \eta_{\Omega} \fnorm{(\nabla_{\Omega} f(\Gammastar, \Omegat) - \nabla_{\Omega} f_n(\Gammat, \Omegat))_{\calT}} \big) \nonumber \\
	&\leq \sqrt{1 + \frac{2\sqrt{s_{\Omega}^{\star}}}{\sqrt{s_{\Omega} - s_{\Omega}^{\star}}}} \big(\fnorm{\Omegat - \eta_{\Omega} \nabla_{\Omega} f(\Gammastar, \Omegat) - \Omegastar} + \eta_{\Omega} \fnorm{(\nabla_{\Omega} f(\Gammastar, \Omegat) - \nabla_{\Omega} f_n(\Gammat, \Omegat))_{\calT}} \big), \label{Iteration_Omega}
\end{align}
where the first inequality is based on Lemma \ref{Contraction_HardTh}.
%and the last inequality is from the Cauchy–Schwarz inequality.

The first term of (\ref{Iteration_Omega}) could be bounded by the strong convexity and the smoothness of the population function $f(\Gammastar, \bmOmega)$ about $\bmOmega$ in Lemma \ref{SSOmega} and the corresponding convergence result in Lemma \ref{StrongConvexContrac}
\begin{equation}
	\fnorm{\Omegat - \eta_{\Omega} \nabla_{\Omega} f(\Gammastar, \Omegat) - \Omegastar} \leq \frac{16 \nu_{\max}^2 - \nu_{\min}^2 }{16 \nu_{\max}^2 + \nu_{\min}^2 } \fnorm{\Omegat - \Omegastar}. \label{StrongConvexityOmega}
\end{equation}

The second term of (\ref{Iteration_Omega}) could be rewritten as
\begin{equation}
	\fnorm{(\nabla_{\Omega} f(\Gammastar, \Omegat) - \nabla_{\Omega} f_n(\Gammat, \Omegat))_{\calT}} \leq \fnorm{\nabla_{\Omega} f(\Gammastar, \Omegat) - \nabla_{\Omega} f(\Gammat, \Omegat)} + \fnorm{(\nabla_{\Omega} f(\Gammat, \Omegat) - \nabla_{\Omega} f_n(\Gammat, \Omegat))_{\calT}}.
\end{equation}

The first part could be bounded by the Lipschitz property of $\nabla_{\Omega} f(\cdot, \bmOmega)$ about $\bmGamma$ around $\Gammastar$ in Lemma \ref{LipschitzGradient}
\begin{equation}
	\fnorm{\nabla_{\Omega} f(\Gammastar, \Omegat) - \nabla_{\Omega} f(\Gammat, \Omegat)} \leq \tau_{\max} R \fnorm{\Gammat - \Gammastar}. \label{LipschitzGradientOmega}
\end{equation}

The second part is associated with the sample loss function $f_n(\bmGamma, \bmOmega)$ and needs the sample-based analysis in the following lemma.
\begin{lemma} \label{Sample_Omega}
	Under the same condition as Lemma \ref{Sample_Gamma}, for any $\Gammat \in \ball{F}(\Gammastar, R)$ and $\Omegat \in \ball{F}(\Omegastar, R)$, the difference $(\nabla_{\Omega} f(\Gammat, \Omegat) - \nabla_{\Omega} f_n(\Gammat, \Omegat))_{\calT}$ could be bounded by
	\begin{equation}
		\begin{split}
			%			&\eta_{\Omega} \usup{\mV \in \calC_{2s_{\Omega}} \cap \calS^{m^2 - 1}} \iprod{\mV}{\nabla_{\Omega} f(\Gammat, \Omegat) - \nabla_{\Omega} f_n(\Gammat, \Omegat)} \\
			&\eta_{\Omega} \fnorm{(\nabla_{\Omega} f(\Gammat, \Omegat) - \nabla_{\Omega} f_n(\Gammat, \Omegat))_{\calT}} \\
			&\leq C_{\Omega, 1} \frac{8 \nu_{\max}^2 \nu_{\min}^2 }{16 \nu_{\max}^2 + \nu_{\min}^2 } (\tau_{\max} R \frac{\omega_{\Gamma} + \omega_{\Omega} + u}{\sqrt{n}} \fnorm{\Gammat - \Gammastar} + \frac{\sqrt{\tau_{\max}} }{\sqrt{\nu_{\min}}} \frac{\omega_{\Gamma} + \omega_{\Omega} + u}{\sqrt{n}}\fnorm{\Gammat - \Gammastar} + \frac{1}{\nu_{\min}} \frac{\omega_{\Omega} + u}{\sqrt{n}}),
		\end{split}
	\end{equation}
	with probability at least $1 - 6 \exp(- u^2)$, when $n \geq  (\omega_{\Gamma} + \omega_{\Omega} + u)^2$.
\end{lemma}

\subsection{Analysis of the whole convergence result}
We define the convergence parameter $\rho_{\text{pop}}$ associated with the population loss function as
\begin{equation}
	\begin{split}
		\rho_{\text{pop}} &= \max(1 - \frac{2 \tau_{\min} \nu_{\min} - 2 \tau_{\max} R}{\tau_{\max} \nu_{\max} + \tau_{\min} \nu_{\min}}, 1 - \frac{2 \nu_{\min}^2 - 8 \nu_{\max}^2 \nu_{\min}^2 \tau_{\max} R}{16 \nu_{\max}^2 + \nu_{\min}^2}) \\
		&\leq \max(1 - \frac{\tau_{\min} \nu_{\min}}{\tau_{\max} \nu_{\max} + \tau_{\min} \nu_{\min}}, 1 - \frac{\nu_{\min}^2}{16 \nu_{\max}^2 + \nu_{\min}^2}),
	\end{split} \label{Def_Popular_Rate}
\end{equation}
where the inequality is from $R \leq \min(\frac{\tau_{\min} \nu_{\min}}{2 \tau_{\max}}, \frac{1}{8 \tau_{\max} \nu_{\max}^2})$, which guarantees $\rho_{\text{pop}} < 1$.

By the assumptions $s_{\Gamma} \geq (1 + 4(1 / \rho_{\text{pop}} - 1)^2) s_{\Gamma}^{\star}$ and $s_{\Omega} \geq (1 + 4(1 / \rho_{\text{pop}} - 1)^2) s_{\Omega}^{\star}$, we could bound the two parameters associated with the hard thresholding operation by
\begin{equation}
	\max(\sqrt{1 + \frac{2\sqrt{s_{\Gamma}^{\star}}}{\sqrt{s_{\Gamma} - s_{\Gamma}^{\star}}}}, \sqrt{1 + \frac{2\sqrt{s_{\Omega}^{\star}}}{\sqrt{s_{\Omega} - s_{\Omega}^{\star}}}}) \leq \frac{1}{\sqrt{\rho_{\text{pop}}}}. \label{Sparsity_Constant_HT}
\end{equation}

Then, we consider all components of $\fnorm{\Gammatp - \Gammastar}$. Taking (\ref{StrongConvexityGamma}), (\ref{LipschitzGradientGamma}) and Lemma \ref{Sample_Gamma} into (\ref{Iteration_Gamma}), we could derive
\begin{align}
	&\fnorm{\Gammatp - \Gammastar} \nonumber \\
	&\leq \sqrt{1 + \frac{2\sqrt{s_{\Gamma}^{\star}}}{\sqrt{s_{\Gamma} - s_{\Gamma}^{\star}}}} \cdot \bBa{\frac{\tau_{\max} \nu_{\max} - \tau_{\min} \nu_{\min}}{\tau_{\max} \nu_{\max} + \tau_{\min} \nu_{\min}} \fnorm{\Gammat - \Gammastar}+ \frac{2 \tau_{\max} R}{\tau_{\max} \nu_{\max} + \tau_{\min} \nu_{\min}} \fnorm{\Omegat - \Omegastar} \nonumber \\
		&\qquad + \eta_{\Gamma} \usup{\mV \in \calC_{2s_{\Gamma}} \cap \calS^{dm - 1}} \iprod{\mV}{\nabla_{\Gamma} f(\Gammat, \Omegat) - \nabla_{\Gamma} f_n(\Gammat, \Omegat)}} \nonumber \\
	&\leq (\frac{\rho_{\Gamma, \text{pop}}}{\sqrt{\rho_{\text{pop}}}} + \rho_{\Gamma, \text{sam}})\max(\fnorm{\Gammat - \Gammastar}, \fnorm{\Omegat - \Omegastar}) + \epsilon_{\Gamma} \nonumber \\
	&\leq (\sqrt{\rho_{\text{pop}}} + \rho_{\Gamma, \text{sam}})\max(\fnorm{\Gammat - \Gammastar}, \fnorm{\Omegat - \Omegastar}) + \epsilon_{\Gamma},
\end{align}
where the second inequality is based on the assumption of $s_{\Gamma}$ in \eqref{Sparsity_Constant_HT} and the third inequality is from \eqref{Def_Popular_Rate}.

Here
\begin{align}
	\rho_{\Gamma, \text{pop}} &= \frac{\tau_{\max} \nu_{\max} - \tau_{\min} \nu_{\min}}{\tau_{\max} \nu_{\max} + \tau_{\min} \nu_{\min}} + \frac{2 \tau_{\max} R}{\tau_{\max} \nu_{\max} + \tau_{\min} \nu_{\min}} = 1 - \frac{2 \tau_{\min} \nu_{\min} - 2 \tau_{\max} R}{\tau_{\max} \nu_{\max} + \tau_{\min} \nu_{\min}} \\
	\rho_{\Gamma, \text{sam}} &= \frac{C_{\Gamma, 1}}{\sqrt{\rho_{\text{pop}}}} \frac{\nu_{\max} \tau_{\max}}{\nu_{\max} \tau_{\max} + \nu_{\min} \tau_{\min}} (\frac{R}{\nu_{\max}} \frac{\omega_{\Gamma} + \omega_{\Omega} + u}{\sqrt{n}} + \frac{\omega_{\Gamma} + u}{\sqrt{n}} + \frac{1}{\sqrt{\tau_{\max} \nu_{\min}} \nu_{\max}} \frac{\omega_{\Gamma} + \omega_{\Omega} + u}{\sqrt{n}}) \nonumber \\
	&\leq \frac{C_{\Gamma, 2}}{\sqrt{\rho_{\text{pop}}}} \frac{\omega_{\Gamma} + \omega_{\Omega} + u}{\sqrt{n}} \\
	\epsilon_{\Gamma} &= \frac{C_{\Gamma, 1}}{\sqrt{\rho_{\text{pop}}}} \frac{\nu_{\max} \tau_{\max}}{\nu_{\max} \tau_{\max} + \nu_{\min} \tau_{\min}} \frac{1}{\sqrt{\tau_{\max} \nu_{\max}}} \frac{\omega_{\Gamma} + u}{\sqrt{n}} \leq \frac{C_{\Gamma, 1}}{\sqrt{\rho_{\text{pop}}}} \frac{1}{\sqrt{\tau_{\max} \nu_{\max}}} \frac{\omega_{\Gamma} + u}{\sqrt{n}}.
\end{align}

%When we set $R \leq 1$, we have
%\begin{align}
%	\rho_{\Gamma, \text{sam}} &\leq C (\frac{1}{\nu} \frac{\omega_{\Gamma} + \omega_{\Omega} + u}{\sqrt{n}} + \frac{\omega_{\Gamma} + u}{\sqrt{n}}) \nonumber \\
%	\epsilon_{\Gamma} &\leq C (\sqrt{\frac{1}{\nu \tau}} \frac{\omega_{\Gamma} + \omega_{\Omega} + u}{\sqrt{n}} + \sqrt{\frac{1}{\nu \tau}} \frac{\omega_{\Gamma} + u}{\sqrt{n}}).
%\end{align}
If we want $\fnorm{\Gammatp - \Gammastar} \leq R$, we need to guarantee
\begin{equation}
	(\sqrt{\rho_{\text{pop}}} + \rho_{\Gamma, \text{sam}}) R + \epsilon_{\Gamma} \leq R
\end{equation}
or
\begin{equation}
	\epsilon_{\Gamma} + \rho_{\Gamma, \text{sam}} R \leq (1 - \sqrt{\rho_{\text{pop}}}) R.
\end{equation}

Then, we could derive
\begin{align*}
	\epsilon_{\Gamma} + \rho_{\Gamma, \text{sam}} R &\leq \frac{C_{\Gamma, 1}}{\sqrt{\rho_{\text{pop}}}} \frac{\nu_{\max} \tau_{\max}}{\nu_{\max} \tau_{\max} + \nu_{\min} \tau_{\min}} ((\frac{R}{\nu_{\max}} \frac{\omega_{\Gamma} + \omega_{\Omega} + u}{\sqrt{n}} + \frac{\omega_{\Gamma} + u}{\sqrt{n}} + \frac{1}{\sqrt{\tau_{\max} \nu_{\min}} \nu_{\max}} \frac{\omega_{\Gamma} + \omega_{\Omega} + u}{\sqrt{n}}) R \\
	&\qquad + \frac{1}{\sqrt{\tau_{\max} \nu_{\max}}} \frac{\omega_{\Gamma} + u}{\sqrt{n}}) \\
	&\leq \frac{C_{\Gamma, 3}}{\sqrt{\rho_{\text{pop}}}} \frac{\omega_{\Gamma} + \omega_{\Omega} + u}{\sqrt{n}}.
\end{align*}
%which could be rewritten as
%\begin{align}
%	\begin{split}
%		\epsilon_{\Gamma} + \rho_{\Gamma, \text{sam}} R &\leq C (\sqrt{\frac{1}{\nu \tau}} \frac{\omega_{\Gamma} + \omega_{\Omega} + u}{\sqrt{n}} + \sqrt{\frac{1}{\nu \tau}} \frac{\omega_{\Gamma} + u}{\sqrt{n}} + (\frac{1}{\nu} \frac{\omega_{\Gamma} + \omega_{\Omega} + u}{\sqrt{n}} + \frac{\omega_{\Gamma} + u}{\sqrt{n}})R) \\
%		&\leq \sqrt{\rho_{\text{pop}}}(1 - \sqrt{\rho_{\text{pop}}}) R.
%	\end{split}
%\end{align}
When the number of measurements satisfies
%\begin{equation}
%	n \geq C \frac{\frac{1}{\nu \tau}(\omega_{\Gamma} + \omega_{\Omega} + u)^2}{\rho_{\text{pop}}(1 - \sqrt{\rho_{\text{pop}}})^2 R^2},
%\end{equation}
\begin{equation}
	\sqrt{n} \geq C_{\Gamma, 3} \frac{\omega_{\Gamma} + \omega_{\Omega} + u}{\sqrt{\rho_{\text{pop}}} (1 - \sqrt{\rho_{\text{pop}}}) R}
\end{equation}
we could guarantee $\fnorm{\Gammatp - \Gammastar} \leq R$.

Next, we consider all components of $\fnorm{\Omegatp - \Omegastar}$. Taking (\ref{StrongConvexityOmega}), (\ref{LipschitzGradientOmega}) and Lemma \ref{Sample_Omega} into (\ref{Iteration_Omega}), we could derive
\begin{align}
	&\fnorm{\Omegatp - \Omegastar} \nonumber \\
	&\leq \sqrt{1 + \frac{2\sqrt{s_{\Omega}^{\star}}}{\sqrt{s_{\Omega} - s_{\Omega}^{\star}}}} \cdot \bBa{\frac{16 \nu_{\max}^2 - \nu_{\min}^2 }{16 \nu_{\max}^2 + \nu_{\min}^2 } \fnorm{\Omegat - \Omegastar} + \frac{8 \nu_{\max}^2 \nu_{\min}^2 \tau_{\max} R}{16 \nu_{\max}^2 + \nu_{\min}^2 } \fnorm{\Gammat - \Gammastar} \nonumber \\
		&\qquad + \eta_{\Omega} \usup{\mV \in \calC_{2s_{\Omega}} \cap \calS^{m^2 - 1}} \iprod{\mV}{\nabla_{\Omega} f(\Gammat, \Omegat) - \nabla_{\Omega} f_n(\Gammat, \Omegat)}} \nonumber \\
	&\leq (\frac{\rho_{\Omega, \text{pop}}}{\sqrt{\rho_{\text{pop}}}} + \rho_{\Omega, \text{sam}}) \max(\fnorm{\Gammat - \Gammastar}, \fnorm{\Omegat - \Omegastar}) + \epsilon_{\Omega} \nonumber \\
	&\leq (\sqrt{\rho_{\text{pop}}} + \rho_{\Omega, \text{sam}}) \max(\fnorm{\Gammat - \Gammastar}, \fnorm{\Omegat - \Omegastar}) + \epsilon_{\Omega},
\end{align}
where the second inequality is based on the assumption of $s_{\Omega}$ in \eqref{Sparsity_Constant_HT} and the third inequality is from \eqref{Def_Popular_Rate}.

Here
\begin{align}
	\rho_{\Omega, \text{pop}} &= \frac{16 \nu_{\max}^2 - \nu_{\min}^2 }{16 \nu_{\max}^2 + \nu_{\min}^2 } + \frac{8 \nu_{\max}^2 \nu_{\min}^2 \tau_{\max} R}{16 \nu_{\max}^2 + \nu_{\min}^2 } = 1 - \frac{2 \nu_{\min}^2 - 8 \nu_{\max}^2 \nu_{\min}^2 \tau_{\max} R}{16 \nu_{\max}^2 + \nu_{\min}^2} \\
	\rho_{\Omega, \text{sam}} &= \frac{C_{\Omega, 1}}{\sqrt{\rho_{\text{pop}}}} \frac{8 \nu_{\max}^2 \nu_{\min}^2 }{16 \nu_{\max}^2 + \nu_{\min}^2 } (\tau_{\max} R \frac{\omega_{\Gamma} + \omega_{\Omega} + u}{\sqrt{n}} + \frac{\sqrt{\tau_{\max}} }{\sqrt{\nu_{\min}}} \frac{\omega_{\Gamma} + \omega_{\Omega} + u}{\sqrt{n}}) \nonumber \\
	&\leq \frac{C_{\Omega, 2}}{\sqrt{\rho_{\text{pop}}}} \frac{\omega_{\Gamma} + \omega_{\Omega} + u}{\sqrt{n}} \\
	\epsilon_{\Omega} &= \frac{C_{\Omega, 1}}{\sqrt{\rho_{\text{pop}}}} \frac{8 \nu_{\max}^2 \nu_{\min}^2 }{16 \nu_{\max}^2 + \nu_{\min}^2 } \frac{1}{\nu_{\min}} \frac{\omega_{\Omega} + u}{\sqrt{n}} \leq \frac{C_{\Omega, 1}}{\sqrt{\rho_{\text{pop}}}} \nu_{\min} \frac{\omega_{\Omega} + u}{\sqrt{n}}.
\end{align}

If we want $\fnorm{\Omegatp - \Omegastar} \leq R$, we need to guarantee
\begin{equation}
	(\sqrt{\rho_{\text{pop}}} + \rho_{\Omega, \text{sam}}) R + \epsilon_{\Omega} \leq R
\end{equation}
or
\begin{equation}
	\epsilon_{\Omega} + \rho_{\Omega, \text{sam}} R \leq (1 - \sqrt{\rho_{\text{pop}}}) R.
\end{equation}
Then, we could derive
\begin{align*}
	\epsilon_{\Omega} + \rho_{\Omega, \text{sam}} R &\leq \frac{C_{\Omega, 1}}{\sqrt{\rho_{\text{pop}}}} \frac{8 \nu_{\max}^2 \nu_{\min}^2 }{16 \nu_{\max}^2 + \nu_{\min}^2 } ((\tau_{\max} R \frac{\omega_{\Gamma} + \omega_{\Omega} + u}{\sqrt{n}} + \frac{\sqrt{\tau_{\max}} }{\sqrt{\nu_{\min}}} \frac{\omega_{\Gamma} + \omega_{\Omega} + u}{\sqrt{n}})R + \frac{1}{\nu_{\min}} \frac{\omega_{\Omega} + u}{\sqrt{n}}) \\
	&\leq \frac{C_{\Omega, 3}}{\sqrt{\rho_{\text{pop}}}} \frac{\omega_{\Gamma} + \omega_{\Omega} + u}{\sqrt{n}}.
\end{align*}

When the number of measurements satisfies
\begin{equation}
	\sqrt{n} \geq C_{\Omega, 3} \frac{\omega_{\Gamma} + \omega_{\Omega} + u}{\sqrt{\rho_{\text{pop}}}(1 - \sqrt{\rho_{\text{pop}}})R},
\end{equation}
we could guarantee $\fnorm{\Omegatp - \Omegastar} \leq R$.

Finally, we consider $\fnorm{\Gammatp - \Gammastar}$ and $\fnorm{\Omegatp - \Omegastar}$ as a whole and derive
\begin{align}
	&\max(\fnorm{\Gammatp - \Gammastar}, \fnorm{\Omegatp - \Omegastar}) \nonumber \\
	&\leq (\sqrt{\rho_{\text{pop}}} + \max(\rho_{\Gamma, \text{sam}}, \rho_{\Omega, \text{sam}})) \max(\fnorm{\Gammat - \Gammastar}, \fnorm{\Omegat - \Omegastar}) + \max(\epsilon_{\Gamma}, \epsilon_{\Omega}).
\end{align}

%We define the convergence parameter $\rho_{\text{pop}}$ associated with the population loss function as
%\begin{equation}
%	\begin{split}
%		\rho_{\text{pop}} &= \max(\rho_{\Gamma, \text{pop}}, \rho_{\Omega, \text{pop}}) \\
%		&= \max(\frac{\tau_{\max} \nu_{\max} - \tau_{\min} \nu_{\min}}{\tau_{\max} \nu_{\max} + \tau_{\min} \nu_{\min}} + \frac{2 \tau_{\max} R}{\tau_{\max} \nu_{\max} + \tau_{\min} \nu_{\min}}, \frac{16 \nu_{\max}^2 - \nu_{\min}^2 }{16 \nu_{\max}^2 + \nu_{\min}^2 } + \frac{8 \nu_{\max}^2 \nu_{\min}^2 \tau_{\max} R}{16 \nu_{\max}^2 + \nu_{\min}^2 }) \\
%		&= \max(1 - \frac{2 \tau_{\min} \nu_{\min} - 2 \tau_{\max} R}{\tau_{\max} \nu_{\max} + \tau_{\min} \nu_{\min}}, 1 - \frac{2 \nu_{\min}^2 - 8 \nu_{\max}^2 \nu_{\min}^2 \tau_{\max} R}{16 \nu_{\max}^2 + \nu_{\min}^2}) \\
%		&\leq \max(1 - \frac{\tau_{\min} \nu_{\min}}{\tau_{\max} \nu_{\max} + \tau_{\min} \nu_{\min}}, 1 - \frac{\nu_{\min}^2}{16 \nu_{\max}^2 + \nu_{\min}^2}),
%	\end{split}
%\end{equation}
%where the last inequality is from $R \leq \min(\frac{\tau_{\min} \nu_{\min}}{2 \tau_{\max}}, \frac{1}{8 \tau_{\max} \nu_{\max}^2})$, which guarantees $\rho_{\text{pop}} < 1$.

We also define $\rho_{\text{sam}} = \max(\rho_{\Gamma, \text{sam}}, \rho_{\Omega, \text{sam}})$ and $\epsilon = \max(\epsilon_{\Gamma}, \epsilon_{\Omega})$.

\subsection{Analysis of initialization (Proof of Theorem 2)}
The initialization of $\bmGamma$ is derived from the following optimization problem
\begin{align}
	&\umin{\bmGamma} \frac{1}{2} \fnorm{\bmY - \bmX \bmGamma}^2 \nonumber \\
	&\st \quad \lzeronorm{\Vect{\mGamma^T}} \leq s_{\Gamma}.
\end{align}
The initialization of $\bmOmega$ is derived from the following optimization problem
\begin{align}
	&\umin{\bmOmega} \frac{1}{2} \fnorm{\bmOmega - \bmS^{-1}}^2 \nonumber \\
	&\st \quad \lzeronorm{\Vect{\mOmega^T}} \leq s_{\Omega},
\end{align}
where $\bmS = (\bmY - \bmX \bmGamma_{\text{ini}})^T(\bmY - \bmX \bmGamma_{\text{ini}}) / n$.

The error $\fnorm{\Gammaini - \Gammastar}$ is analyzed as the Lasso.
\begin{lemma} \label{Initialization_Gamma_IHT}
	When $\sqrt{n} \geq C_{\Gamma, 4} \frac{\tau_{\max}}{\tau_{\min}} (\omega_{\Gamma} + u)$, we could derive
	\begin{equation}
		\fnorm{\Gammaini - \Gammastar} \leq C_{\Gamma, 5} \frac{\omega_{\Gamma} + u}{\sqrt{n}} \frac{\tau_{\max}^{\frac{1}{2}}}{\tau_{\min} \nu_{\min}^{\frac{1}{2}}},
	\end{equation}
	with probability at least $1 - 4 \exp(-u^2)$.
\end{lemma}
When $n \geq C_{\Gamma, 6} (\omega_{\Gamma} + u)^2 / R^2$, we could derive
\begin{equation}
	\fnorm{\Gammaini - \Gammastar} \leq R.
\end{equation}
The analysis of $\fnorm{\Omegaini - \Omegastar}$ is more complicated.
\begin{lemma} \label{Initialization_Omega_IHT}
	When $\sqrt{n} > C_{\Omega, 4} \frac{\tau_{\max} \nu_{\max}}{\tau_{\min} \nu_{\min}} (m + \omega_{\Gamma} + u)$, we could derive
	\begin{equation}
		\fnorm{\Omegaini - \Omegastar} \leq C_{\Omega, 5} \frac{\tau_{\max}^2 \nu_{\max}^2}{\tau_{\min}^2 \nu_{\min}} \frac{m + \omega_{\Gamma} + u}{\sqrt{n}},
	\end{equation}
	with probability at least $1 - 18 \exp(-u^2)$.
\end{lemma}
When $n > C_{\Omega, 6} (m + \omega_{\Gamma} + u)^2 / R^2$, we could derive $\fnorm{\bmOmega_{\text{ini}} - \Omegastar} \leq R$.

\section{Analysis of alternating projected gradient descent for general convex regularizers (Proof of Theorem \ref{ConvergenceAltPGD})}
Our analysis is based on the facts $\Gammat \in \ball{F}(\Gammastar, R)$ and $\Omegat \in \ball{F}(\Omegastar, R)$.
\subsection{Analysis of the iteration about $\bmGamma$}

With the following lemma, we could bound the distance between the point after projection and the point in the constraint by a supremum of a series of inner products.
\begin{lemma}\label{Difference_Projection}
	Suppose $\bar{\vx} = \calP_{\calK}(\vy)$, where $\calK = \ba{\vx \mid \Reg{\vx} \leq \Reg{\xreal}}$ and $\Reg{\cdot}$ is a convex function. Then we could bound $\ltwonorm{\bar{\vx} - \xreal}$ as
	\begin{equation}
		\ltwonorm{\bar{\vx} - \xreal} \leq \usup{\vv \in \calC \cap \SS_2} \iprod{\vv}{\vy - \xreal},
	\end{equation}
	where $\calC = \cone(\calD)$ is the decent cone, $\calD = \calK - \ba{\xreal}$ is the descent set and $\SS_2$ is the sphere with unit Euclidean norm.
\end{lemma}

%The following lemma lays a foundation for the analysis of gradient descent iterations.
%\begin{lemma} \cite{Nesterov:2014:ILC:2670022} \label{StrongConvexContrac}
%	Suppose $f(\bmx)$ is $\mu$-strongly convex and $L$-smooth. With the step size $\eta = 2 / (L + \mu)$, the gradient descent iteration would contract as
%	\begin{equation}
%		\ltwonorm{(\bmx - \eta \nabla f(\bmx) - \xreal)} \leq \frac{L - \mu}{L + \mu} \ltwonorm{\bmx - \xreal},
%	\end{equation}
%	where $\xreal$ is the optimal point.
%\end{lemma}
%
%We set the step sizes as
%\begin{equation}
%	\eta_{\Gamma} = \frac{1}{\nu_{\max} \tau_{\max} + \nu_{\min} \tau_{\min}} \quad \text{and} \quad \eta_{\Omega} = \frac{8 \nu_{\max}^2 \nu_{\min}^2 }{16 \nu_{\max}^2 + \nu_{\min}^2 }.
%\end{equation}

Now, we could rewrite $\fnorm{\Gammatp - \Gammastar}$ as
\begin{align}
	&\fnorm{\Gammatp - \Gammastar} \nonumber \\
	&= \fnorm{\calP_{\calK_{\Gamma}}(\Gammat - \eta_{\Gamma} \nabla_{\Gamma} f_n(\Gammat, \Omegat)) - \Gammastar} \\
	&\leq \usup{\mV \in \calC_{\Gamma} \cap \calS^{dm - 1}} \iprod{\mV}{\Gammat - \Gammastar - \eta_{\Gamma} \nabla_{\Gamma} f_n(\Gammat, \Omegat)} \\
	&= \usup{\mV \in \calC_{\Gamma} \cap \calS^{dm - 1}} \iprod{\mV}{\Gammat - \Gammastar - \eta_{\Gamma} \nabla_{\Gamma} f(\Gammat, \Omegastar) \\
		&\qquad + \eta_{\Gamma} (\nabla_{\Gamma} f(\Gammat, \Omegastar) - \nabla_{\Gamma} f(\Gammat, \Omegat)) + \eta_{\Gamma} (\nabla_{\Gamma} f(\Gammat, \Omegat) - \nabla_{\Gamma} f_n(\Gammat, \Omegat))} \\
	&\leq \fnorm{\Gammat - \eta_{\Gamma} \nabla_{\Gamma} f(\Gammat, \Omegastar) - \Gammastar} + \eta_{\Gamma} \fnorm{\nabla_{\Gamma} f(\Gammat, \Omegastar) - \nabla_{\Gamma} f(\Gammat, \Omegat)} \\
	&\qquad + \eta_{\Gamma} \usup{\mV \in \calC_{\Gamma} \cap \calS^{dm - 1}} \iprod{\mV}{\nabla_{\Gamma} f(\Gammat, \Omegat) - \nabla_{\Gamma} f_n(\Gammat, \Omegat)}, \label{Iteration_Gamma_AltPGD}
\end{align}
where the first inequality is based on Lemma \ref{Difference_Projection} and the last inequality is from the Cauchy–Schwarz inequality.

The first and second terms of (\ref{Iteration_Gamma_AltPGD}) have been bounded in the previous analysis.

%by the strong convexity and the smoothness of the population function $f(\bmGamma, \Omegastar)$ about $\bmGamma$ in Lemma \ref{SSGamma} and the corresponding convergence result in Lemma \ref{StrongConvexContrac}
%\begin{equation}
%	\fnorm{\Gammat - \eta_{\Gamma} \nabla_{\Gamma} f(\Gammat, \Omegastar) - \Gammastar} \leq \frac{\tau_{\max} \nu_{\max} - \tau_{\min} \nu_{\min}}{\tau_{\max} \nu_{\max} + \tau_{\min} \nu_{\min}} \fnorm{\Gammat - \Gammastar}. \label{StrongConvexityGamma}
%\end{equation}
%
%
%The second term of (\ref{Iteration_Gamma_AltPGD}) could be bounded by the Lipschitz property of $\nabla_{\Gamma} f(\bmGamma, \cdot)$ about $\bmOmega$ around $\Omegastar$ in Lemma \ref{LipschitzGradient}
%\begin{equation}
%	\fnorm{\nabla_{\Gamma} f(\Gammat, \Omegastar) - \nabla_{\Gamma} f(\Gammat, \Omegat)} \leq 2 \tau_{\max} R \fnorm{\Omegat - \Omegastar}. \label{LipschitzGradientGamma}
%\end{equation}
The third term of (\ref{Iteration_Gamma_AltPGD}) could be analyzed in the same way as Lemma \ref{Sample_Gamma} with a different set $\calC_{\Gamma}$.
\begin{lemma} \label{Sample_Gamma_AltPGD}
	Under Assumption \ref{SingularValue_Omega} and \ref{SingularValue_Sigma}, we set $\eta_{\Gamma} = \frac{1}{\nu_{\max} \tau_{\max} + \nu_{\min} \tau_{\min}}$. For any $\Gammat \in \ball{F}(\Gammastar, R)$ and $\Omegat \in \ball{F}(\Omegastar, R)$, the term $\eta_{\Gamma} \usup{\mV \in \calC_{\Gamma} \cap \calS^{dm - 1}} \iprod{\mV}{\nabla_{\Gamma} f(\Gammat, \Omegat) - \nabla_{\Gamma} f_n(\Gammat, \Omegat)}$ could be bounded by
	\begin{equation}
		\begin{split}
			&\eta_{\Gamma} \usup{\mV \in \calC_{\Gamma} \cap \calS^{dm - 1}} \iprod{\mV}{\nabla_{\Gamma} f(\Gammat, \Omegat) - \nabla_{\Gamma} f_n(\Gammat, \Omegat)}  \\
			& \leq C_{\Gamma, 1} \frac{\nu_{\max} \tau_{\max}}{\nu_{\max} \tau_{\max} + \nu_{\min} \tau_{\min}} (\frac{R}{\nu_{\max}} \frac{\bar{\omega}_{\Gamma} + \bar{\omega}_{\Omega} + u}{\sqrt{n}} \fnorm{\Omegat - \Omegastar} + \frac{\bar{\omega}_{\Gamma} + u}{\sqrt{n}} \fnorm{\Gammat - \Gammastar} \\
			&  + \frac{1}{\sqrt{\tau_{\max} \nu_{\min}} \nu_{\max}} \frac{\bar{\omega}_{\Gamma} + \bar{\omega}_{\Omega} + u}{\sqrt{n}} \fnorm{\Omegat - \Omegastar} + \frac{1}{\sqrt{\tau_{\max} \nu_{\max}}} \frac{\bar{\omega}_{\Gamma} + u}{\sqrt{n}}),
		\end{split}
	\end{equation}
	with probability at least $1 - 8 \exp(- u^2)$, when $n \geq  (\bar{\omega}_{\Gamma} + \bar{\omega}_{\Omega} + u)^2$.
\end{lemma}

\subsection{Analysis of the iteration about $\bmOmega$}
%Similarly, we write $\calT = \calT_{t+1} \cup \calT_{\star}$, where $\calT_{t+1}$ and $\calT_{\star}$ are the support sets of $\Omegatp$ and $\Omegastar$, respectively.

First, we could rearrange $\fnorm{\Omegatp - \Omegastar}$ as
\begin{align}
	&\fnorm{\Omegatp - \Omegastar} \nonumber \\
	&= \fnorm{\calP_{\calK_{\Omega}}(\Omegat - \eta_{\Omega} \nabla_{\Omega} f_n(\Gammat, \Omegat)) - \Omegastar} \\
	&\leq \usup{\mV \in \calC_{\Omega} \cap \calS^{m^2 - 1}} \iprod{\mV}{\Omegat - \Omegastar - \eta_{\Omega} \nabla_{\Omega} f_n(\Gammat, \Omegat)} \\
	&= \usup{\mV \in \calC_{\Omega} \cap \calS^{m^2 - 1}} \iprod{\mV}{\Omegat - \Omegastar - \eta_{\Omega} \nabla_{\Omega} f(\Gammastar, \Omegat) \\
		&\qquad + \eta_{\Omega} (\nabla_{\Omega} f(\Gammastar, \Omegat) - \nabla_{\Omega} f(\Gammat, \Omegat)) + \eta_{\Omega} (\nabla_{\Omega} f(\Gammat, \Omegat) - \nabla_{\Omega} f_n(\Gammat, \Omegat))} \\
	&\leq \fnorm{\Omegat - \eta_{\Omega} \nabla_{\Omega} f(\Gammastar, \Omegat) - \Omegastar} + \eta_{\Omega} \fnorm{\nabla_{\Omega} f(\Gammastar, \Omegat) - \nabla_{\Omega} f(\Gammat, \Omegat)} \\
	&\qquad + \eta_{\Omega} \usup{\mV \in \calC_{\Omega} \cap \calS^{m^2 - 1}} \iprod{\mV}{\nabla_{\Omega} f(\Gammat, \Omegat) - \nabla_{\Omega} f_n(\Gammat, \Omegat)}, \label{Iteration_Omega_AltPGD}
\end{align}
where the first inequality is based on Lemma \ref{Difference_Projection} and the last inequality is from the Cauchy–Schwarz inequality.

The first and second terms of (\ref{Iteration_Omega_AltPGD}) have been bounded in the previous analysis.

%by the strong convexity and the smoothness of the population function $f(\Gammastar, \bmOmega)$ about $\bmOmega$ in Lemma \ref{SSOmega} and the corresponding convergence result in Lemma \ref{StrongConvexContrac}
%\begin{equation}
%	\fnorm{\Omegat - \eta_{\Omega} \nabla_{\Omega} f(\Gammastar, \Omegat) - \Omegastar} \leq \frac{16 \nu_{\max}^2 - \nu_{\min}^2 }{16 \nu_{\max}^2 + \nu_{\min}^2 } \fnorm{\Omegat - \Omegastar}. \label{StrongConvexityOmega}
%\end{equation}
%The second term of (\ref{Iteration_Omega_AltPGD}) could be bounded by the Lipschitz property of $\nabla_{\Omega} f(\cdot, \bmOmega)$ about $\bmGamma$ around $\Gammastar$ in Lemma \ref{LipschitzGradient}
%\begin{equation}
%	\fnorm{\nabla_{\Omega} f(\Gammastar, \Omegat) - \nabla_{\Omega} f(\Gammat, \Omegat)} \leq \tau_{\max} R \fnorm{\Gammat - \Gammastar}. \label{LipschitzGradientOmega}
%\end{equation}
The third term of (\ref{Iteration_Omega_AltPGD}) could be analyzed in the same way as Lemma \ref{Sample_Omega} with a different set $\calC_{\Omega}$.
\begin{lemma} \label{Sample_Omega_AltPGD}
	Under the same condition as Lemma \ref{Sample_Gamma_AltPGD}. For any $\Gammat \in \ball{F}(\Gammastar, R)$ and $\Omegat \in \ball{F}(\Omegastar, R)$, the term \\$\eta_{\Omega} \usup{\mV \in \calC_{\Omega} \cap \calS^{m^2 - 1}} \iprod{\mV}{\nabla_{\Omega} f(\Gammat, \Omegat) - \nabla_{\Omega} f_n(\Gammat, \Omegat)}$ could be bounded by
	\begin{equation}
		\begin{split}
			&\eta_{\Omega} \usup{\mV \in \calC_{\Omega} \cap \calS^{m^2 - 1}} \iprod{\mV}{\nabla_{\Omega} f(\Gammat, \Omegat) - \nabla_{\Omega} f_n(\Gammat, \Omegat)} \\
			&\leq C_{\Omega, 1} \frac{8 \nu_{\max}^2 \nu_{\min}^2 }{16 \nu_{\max}^2 + \nu_{\min}^2 } (\tau_{\max} R \frac{\bar{\omega}_{\Gamma} + \bar{\omega}_{\Omega} + u}{\sqrt{n}} \fnorm{\Gammat - \Gammastar} + \frac{\sqrt{\tau_{\max}} }{\sqrt{\nu_{\min}}} \frac{\bar{\omega}_{\Gamma} + \bar{\omega}_{\Omega} + u}{\sqrt{n}}\fnorm{\Gammat - \Gammastar} + \frac{1}{\nu_{\min}} \frac{\bar{\omega}_{\Omega} + u}{\sqrt{n}}),
		\end{split}
	\end{equation}
	with probability at least $1 - 6 \exp(- u^2)$, when $n \geq  (\bar{\omega}_{\Gamma} + \bar{\omega}_{\Omega} + u)^2$.
\end{lemma}

\subsection{Analysis of the whole convergence result}

Then, we consider all components of $\fnorm{\Gammatp - \Gammastar}$ and derive
\begin{align}
	&\fnorm{\Gammatp - \Gammastar} \nonumber \\
	&\leq \frac{\tau_{\max} \nu_{\max} - \tau_{\min} \nu_{\min}}{\tau_{\max} \nu_{\max} + \tau_{\min} \nu_{\min}} \fnorm{\Gammat - \Gammastar}+ \frac{2 \tau_{\max} R}{\tau_{\max} \nu_{\max} + \tau_{\min} \nu_{\min}} \fnorm{\Omegat - \Omegastar} \nonumber \\
	&\qquad + \eta_{\Gamma} \usup{\mV \in \calC_{\Gamma} \cap \calS^{dm - 1}} \iprod{\mV}{\nabla_{\Gamma} f(\Gammat, \Omegat) - \nabla_{\Gamma} f_n(\Gammat, \Omegat)} \nonumber \\
	&\leq (\rho_{\Gamma, \text{pop}} + \rho_{\Gamma, \text{sam}})\max(\fnorm{\Gammat - \Gammastar}, \fnorm{\Omegat - \Omegastar}) + \epsilon_{\Gamma},
\end{align}
where
\begin{align}
	\rho_{\Gamma, \text{pop}} &= \frac{\tau_{\max} \nu_{\max} - \tau_{\min} \nu_{\min}}{\tau_{\max} \nu_{\max} + \tau_{\min} \nu_{\min}} + \frac{2 \tau_{\max} R}{\tau_{\max} \nu_{\max} + \tau_{\min} \nu_{\min}} = 1 - \frac{2 \tau_{\min} \nu_{\min} - 2 \tau_{\max} R}{\tau_{\max} \nu_{\max} + \tau_{\min} \nu_{\min}} \\
	\rho_{\Gamma, \text{sam}} &= C_{\Gamma, 1} \frac{\nu_{\max} \tau_{\max}}{\nu_{\max} \tau_{\max} + \nu_{\min} \tau_{\min}} (\frac{R}{\nu_{\max}} \frac{\bar{\omega}_{\Gamma} + \bar{\omega}_{\Omega} + u}{\sqrt{n}} + \frac{\bar{\omega}_{\Gamma} + u}{\sqrt{n}} + \frac{1}{\sqrt{\tau_{\max} \nu_{\min}} \nu_{\max}} \frac{\bar{\omega}_{\Gamma} + \bar{\omega}_{\Omega} + u}{\sqrt{n}}) \nonumber \\
	&\leq C_{\Gamma, 2} \frac{\bar{\omega}_{\Gamma} + \bar{\omega}_{\Omega} + u}{\sqrt{n}} \\
	\epsilon_{\Gamma} &= C_{\Gamma, 1} \frac{\nu_{\max} \tau_{\max}}{\nu_{\max} \tau_{\max} + \nu_{\min} \tau_{\min}} \frac{1}{\sqrt{\tau_{\max} \nu_{\max}}} \frac{\bar{\omega}_{\Gamma} + u}{\sqrt{n}} \leq C_{\Gamma, 1} \frac{1}{\sqrt{\tau_{\max} \nu_{\max}}} \frac{\bar{\omega}_{\Gamma} + u}{\sqrt{n}}.
\end{align}

%When we set $R \leq 1$, we have
%\begin{align}
%	\rho_{\Gamma, \text{sam}} &\leq C (\frac{1}{\nu} \frac{\bar{\omega}_{\Gamma} + \bar{\omega}_{\Omega} + u}{\sqrt{n}} + \frac{\bar{\omega}_{\Gamma} + u}{\sqrt{n}}) \nonumber \\
%	\epsilon_{\Gamma} &\leq C (\sqrt{\frac{1}{\nu \tau}} \frac{\bar{\omega}_{\Gamma} + \bar{\omega}_{\Omega} + u}{\sqrt{n}} + \sqrt{\frac{1}{\nu \tau}} \frac{\bar{\omega}_{\Gamma} + u}{\sqrt{n}}).
%\end{align}
If we want $\fnorm{\Gammatp - \Gammastar} \leq R$, we need to guarantee
\begin{equation}
	(\rho_{\Gamma, \text{pop}} + \rho_{\Gamma, \text{sam}}) R + \epsilon_{\Gamma} \leq R
\end{equation}
or
\begin{equation}
	\epsilon_{\Gamma} + \rho_{\Gamma, \text{sam}} R \leq (1 - \rho_{\text{pop}}) R.
\end{equation}

Then, we could derive
\begin{align*}
	\epsilon_{\Gamma} + \rho_{\Gamma, \text{sam}} R &\leq C_{\Gamma, 1} \frac{\nu_{\max} \tau_{\max}}{\nu_{\max} \tau_{\max} + \nu_{\min} \tau_{\min}} ((\frac{R}{\nu_{\max}} \frac{\bar{\omega}_{\Gamma} + \bar{\omega}_{\Omega} + u}{\sqrt{n}} + \frac{\bar{\omega}_{\Gamma} + u}{\sqrt{n}} + \frac{1}{\sqrt{\tau_{\max} \nu_{\min}} \nu_{\max}} \frac{\bar{\omega}_{\Gamma} + \bar{\omega}_{\Omega} + u}{\sqrt{n}}) R \\
	&\qquad + \frac{1}{\sqrt{\tau_{\max} \nu_{\max}}} \frac{\bar{\omega}_{\Gamma} + u}{\sqrt{n}}) \\
	&\leq C_{\Gamma, 3} \frac{\bar{\omega}_{\Gamma} + \bar{\omega}_{\Omega} + u}{\sqrt{n}}.
\end{align*}
%which could be rewritten as
%\begin{align}
%	\begin{split}
%		\epsilon_{\Gamma} + \rho_{\Gamma, \text{sam}} R &\leq C (\sqrt{\frac{1}{\nu \tau}} \frac{\bar{\omega}_{\Gamma} + \bar{\omega}_{\Omega} + u}{\sqrt{n}} + \sqrt{\frac{1}{\nu \tau}} \frac{\bar{\omega}_{\Gamma} + u}{\sqrt{n}} + (\frac{1}{\nu} \frac{\bar{\omega}_{\Gamma} + \bar{\omega}_{\Omega} + u}{\sqrt{n}} + \frac{\bar{\omega}_{\Gamma} + u}{\sqrt{n}})R) \\
%		&\leq \sqrt{\rho_{\text{pop}}}(1 - \sqrt{\rho_{\text{pop}}}) R.
%	\end{split}
%\end{align}
When the number of measurements satisfies
%\begin{equation}
%	n \geq C \frac{\frac{1}{\nu \tau}(\bar{\omega}_{\Gamma} + \bar{\omega}_{\Omega} + u)^2}{\rho_{\text{pop}}(1 - \sqrt{\rho_{\text{pop}}})^2 R^2},
%\end{equation}
\begin{equation}
	\sqrt{n} \geq C_{\Gamma, 3} \frac{\bar{\omega}_{\Gamma} + \bar{\omega}_{\Omega} + u}{(1 - \rho_{\Gamma, \text{pop}}) R}
\end{equation}
we could guarantee $\fnorm{\Gammatp - \Gammastar} \leq R$.

Next, we consider all components of $\fnorm{\Omegatp - \Omegastar}$ and derive
\begin{align}
	&\fnorm{\Omegatp - \Omegastar} \nonumber \\
	&\leq \frac{16 \nu_{\max}^2 - \nu_{\min}^2 }{16 \nu_{\max}^2 + \nu_{\min}^2 } \fnorm{\Omegat - \Omegastar} + \frac{8 \nu_{\max}^2 \nu_{\min}^2 \tau_{\max} R}{16 \nu_{\max}^2 + \nu_{\min}^2 } \fnorm{\Gammat - \Gammastar} \nonumber \\
	&\qquad + \eta_{\Omega} \usup{\mV \in \calC_{\Omega} \cap \calS^{m^2 - 1}} \iprod{\mV}{\nabla_{\Omega} f(\Gammat, \Omegat) - \nabla_{\Omega} f_n(\Gammat, \Omegat)} \nonumber \\
	&\leq (\rho_{\Omega, \text{pop}} + \rho_{\Omega, \text{sam}}) \max(\fnorm{\Gammat - \Gammastar}, \fnorm{\Omegat - \Omegastar}) + \epsilon_{\Omega},
\end{align}
where
\begin{align}
	\rho_{\Omega, \text{pop}} &= \frac{16 \nu_{\max}^2 - \nu_{\min}^2 }{16 \nu_{\max}^2 + \nu_{\min}^2 } + \frac{8 \nu_{\max}^2 \nu_{\min}^2 \tau_{\max} R}{16 \nu_{\max}^2 + \nu_{\min}^2 } = 1 - \frac{2 \nu_{\min}^2 - 8 \nu_{\max}^2 \nu_{\min}^2 \tau_{\max} R}{16 \nu_{\max}^2 + \nu_{\min}^2} \\
	\rho_{\Omega, \text{sam}} &= C_{\Omega, 1} \frac{8 \nu_{\max}^2 \nu_{\min}^2 }{16 \nu_{\max}^2 + \nu_{\min}^2 } (\tau_{\max} R \frac{\bar{\omega}_{\Gamma} + \bar{\omega}_{\Omega} + u}{\sqrt{n}} + \frac{\sqrt{\tau_{\max}} }{\sqrt{\nu_{\min}}} \frac{\bar{\omega}_{\Gamma} + \bar{\omega}_{\Omega} + u}{\sqrt{n}}) \nonumber \\
	&\leq C_{\Omega, 2} \frac{\bar{\omega}_{\Gamma} + \bar{\omega}_{\Omega} + u}{\sqrt{n}} \\
	\epsilon_{\Omega} &= C_{\Omega, 1} \frac{8 \nu_{\max}^2 \nu_{\min}^2 }{16 \nu_{\max}^2 + \nu_{\min}^2 } \frac{1}{\nu_{\min}} \frac{\bar{\omega}_{\Omega} + u}{\sqrt{n}} \leq C_{\Omega, 1} \nu_{\min} \frac{\bar{\omega}_{\Omega} + u}{\sqrt{n}}.
\end{align}

If we want $\fnorm{\Omegatp - \Omegastar} \leq R$, we need to guarantee
\begin{equation}
	(\rho_{\Omega, \text{pop}} + \rho_{\Omega, \text{sam}}) R + \epsilon_{\Omega} \leq R
\end{equation}
or
\begin{equation}
	\epsilon_{\Omega} + \rho_{\Omega, \text{sam}} R \leq (1 - \rho_{\Omega, \text{pop}}) R.
\end{equation}
Then, we could derive
\begin{align*}
	\epsilon_{\Omega} + \rho_{\Omega, \text{sam}} R &\leq C_{\Omega, 1} \frac{8 \nu_{\max}^2 \nu_{\min}^2 }{16 \nu_{\max}^2 + \nu_{\min}^2 } ((\tau_{\max} R \frac{\bar{\omega}_{\Gamma} + \bar{\omega}_{\Omega} + u}{\sqrt{n}} + \frac{\sqrt{\tau_{\max}} }{\sqrt{\nu_{\min}}} \frac{\bar{\omega}_{\Gamma} + \bar{\omega}_{\Omega} + u}{\sqrt{n}})R + \frac{1}{\nu_{\min}} \frac{\bar{\omega}_{\Omega} + u}{\sqrt{n}}) \\
	&\leq C_{\Omega, 3} \frac{\bar{\omega}_{\Gamma} + \bar{\omega}_{\Omega} + u}{\sqrt{n}}.
\end{align*}

When the number of measurements satisfies
\begin{equation}
	\sqrt{n} \geq C_{\Omega, 3} \frac{\bar{\omega}_{\Gamma} + \bar{\omega}_{\Omega} + u}{(1 - \rho_{\Omega, \text{pop}})R},
\end{equation}
we could guarantee $\fnorm{\Omegatp - \Omegastar} \leq R$.

Finally, we consider $\fnorm{\Gammatp - \Gammastar}$ and $\fnorm{\Omegatp - \Omegastar}$ as a whole and derive
\begin{align}
	&\max(\fnorm{\Gammatp - \Gammastar}, \fnorm{\Omegatp - \Omegastar}) \nonumber \\
	&\leq (\max(\rho_{\Gamma, \text{pop}}, \rho_{\Omega, \text{pop}}) + \max(\rho_{\Gamma, \text{sam}}, \rho_{\Omega, \text{sam}})) \max(\fnorm{\Gammat - \Gammastar}, \fnorm{\Omegat - \Omegastar}) + \max(\epsilon_{\Gamma}, \epsilon_{\Omega}).
\end{align}

We define the convergence parameter $\rho_{\text{pop}}$ associated with the population loss function as
\begin{equation}
	\begin{split}
		\rho_{\text{pop}} &= \max(\rho_{\Gamma, \text{pop}}, \rho_{\Omega, \text{pop}}) \\
		&= \max(\frac{\tau_{\max} \nu_{\max} - \tau_{\min} \nu_{\min}}{\tau_{\max} \nu_{\max} + \tau_{\min} \nu_{\min}} + \frac{2 \tau_{\max} R}{\tau_{\max} \nu_{\max} + \tau_{\min} \nu_{\min}}, \frac{16 \nu_{\max}^2 - \nu_{\min}^2 }{16 \nu_{\max}^2 + \nu_{\min}^2 } + \frac{8 \nu_{\max}^2 \nu_{\min}^2 \tau_{\max} R}{16 \nu_{\max}^2 + \nu_{\min}^2 }) \\
		&= \max(1 - \frac{2 \tau_{\min} \nu_{\min} - 2 \tau_{\max} R}{\tau_{\max} \nu_{\max} + \tau_{\min} \nu_{\min}}, 1 - \frac{2 \nu_{\min}^2 - 8 \nu_{\max}^2 \nu_{\min}^2 \tau_{\max} R}{16 \nu_{\max}^2 + \nu_{\min}^2}) \\
		&\leq \max(1 - \frac{\tau_{\min} \nu_{\min}}{\tau_{\max} \nu_{\max} + \tau_{\min} \nu_{\min}}, 1 - \frac{\nu_{\min}^2}{16 \nu_{\max}^2 + \nu_{\min}^2}),
	\end{split}
\end{equation}
where the last inequality is from $R \leq \min(\frac{\tau_{\min} \nu_{\min}}{2 \tau_{\max}}, \frac{1}{8 \tau_{\max} \nu_{\max}^2})$, which guarantees $\rho_{\text{pop}} < 1$.

We also define $\rho_{\text{sam}} = \max(\rho_{\Gamma, \text{sam}}, \rho_{\Omega, \text{sam}})$ and $\epsilon = \max(\epsilon_{\Gamma}, \epsilon_{\Omega})$.

The proof of Corollary \ref{General_Initialization} is the same as Theorem \ref{InitializationIHT} apart from a different set $\calC_{\Gamma}$.

\section{Analysis of ordinary projected gradient descent (Proof of Corollary \ref{PGD_Sparse})}
In this condition, the loss function becomes
\begin{equation}
	f_n(\bmGamma) = \frac{1}{2n} \tr((\bmY - \bmX \bmGamma) \Omegastar (\bmY - \bmX \bmGamma)^T).
\end{equation}
The corresponding gradients and Hessian matrix are
\begin{align}
	\nabla f_n(\bmGamma) &= \frac{1}{n} \bmX^T \bmX (\bmGamma - \Gammastar) \Omegastar - \frac{1}{n} \bmX^T \bmE \Omegastar \\
	\nabla f (\bmGamma) &= \bmSigma_{\bmX} (\bmGamma - \Gammastar) \Omegastar \\
	\nabla^{2} f (\bmGamma) &= \Omegastar \otimes  \bmSigma_{\bmX}. \label{HessianIHT}
\end{align}
We set the step sizes as
\begin{equation}
	\eta_{\Gamma} = \frac{2}{\tau_{\max} \nu_{\max} + \tau_{\min} \nu_{\min}}.
\end{equation}

We could write the projected gradient descent iteration as
\begin{align*}
	\fnorm{\Gammatp - \Gammastar} &= \fnorm{\calP_{\calK_{\Gamma}}(\Gammat - \eta_{\Gamma} \nabla f_n(\Gammat)) - \Gammastar} \\
	&\leq \usup{\mV \in \calC_{\Gamma} \cap \calS^{dm - 1}} \iprod{\mV}{\Gammat - \Gammastar - \eta_{\Gamma} \nabla f_n(\Gammat)}  \nonumber \\
	&\leq \fnorm{\Gammat - \Gammastar - \eta_{\Gamma} \nabla f(\Gammat) } + \eta_{\Gamma} \usup{\mV \in \calC_{\Gamma} \cap \calS^{dm - 1}} \iprod{\mV}{\nabla f(\Gammat) - \nabla f_n(\Gammat)} ,
\end{align*}
where the first inequality is based on Lemma \ref{Difference_Projection} and the second inequality is from the Cauchy–Schwarz inequality.

The first term could be bounded by the strong convexity and the smoothness of $f(\bmGamma)$, which could be derived from the Hessian matrix $\nabla^2 f(\bmGamma)$ (\ref{HessianIHT}) and Assumption \ref{SingularValue_Omega}, \ref{SingularValue_Sigma}. With Lemma \ref{StrongConvexContrac}, we have
\begin{equation}
	\fnorm{\Gammat - \eta_{\Gamma} \nabla f(\Gammat) - \Gammastar} \leq \frac{\tau_{\max} \nu_{\max} - \tau_{\min} \nu_{\min}}{\tau_{\max} \nu_{\max} + \tau_{\min} \nu_{\min}} \fnorm{\Gammat - \Gammastar}.
\end{equation}

The second term could be rewritten as
\begin{equation}
	\begin{split}
		&\eta_{\Gamma} \usup{\mV \in \calC_{\Gamma} \cap \calS^{dm - 1}} \iprod{\mV}{\nabla f(\Gammat) - \nabla f_n(\Gammat)} = \eta_{\Gamma} \usup{\mV \in \calC_{\Gamma} \cap \calS^{dm - 1}} \iprod{\mV}{(\bmSigma_{\bmX} - \frac{1}{n} \bmX^T \bmX) (\Gammat - \Gammastar) \Omegastar + \frac{1}{n} \bmX^T \bmE \Omegastar} \\
		&\leq \eta_{\Gamma} \usup{\mV \in \calC_{\Gamma} \cap \calS^{dm - 1}} \iprod{\mV}{(\bmSigma_{\bmX} - \frac{1}{n} \bmX^T \bmX) (\Gammat - \Gammastar) \Omegastar} + \eta_{\Gamma} \usup{\mV \in \calC_{\Gamma} \cap \calS^{dm - 1}} \iprod{\mV}{\frac{1}{n} \bmX^T \bmE \Omegastar}.
	\end{split}
\end{equation}
These two parts have been analyzed in Lemma \ref{Sample_Gamma}. The next two lemmas follow the same procedures as Lemma \ref{SampleGammaLemmaTwo} and Lemma \ref{SampleGammaLemmaFour}.
\begin{lemma}
	Under the condition of $n \geq  (\bar{\omega}_{\Gamma} + u)^2$, we could derive
	\begin{equation}
		P(\usup{\bmU, \bmV \in \calC_{\Gamma} \cap \calS^{dm - 1}} \iprod{\bmV}{(\bmSigma_{\bmX} - \frac{\bmX^T \bmX}{n}) \bmU \Omegastar} > C_{\Gamma, 7} \norm{\bmSigma_{\bmX}} \norm{\Omegastar} (\frac{\bar{\omega}_{\Gamma} + u}{\sqrt{n}})) \leq 2 \exp(- u^2).
	\end{equation}
\end{lemma}
\begin{lemma}
	Under the condition of $n \geq  (\bar{\omega}_{\Gamma} + u)^2$, we could derive
	\begin{equation}
		P(\usup{\bmV \in \calC_{\Gamma} \cap \calS^{dm - 1}} \iprod{\bmV}{\frac{1}{n} \bmX^T \bmE \Omegastar} > C_{\Gamma, 8} \norm{\bmSigma_{\bmX}^{\frac{1}{2}}} \norm{\Omegastar^{\frac{1}{2}}} (\frac{\bar{\omega}_{\Gamma} + u}{\sqrt{n}})) \leq 2 \exp(- u^2).
	\end{equation}
\end{lemma}
We set
\begin{equation}
	\rho_{\text{pop}} = \frac{\tau_{\max} \nu_{\max} - \tau_{\min} \nu_{\min}}{\tau_{\max} \nu_{\max} + \tau_{\min} \nu_{\min}} = 1 - \frac{2 \tau_{\min} \nu_{\min}}{\tau_{\max} \nu_{\max} + \tau_{\min} \nu_{\min}}.
\end{equation}
%By the assumption $s_{\Gamma} \geq (1 + 4(1 / \rho - 1)^2) s_{\Gamma}^{\star}$, we could derive
%\begin{equation}
%	\sqrt{1 + \frac{2\sqrt{s_{\Gamma}^{\star}}}{\sqrt{s_{\Gamma} - s_{\Gamma}^{\star}}}} \leq \frac{1}{\sqrt{\rho_{\text{pop}}}}.
%\end{equation}
When $n \geq (\bar{\omega}_{\Gamma} + u)^2$, we could derive
\begin{align*}
	&\fnorm{\Gammatp - \Gammastar} \\
	&\leq (\rho_{\text{pop}} + 2 C_{\Gamma, 7} \frac{\tau_{\max} \nu_{\max}}{\tau_{\max} \nu_{\max} + \tau_{\min} \nu_{\min}} \frac{\bar{\omega}_{\Gamma} + u}{\sqrt{n}}) \fnorm{\Gammat - \Gammastar} + 2C_{\Gamma, 8}  \frac{\tau_{\max} \nu_{\max}}{\tau_{\max} \nu_{\max} + \tau_{\min} \nu_{\min}} \frac{1}{\sqrt{\tau_{\max} \nu_{\max}}} \frac{\bar{\omega}_{\Gamma} + u}{\sqrt{n}}\\
	&\leq (\rho_{\text{pop}} + \rho_{\Gamma, \text{sam}}) \fnorm{\Gammat - \Gammastar} + \epsilon,
\end{align*}
with probability at least $1 - 4 \exp(-u^2)$.

Here we define
\begin{align}
	\rho_{\Gamma, \text{sam}} = 2 C_{\Gamma, 7} \frac{\tau_{\max} \nu_{\max}}{\tau_{\max} \nu_{\max} + \tau_{\min} \nu_{\min}} \frac{\bar{\omega}_{\Gamma} + u}{\sqrt{n}} \le C_{\Gamma, 9} \frac{\bar{\omega}_{\Gamma} + u}{\sqrt{n}},
\end{align}
and
\begin{align}
	\epsilon = 2C_{\Gamma, 8}  \frac{\tau_{\max} \nu_{\max}}{\tau_{\max} \nu_{\max} + \tau_{\min} \nu_{\min}} \frac{1}{\sqrt{\tau_{\max} \nu_{\max}}} \frac{\bar{\omega}_{\Gamma} + u}{\sqrt{n}} \leq C_{\Gamma, 10} \frac{1}{\sqrt{\tau_{\max} \nu_{\max}}} \frac{\bar{\omega}_{\Gamma} + u}{\sqrt{n}}.
\end{align}

\section{Proof of technical lemmas}
We use $C$ and $c$ to denote positive constants which might change from line to line throughout this part.

\subsection{Proof of Lemma \ref{Quadratic_form_XX}}
This lemma could be viewed as a proposition of the Hanson-Wright inequality.
\begin{lemma}[Hanson-Wright inequality \cite{Rudelson2013HansonWri}] \label{Hanson_Wright_Inequality}
	Suppose $\bmx$ is a random vector with independent sub-Gaussian components $\bmx_i$ satisfying $\EE[\bmx_i] = 0$ and $\subgnorm{\bmx_i} \leq K$. $\bmA \in \RR^{n \times n}$ is a fixed matrix. For $u > 0$, we could get
	\begin{equation}
		P(\abs{\bmx^T \bmA \bmx - \EE \bmx^T \bmA \bmx} > u) \leq 2 \exp(-c \min(-\frac{u^2}{K^4 \fnorm{\bmA}^2}, \frac{u}{K^2 \norm{\bmA}})),
	\end{equation}
	where $c > 0$ is a constant.
\end{lemma}
We could rearrange
\begin{equation}
	\tr(\bmX \bmU \bmX^T) = \Vect{\bmX^T}^T(\bmI_n \otimes \bmU) \Vect{\bmX^T} = \Vect{\bmX^T}^T\bmUpsilon_{\bmX}^{-\frac{1}{2}}\bmUpsilon_{\bmX}^{\frac{1}{2}}(\bmI_n \otimes \bmU) \bmUpsilon_{\bmX}^{\frac{1}{2}} \bmUpsilon_{\bmX}^{-\frac{1}{2}} \Vect{\bmX^T}.
\end{equation}
In this way, $\bmUpsilon_{\bmX}^{-\frac{1}{2}} \Vect{\bmX^T}$ becomes an isotropic Gaussian vector. Combining the rotation invariance of Gaussian vectors, we could derive
\begin{align*}
	P(\abs{\tr(\bmX \bmU \bmX^T) - \EE \tr(\bmX \bmU \bmX^T)} > u) &= P(\abs{\bmg^T \bmUpsilon_{\bmX}^{\frac{1}{2}}(\bmI_n \otimes \bmU) \bmUpsilon_{\bmX}^{\frac{1}{2}} \bmg - \EE \bmg^T \bmUpsilon_{\bmX}^{\frac{1}{2}}(\bmI_n \otimes \bmU) \bmUpsilon_{\bmX}^{\frac{1}{2}} \bmg} > u) \\
	&\leq 2 \exp(- c \min (\frac{u^2}{ \fnorm{\bmUpsilon_{\bmX}^{\frac{1}{2}}(\bmI_n \otimes \bmU) \bmUpsilon_{\bmX}^{\frac{1}{2}}}^2}, \frac{u}{\norm{\bmUpsilon_{\bmX}^{\frac{1}{2}}(\bmI_n \otimes \bmU) \bmUpsilon_{\bmX}^{\frac{1}{2}}}})) \\
	&\leq 2 \exp(- c \min (\frac{u^2}{n \norm{\bmUpsilon_{\bmX}}^2 \fnorm{\bmU}^2}, \frac{u}{\norm{\bmUpsilon_{\bmX}} \fnorm{\bmU}})),
\end{align*}
where $\bmg$ is a vector with independent standard Gaussian entries and the first inequality is based on Lemma \ref{Hanson_Wright_Inequality}. In the second inequality, we use $\fnorm{\bmA \bmB} \leq \norm{\bmA} \fnorm{\bmB}$, $\norm{\bmA \bmB} \leq \norm{\bmA} \norm{\bmB}$ and $\norm{\bmA} \leq \fnorm{\bmA}$ for two matrices $\bmA$ and $\bmB$.

\subsection{Proof of Lemma \ref{Quadratic_form_Independent}}
This lemma could be viewed as an extension of the Bernstein's inequality (Theorem 2.8.1 in \cite{vershynin2018HDP}).
%And we note the proof is similar to the one for the Hanson-Wright inequality after decoupling (Theorem 6.2.1 in \cite{vershynin2018HDP}).

From the independence between $\bmX$ and $\bmE$ and the rotation invariance of Gaussian vectors, we could derive
\begin{align*}
	P(\abs{\tr(\bmE \bmU \bmX^T)} > u) &= P(\abs{\Vect{\bmX^T}^T(\bmI_n \otimes \bmU^T) \Vect{\bmE^T}} > u) = P(\abs{\bmg_{\bmX}^T\bmUpsilon_{\bmX}^{\frac{1}{2}}(\bmI_n \otimes \bmU^T) \bmUpsilon_{\bmE}^{\frac{1}{2}} \bmg_{\bmE}} > u),
\end{align*}
where $\bmg_{\bmE}$ and $\bmg_{\bmX}$ are two independent vectors with independent standard Gaussian entries.

We set $\bmQ = \bmUpsilon_{\bmX}^{\frac{1}{2}}(\bmI_n \otimes \bmU^T) \bmUpsilon_{\bmE}^{\frac{1}{2}}$ with the singular value decomposition $\bmU_{\bmQ} \bmSigma_{\bmQ} \bmV_{\bmQ}$, where $\bmU_{\bmQ}$ and $\bmV_{\bmQ}$ are two unitary matrices.

We adopt the rotation invariance of Gaussian vectors again and derive
\begin{align*}
	P(\abs{\bmg_{\bmX}^T\bmUpsilon_{\bmX}^{\frac{1}{2}}(\bmI_n \otimes \bmU^T) \bmUpsilon_{\bmE}^{\frac{1}{2}} \bmg_{\bmE}} > u) &= P(\abs{\bmg_{\bmX}^T \bmU_{\bmQ} \bmSigma_{\bmQ} \bmV_{\bmQ} \bmg_{\bmE}} > u) \\
	&= P(\abs{\bar{\bmg}_{\bmX}^T \bmSigma_{\bmQ} \bar{\bmg}_{\bmE}} > u) \\
	&= P(\abs{\sum_{i = 1}^{nm} \sigma_{i} \bar{g}_{x}^{i} \bar{g}_{e}^{i}}> u) \\
	&\leq 2 \exp(- c \min (\frac{u^2}{ \fnorm{\bmSigma_{\bmQ}}^2}, \frac{u}{\norm{\bmSigma_{\bmQ}}})) \\
	&\leq 2 \exp(- c \min (\frac{u^2}{n \norm{\bmUpsilon_{\bmX}^{\frac{1}{2}}}^2 \norm{\bmUpsilon_{\bmE}^{\frac{1}{2}}}^2 \fnorm{\bmU}^2}, \frac{u}{\norm{\bmUpsilon_{\bmX}^{\frac{1}{2}}} \norm{\bmUpsilon_{\bmE}^{\frac{1}{2}}} \fnorm{\bmU}})),
\end{align*}
where $\bar{\bmg}_{\bmE}$ and $\bar{\bmg}_{\bmX}$ are two independent vectors with independent standard Gaussian entries, $\ba{\bar{g}_{x}^{i}}$ and $\ba{\bar{g}_{e}^{i}}$ are entries of $\bar{\bmg}_{\bmE}$ and $\bar{\bmg}_{\bmX}$ respectively, $\ba{\sigma_{i}}$ are singular values of $\mQ$, for $i = 1, \cdots, nm$. Here, we suppose $m < d$. In the second equality, we use the rotation invariance of Gaussian vectors. The first inequality is based on the Bernstein's inequality for the sum of the product of independent Gaussian variables. We also use $\fnorm{\mQ} = \fnorm{\mSigma_{\mQ}}$, $\norm{\mQ} = \norm{\mSigma_{\mQ}}$ and $\fnorm{\bmA \bmB} \leq \norm{\bmA} \fnorm{\bmB}$, $\norm{\bmA \bmB} \leq \norm{\bmA} \norm{\bmB}$, $\norm{\bmA} \leq \fnorm{\bmA}$ for two matrices $\bmA$ and $\bmB$ in the last inequality.

\subsection{Proof of Lemma \ref{Difference_Projection}}
From the definition of projection, $\bar{\vx}$ is the optimal solution of the following optimization problem
\begin{align}
	\bar{\vx} = \uargmin{\vx} \iota_{\calK}(\vx) + \frac{1}{2} \ltwonorm{\vx - \vy}^2,
\end{align}
where $\iota_{\calK}(\cdot)$ is the indicator function defined as
\begin{align}
	\iota_{\calK}(\vx) = \left\{
	\begin{array}{cc}
		0 & \text{if } \vx \in \calK, \\
		\infty & \text{otherwise}.
	\end{array}
	\right.
\end{align}

According to the fact that $\bar{\vx}$ is the optimal solution, we could derive
\begin{equation}
	\vzero \in \partial \iota_{\calK}(\bar{\vx}) + \bar{\vx} - \vy = \partial \iota_{\calK}(\bar{\vx}) + \bar{\vx} -\xreal + \xreal - \vy.
\end{equation}

After reformulation, we could derive
\begin{equation}
	-(\bar{\vx} -\xreal + \xreal - \vy) \in \partial \iota_{\calK}(\bar{\vx}) = N(\bar{\vx}; \calK),
\end{equation}
where $N(\bar{\vx}; \calK)$ is the normal cone of $\calK$ at $\bar{\vx}$. Here we adopt the fact that $\partial \iota_{\calK}(\bar{\vx}) = N(\bar{\vx}; \calK)$ from \cite[Example 2.32]{Mordukhovich2013AnEasyP} and the normal cone at $\bar{\vx} \in \calK$ is defined in \cite[Definition 9]{Mordukhovich2013AnEasyP} as
\begin{equation}\label{Definition_NormalCone}
	N(\bar{\vx}; \calK) \coloneqq \ba{\vv \mid \iprod{\vv}{\vx - \bar{\vx}} \leq 0,~\forall \vx \in \calK}.
\end{equation}

Combining with the definition of normal cone \eqref{Definition_NormalCone}, we could get
\begin{equation}
	\iprod{-(\bar{\vx} -\xreal + \xreal - \vy)}{\xreal - \bar{\vx}} \leq 0,
\end{equation}
where we use the fact $\xreal \in \calK$.

Then it is easy to verify that
\begin{equation}
	\ltwonorm{\bar{\vx} - \xreal}^2 \leq \iprod{\bar{\vx} - \xreal}{\vy - \xreal} \leq \usup{\vv \in \calC \cap \SS_{2}} \iprod{\vv}{\vy - \xreal} \ltwonorm{\bar{\vx} - \xreal},
\end{equation}
where the second inequality is from $(\bar{\vx} - \xreal) / \ltwonorm{\bar{\vx} - \xreal} \in \calC \cap \SS_{2}$.

\subsection{Proof of Lemma \ref{Sample_Gamma}}
We first rewrite $\nabla_{\Gamma} f(\Gammat, \Omegat) - \nabla_{\Gamma} f_n(\Gammat, \Omegat)$ as
\begin{equation}
	\begin{split}
		&\nabla_{\Gamma} f(\Gammat, \Omegat) - \nabla_{\Gamma} f_n(\Gammat, \Omegat)  \\
		&= 2\bmSigma_{\bmX} (\Gammat - \Gammastar) \Omegat - \frac{2}{n} \bmX^T \bmX (\Gammat - \Gammastar) \Omegat - \frac{2}{n} \bmX^T \bmE \Omegat  \\
		&= 2(\bmSigma_{\bmX} - \frac{\bmX^T \bmX}{n}) (\Gammat - \Gammastar) (\Omegat - \Omegastar) + 2(\bmSigma_{\bmX} - \frac{\bmX^T \bmX}{n}) (\Gammat - \Gammastar) \Omegastar - \frac{2}{n} \bmX^T \bmE (\Omegat - \Omegastar) - \frac{2}{n} \bmX^T \bmE \Omegastar.
	\end{split} \label{Sample_Difference_Gamma}
\end{equation}

With the definition of $\calC_{2 s_{\Gamma}}$, we could derive
\begin{equation}
	\fnorm{(\nabla_{\Gamma} f(\Gammat, \Omegat) - \nabla_{\Gamma} f_n(\Gammat, \Omegat))_{\calI}} \leq \usup{\bmV \in \calC_{2 s_{\Gamma}} \cap \calS^{dm - 1}} \iprod{\bmV}{\nabla_{\Gamma} f(\Gammat, \Omegat) - \nabla_{\Gamma} f_n(\Gammat, \Omegat)},
\end{equation}
where we use the fact $\Card(\calI) \leq 2 s_{\Gamma}$.

In this way,  to bound $\fnorm{(\nabla_{\Gamma} f(\Gammat, \Omegat) - \nabla_{\Gamma} f_n(\Gammat, \Omegat))_{\calI}}$, we need to deal with four suprema of random processes.

%In this way,  to bound $\usup{\bmV \in \calC_{2s_{\Gamma}} \cap \calS^{dm - 1}} \iprod{\bmV}{\nabla_{\Gamma} f(\Gammat, \Omegat) - \nabla_{\Gamma} f_n(\Gammat, \Omegat)}$, we need to deal with the four suprema of random processes.

The supreme of the random process associated with the first term of (\ref{Sample_Difference_Gamma}) could be bounded by Lemma \ref{Quadratic_form_XX} and \ref{Suprema}. We need to verify it has a mixed tail. We rewrite the random process as
\begin{equation}
	\iprod{\bmV}{2(\bmSigma_{\bmX} - \frac{\bmX^T \bmX}{n}) (\Gammat - \Gammastar) (\Omegat - \Omegastar)} = \iprod{\bmV}{2(\bmSigma_{\bmX} - \frac{\bmX^T \bmX}{n}) \bmP \bmU} \fnorm{\Gammat - \Gammastar} \fnorm{\Omegat - \Omegastar},
\end{equation}
where $\bmP, \bmV \in \calC_{2s_{\Gamma}} \cap \calS^{dm - 1}$ and $\bmU \in \calC_{2s_{\Omega}} \cap \calS^{m^2 - 1}$.

Then we could rearrange the increment as
\begin{align*}
	&X_{\bmU, \bmV, \bmP} - X_{\bmW, \bmZ, \bmQ} \nonumber \\
	&= \iprod{\bmV}{2(\bmSigma_{\bmX} - \frac{\bmX^T \bmX}{n}) \bmP \bmU} - \iprod{\bmZ}{2(\bmSigma_{\bmX} - \frac{\bmX^T \bmX}{n}) \bmQ \bmW} \nonumber \\
	&= \EE [\frac{2}{n} \Vect{\bmX^T}^T (\bmI_n \otimes (\bmP \bmU \bmV^T - \bmQ \bmW \bmZ^T)) \Vect{\bmX^T}] - \frac{2}{n} \Vect{\bmX^T}^T (\bmI_n \otimes (\bmP \bmU \bmV^T - \bmQ \bmW \bmZ^T)) \Vect{\bmX^T}.
\end{align*}
%where $\bmP, \bmV \in \calC_{2s_{\Gamma}} \cap \calS^{dm - 1}$ and $\bmU \in \calC_{2s_{\Omega}} \cap \calS^{m^2 - 1}$.

We could further rearrange $\bmP \bmU \bmV^T - \bmQ \bmW \bmZ^T$ as
\begin{equation}
	\bmP \bmU \bmV^T - \bmQ \bmW \bmZ^T = \frac{1}{2} \bmP (\bmU - \bmW)(\bmV + \bmZ)^T + \frac{1}{2} \bmP (\bmU + \bmW)(\bmV - \bmZ)^T + (\bmP - \bmQ) \bmW \bmZ^T.
\end{equation}
Its Frobenius norm could be bounded as
\begin{equation}
	\fnorm{\bmP \bmU \bmV^T - \bmQ \bmW \bmZ^T}^2 \leq 4 \fnorm{\bmU- \bmW}^2 + 4 \fnorm{\bmV - \bmZ}^2 + 2 \fnorm{\bmP - \bmQ}^2 \leq 4 \fnorm{(\begin{smallmatrix}
			\bmU \\ \bmV \\ \bmP
		\end{smallmatrix}) - (\begin{smallmatrix}
			\bmW \\ \bmZ \\ \bmQ
		\end{smallmatrix})}^2.
\end{equation}
Combing Lemma \ref{Quadratic_form_XX} with $X_{\bmU, \bmV, \bmP} - X_{\bmW, \bmZ, \bmQ}$, we could derive the mixed tail
\begin{align}
	&P(\abs{\iprod{\bmV}{2(\bmSigma_{\bmX} - \frac{\bmX^T \bmX}{n}) \bmP \bmU} - \iprod{\bmZ}{2(\bmSigma_{\bmX} - \frac{\bmX^T \bmX}{n}) \bmQ \bmW}} > u) \\
	&\qquad \leq 2 \exp(- c \min(\frac{u^2}{\frac{16}{n}\norm{\bmSigma_{\bmX}}^2\fnorm{(\begin{smallmatrix}
				\bmU \\ \bmV \\ \bmP
			\end{smallmatrix}) - (\begin{smallmatrix}
				\bmW \\ \bmZ \\ \bmQ
			\end{smallmatrix})}^2}, \frac{u}{\frac{4}{n}\norm{\bmSigma_{\bmX}}\fnorm{(\begin{smallmatrix}
				\bmU \\ \bmV \\ \bmP
			\end{smallmatrix}) - (\begin{smallmatrix}
				\bmW \\ \bmZ \\ \bmQ
			\end{smallmatrix})}})),
\end{align}
where we use $\norm{\bmUpsilon_{\bmX}} = \norm{\bmSigma_{\bmX}}$ under Assumption \ref{SingularValue_Sigma}.

This means the increment has a mixed tail with $d_2 = 4 \norm{\bmSigma_{\bmX}} \fnorm{\cdot} / \sqrt{n}$ and $d_1 = 4 \norm{\bmSigma_{\bmX}} \fnorm{\cdot} / n$.

With Lemma \ref{Suprema}, we could derive the event
\begin{align}
	&\usup{\begin{subarray}{c}
			\bmP, \bmV \in \calC_{2s_{\Gamma}} \cap \calS^{dm - 1} \\ \bmU \in \calC_{2s_{\Omega}} \cap \calS^{m^2 - 1}
	\end{subarray}} \abs{\iprod{\bmV}{2(\bmSigma_{\bmX} - \frac{\bmX^T \bmX}{n}) \bmP \bmU}} > C(\gamma_2(T, d_2) + \gamma_1(T, d_1) + u\Delta_2(T) + u^2 \Delta_1(T)) \label{SampleGammaOne}
\end{align}
holds with probability at most $2 \exp(- u^2)$. Here $T = \calC_{2s_{\Omega}} \cap \calS^{m^2 - 1} \times \calC_{2s_{\Gamma}} \cap \calS^{dm - 1} \times \calC_{2s_{\Gamma}} \cap \calS^{dm - 1}$.

We adopt the following lemma to transfer the $\gamma_1$-functional to the $\gamma_2$-functional and deal with the coefficients of metrics.
\begin{lemma} \cite{Melnyk2016EstimatingStructuredVAR}
	For $\gamma_{\alpha}$-functional, we have
	\begin{align}
		\gamma_1(S, \ltwonorm{\cdot}) &\leq \gamma_2^2(S, \ltwonorm{\cdot}) \\
		\gamma_{\alpha}(S, cd) &= c \gamma_{\alpha}(S, d),
	\end{align}
	where $\alpha > 0$, $c > 0$.
\end{lemma}
Combining with the Talagrand's majorizing measure theorem \cite{Talagrand2005TheGenericCh}, we could bound the $\gamma_2$-functional by the Gaussian width
\begin{equation}
	\begin{split}
		\gamma_2(T, \fnorm{\cdot}) \leq C (\gwidth{\calC_{2s_{\Gamma}} \cap \calS^{dm - 1}} + \gwidth{\calC_{2s_{\Omega}} \cap \calS^{m^2 - 1}}),
	\end{split}
\end{equation}
where the Frobenius norm for a matrix is equivalent to the $l_2$ norm for a vector.

Then we could rearrange (\ref{SampleGammaOne}) further and derive the event
\begin{align*}
	&\usup{\begin{subarray}{c}
			\bmP, \bmV \in \calC_{2s_{\Gamma}} \cap \calS^{dm - 1} \\ \bmU \in \calC_{2s_{\Omega}} \cap \calS^{m^2 - 1}
	\end{subarray}} \abs{\iprod{\bmV}{2(\bmSigma_{\bmX} - \frac{\bmX^T \bmX}{n}) \bmP \bmU}} \\
	&> C(4 \norm{\bmSigma_{\bmX}} \frac{\omega_{\Gamma} + \omega_{\Omega}}{\sqrt{n}} + 4 \norm{\bmSigma_{\bmX}} \frac{(\omega_{\Gamma} + \omega_{\Omega})^2}{n} + 4 \norm{\bmSigma_{\bmX}}\frac{u}{\sqrt{n}}\Delta_F(T) + 4 \norm{\bmSigma_{\bmX}} \frac{u^2}{n} \Delta_F(T))
	%&> C \norm{\bmSigma_{\bmX}} (\frac{\omega_{\Gamma} + \omega_{\Omega} + u}{\sqrt{n}} + \frac{(\omega_{\Gamma} + \omega_{\Omega} + u)^2}{n})
\end{align*}
holds with probability at most $2 \exp(- u^2)$.
%Here we use the fact $\Delta_F(T) \leq 6$.
%\begin{align*}
%	P(\usup{\begin{subarray}{c}
%			\bmP, \bmV \in \calC_{2s_{\Gamma}} \cap \calS^{dm - 1} \\ \bmU \in \calC_{2s_{\Omega}} \cap \calS^{m^2 - 1}
%	\end{subarray}} \iprod{\bmV}{2(\bmSigma_{\bmX} - \frac{\bmX^T \bmX}{n}) \bmP \bmU} > C(\norm{\bmSigma_{\bmX}} \frac{\omega_{\Gamma} + \omega_{\Omega}}{\sqrt{n}} + \norm{\bmSigma_{\bmX}}^2 \frac{(\omega_{\Gamma} + \omega_{\Omega})^2}{n} + \frac{u}{\sqrt{n}} + \frac{u^2}{n}))
%\end{align*}

From the facts $(\omega_{\Gamma} + \omega_{\Omega})^2 + u^2 \leq (\omega_{\Gamma} + \omega_{\Omega} + u)^2$ and $\Delta_F(T) \leq 6$, we could derive the following lemma when the term $(\omega_{\Gamma} + \omega_{\Omega} + u) / \sqrt{n}$ is dominant.
\begin{lemma} \label{SampleGammaLemmaOne}
	%	For the Frobenius norm of a matrix could be viewed as the $l_2$ norm of a vector, we could derive
	%	\begin{equation}
	%		\begin{split}
	%			&P(\usup{\begin{subarray}{c}
	%					\bmP, \bmV \in \calC_{2s_{\Gamma}} \cap \calS^{dm - 1} \\ \bmU \in \calC_{2s_{\Omega}} \cap \calS^{m^2 - 1}
	%			\end{subarray}} \iprod{\bmV}{2(\bmSigma_{\bmX} - \frac{\bmX^T \bmX}{n}) \bmP \bmU} > C(\norm{\bmSigma_{\bmX}} \frac{\omega_{\Gamma} + \omega_{\Omega}}{\sqrt{n}} + \norm{\bmSigma_{\bmX}}^2 \frac{(\omega_{\Gamma} + \omega_{\Omega})^2}{n} + \frac{u}{\sqrt{n}} + \frac{u^2}{n})) \\
	%			&\leq 2 \exp(- u^2).
	%		\end{split}
	%	\end{equation}
	%	
	Under the condition of $n \geq (\omega_{\Gamma} + \omega_{\Omega} + u)^2$, we have
	\begin{equation}
		P(\usup{\begin{subarray}{c}
				\bmP, \bmV \in \calC_{2s_{\Gamma}} \cap \calS^{dm - 1} \\ \bmU \in \calC_{2s_{\Omega}} \cap \calS^{m^2 - 1}
		\end{subarray}} \abs{\iprod{\bmV}{2(\bmSigma_{\bmX} - \frac{\bmX^T \bmX}{n}) \bmP \bmU}} > C\norm{\bmSigma_{\bmX}}(\frac{\omega_{\Gamma} + \omega_{\Omega} + u}{\sqrt{n}})) \leq 2 \exp(- u^2).
	\end{equation}
\end{lemma}

The random process associated with the second term of (\ref{Sample_Difference_Gamma}) could be written as
\begin{equation}
	\fnorm{(2(\bmSigma_{\bmX} - \frac{\bmX^T \bmX}{n}) (\Gammat - \Gammastar) \Omegastar)_{\calI}} \leq \usup{\bmU, \bmV \in \calC_{2s_{\Gamma}} \cap \calS^{dm - 1}} \iprod{\bmV}{2(\bmSigma_{\bmX} - \frac{\bmX^T \bmX}{n}) \bmU \Omegastar} \fnorm{\Gammat - \Gammastar}.
\end{equation}
We rearrange the random process $X_{\bmU, \bmV} - X_{\bmZ, \bmW}$ as
\begin{align}
	&X_{\bmU, \bmV} - X_{\bmZ, \bmW} \nonumber \\
	&= \iprod{\bmV}{2(\bmSigma_{\bmX} - \frac{\bmX^T \bmX}{n}) \bmU \Omegastar} - \iprod{\bmW}{2(\bmSigma_{\bmX} - \frac{\bmX^T \bmX}{n}) \bmZ \Omegastar} \nonumber \\
	&= \EE[\frac{2}{n} \Vect{\bmX^T}^T (\bmI_n \otimes (\bmU \Omegastar \bmV^T - \bmZ \Omegastar \bmW^T)) \Vect{\bmX^T}] - \frac{2}{n} \Vect{\bmX^T}^T (\bmI_n \otimes (\bmU \Omegastar \bmV^T - \bmZ \Omegastar \bmW^T)) \Vect{\bmX^T}.
\end{align}
From the facts
\begin{equation}
	\bmU \Omegastar \bmV^T - \bmZ \Omegastar \bmW^T = (\bmU - \bmZ) \Omegastar \bmV^T + \bmZ \Omegastar (\bmV - \bmW)^T
\end{equation}
and
\begin{equation}
	\fnorm{\bmU \Omegastar \bmV^T - \bmZ \Omegastar \bmW^T}^2 \leq 2 \norm{\Omegastar}^2 \fnorm{\bmU - \bmZ}^2 + 2 \norm{\Omegastar}^2 \fnorm{\bmV - \bmW}^2,
\end{equation}
we could derive the mixed tail according to Lemma \ref{Quadratic_form_XX}
\begin{equation}
	\begin{split}
		&P(\abs{\iprod{\bmV}{2(\bmSigma_{\bmX} - \frac{\bmX^T \bmX}{n}) \bmU \Omegastar} - \iprod{\bmW}{2(\bmSigma_{\bmX} - \frac{\bmX^T \bmX}{n}) \bmZ \Omegastar}} > u) \\
		&\qquad \leq 2\exp(- c \min(\frac{u^2}{\frac{8}{n} \norm{\bmSigma_{\bmX}}^2 \norm{\Omegastar}^2 \fnorm{(\begin{smallmatrix}
					\bmU \\ \bmV
				\end{smallmatrix}) - (\begin{smallmatrix}
					\bmZ \\ \bmW
				\end{smallmatrix})}^2}, \frac{u}{\frac{2\sqrt{2}}{n} \norm{\bmSigma_{\bmX}} \norm{\Omegastar}\fnorm{(\begin{smallmatrix}
					\bmU \\ \bmV
				\end{smallmatrix}) - (\begin{smallmatrix}
					\bmZ \\ \bmW
				\end{smallmatrix})}})).
	\end{split}
\end{equation}
Combining with Lemma \ref{Suprema}, we have the following lemma.
\begin{lemma} \label{SampleGammaLemmaTwo}
	When $n \geq (\omega_{\Gamma} + u)^2$, we could derive
	\begin{equation}
		P(\usup{\bmU, \bmV \in \calC_{2s_{\Gamma}} \cap \calS^{dm - 1}} \abs{\iprod{\bmV}{2(\bmSigma_{\bmX} - \frac{\bmX^T \bmX}{n}) \bmU \Omegastar}} > C \norm{\bmSigma_{\bmX}} \norm{\Omegastar}(\frac{\omega_{\Gamma} + u}{\sqrt{n}})) \leq 2 \exp(- u^2).
	\end{equation}
\end{lemma}

The random process associated with the third term of (\ref{Sample_Difference_Gamma}) could be written as
\begin{equation}
	\fnorm{(\frac{2}{n} \bmX^T \bmE (\Omegat - \Omegastar))_{\calI}} \leq \usup{\begin{subarray}{c}
			\bmV \in \calC_{2s_{\Gamma}} \cap \calS^{dm - 1} \\ \bmP \in \calC_{2s_{\Omega}} \cap \calS^{m^2 - 1}
	\end{subarray}} \iprod{\bmV}{\frac{2}{n} \bmX^T \bmE \bmP} \fnorm{\Omegat - \Omegastar}.
\end{equation}
The random process $X_{\bmV, \bmP} - X_{\bmZ, \bmQ}$ could be rearranged as
\begin{equation}
	X_{\bmV, \bmP} - X_{\bmZ, \bmQ} = \iprod{\bmV}{\frac{2}{n} \bmX^T \bmE \bmP} - \iprod{\bmZ}{\frac{2}{n} \bmX^T \bmE \bmQ} = \frac{2}{n} \Vect{\bmE^T}^T (\bmI_n \otimes (\bmP \bmV^T - \bmQ \bmZ^T)) \Vect{\bmX^T}.
\end{equation}
From the facts
\begin{equation}
	\bmP \bmV^T - \bmQ \bmZ^T = (\bmP - \bmQ) \bmV^T + \bmQ (\bmV - \bmZ)^T
\end{equation}
and
\begin{equation}
	\fnorm{\bmP \bmV^T - \bmQ \bmZ^T}^2 = 2\fnorm{\bmP - \bmQ}^2 + 2\fnorm{\bmV - \bmZ}^2,
\end{equation}
we could derive the mixed tail according to Lemma \ref{Quadratic_form_Independent}
\begin{equation}
	\begin{split}
		&P(\abs{\iprod{\bmV}{\frac{2}{n} \bmX^T \bmE \bmP} - \iprod{\bmZ}{\frac{2}{n} \bmX^T \bmE \bmQ}} > t) \\
		&\qquad \leq 2 \exp(- c \min(\frac{t^2}{\frac{8}{n} \norm{\Omegastar^{-\frac{1}{2}}}^2 \norm{\bmSigma_{\bmX}^{\frac{1}{2}}}^2 \fnorm{(\begin{smallmatrix}
					\bmV \\ \bmP
				\end{smallmatrix}) - (\begin{smallmatrix}
					\bmZ \\ \bmQ
				\end{smallmatrix})}^2}, \frac{t}{\frac{2\sqrt{2}}{n} \norm{\Omegastar^{-\frac{1}{2}}} \norm{\bmSigma_{\bmX}^{\frac{1}{2}}} \fnorm{(\begin{smallmatrix}
					\bmV \\ \bmP
				\end{smallmatrix}) - (\begin{smallmatrix}
					\bmZ \\ \bmQ
				\end{smallmatrix})}})).
	\end{split}
\end{equation}
%From the generic chaining \cite{Talagrand2014UpperAndLow}, we could derive the following lemma.
Combining with Lemma \ref{Suprema}, we have the following lemma.
\begin{lemma} \label{SampleGammaLemmaThree}
	Under the condition of $n \geq  (\omega_{\Gamma} + \omega_{\Omega} + u)^2$, we could derive
	\begin{equation}
		P(\usup{\begin{subarray}{c}
				\bmV \in \calC_{2s_{\Gamma}} \cap \calS^{dm - 1} \\ \bmP \in \calC_{2s_{\Omega}} \cap \calS^{m^2 - 1}
		\end{subarray}} \abs{\iprod{\bmV}{\frac{2}{n} \bmX^T \bmE \bmP}} > C \norm{\Omegastar^{-\frac{1}{2}}} \norm{\bmSigma_{\bmX}^{\frac{1}{2}}} (\frac{\omega_{\Gamma} + \omega_{\Omega} + u}{\sqrt{n}})) \leq 2\exp(- u^2).
	\end{equation}
\end{lemma}

The random process associated with the fourth term of (\ref{Sample_Difference_Gamma}) could be written as
\begin{equation}
	\fnorm{(\frac{2}{n} \bmX^T \bmE \Omegastar)_{\calI}} \leq \usup{\bmV \in \calC_{2s_{\Gamma}} \cap \calS^{dm - 1}} \iprod{\bmV}{\frac{2}{n} \bmX^T \bmE \Omegastar}.
\end{equation}
We arrange the random process $X_{\bmV} - X_{\bmZ}$ as
\begin{equation}
	X_{\bmV} - X_{\bmZ} = \iprod{\bmV}{\frac{2}{n} \bmX^T \bmE \Omegastar} - \iprod{\bmZ}{\frac{2}{n} \bmX^T \bmE \Omegastar} = \frac{2}{n} \Vect{\bmE^T}^T (\bmI_n \otimes (\Omegastar \bmV^T - \Omegastar \bmZ^T)) \Vect{\bmX^T}.
\end{equation}
Then we could derive the mixed tail according to Lemma \ref{Quadratic_form_Independent}
\begin{equation}
	\begin{split}
		&P(\abs{\iprod{\bmV}{\frac{2}{n} \bmX^T \bmE \Omegastar} - \iprod{\bmZ}{\frac{2}{n} \bmX^T \bmE \Omegastar}} > u) \\
		&\leq 2 \exp(- c \min( \frac{u^2}{\frac{4}{n} \norm{\Omegastar^{\frac{1}{2}}}^2 \norm{\bmSigma_{\bmX}^{\frac{1}{2}}}^2 \fnorm{\bmV - \bmZ}^2}, \frac{u}{\frac{2}{n} \norm{\Omegastar^{\frac{1}{2}}} \norm{\bmSigma_{\bmX}^{\frac{1}{2}}} \fnorm{\bmV - \bmZ}})),
	\end{split}
\end{equation}
where we use the fact $\Vect{(\bmE \Omegastar)^T} \sim \calN(\bmzero, \bmI_n \otimes \Omegastar)$ under Assumption \ref{SingularValue_Omega}.

Combining with Lemma \ref{Suprema}, we have the following lemma.
\begin{lemma} \label{SampleGammaLemmaFour}
	Under the condition of $n \geq  (\omega_{\Gamma} + u)^2$, we could derive
	\begin{equation}
		P(\usup{\bmV \in \calC_{2s_{\Gamma}} \cap \calS^{dm - 1}} \abs{\iprod{\bmV}{\frac{2}{n} \bmX^T \bmE \Omegastar}} > C \norm{\Omegastar^{\frac{1}{2}}} \norm{\bmSigma_{\bmX}^{\frac{1}{2}}} (\frac{\omega_{\Gamma} + u}{\sqrt{n}})) \leq 2\exp(- u^2).
	\end{equation}
\end{lemma}

Taking Lemma \ref{SampleGammaLemmaOne}, \ref{SampleGammaLemmaTwo}, \ref{SampleGammaLemmaThree} and \ref{SampleGammaLemmaFour} into consideration, we could derive the event
\begin{align*}
	%	&\fnorm{\usup{\mV \in \calC_{2s_{\Gamma}} \cap \calS^{dm - 1}} \iprod{\mV}{\nabla_{\Gamma} f(\Gammat, \Omegat) - \nabla_{\Gamma} f_n(\Gammat, \Omegat)}} \\
	&\fnorm{(\nabla_{\Gamma} f(\Gammat, \Omegat) - \nabla_{\Gamma} f_n(\Gammat, \Omegat))_{\calI}} \\
	&\leq C(\norm{\bmSigma_{\bmX}}(\frac{\omega_{\Gamma} + \omega_{\Omega} + u}{\sqrt{n}}) \fnorm{\Gammat - \Gammastar} \fnorm{\Omegat - \Omegastar} + \norm{\bmSigma_{\bmX}} \norm{\Omegastar}(\frac{\omega_{\Gamma} + u}{\sqrt{n}}) \fnorm{\Gammat - \Gammastar} \\
	&\qquad + \norm{\Omegastar^{-\frac{1}{2}}} \norm{\bmSigma_{\bmX}^{\frac{1}{2}}} (\frac{\omega_{\Gamma} + \omega_{\Omega} + u}{\sqrt{n}}) \fnorm{\Omegat - \Omegastar} + \norm{\Omegastar^{\frac{1}{2}}} \norm{\bmSigma_{\bmX}^{\frac{1}{2}}} (\frac{\omega_{\Gamma} + u}{\sqrt{n}})) \\
	&\leq C(\tau_{\max} R (\frac{\omega_{\Gamma} + \omega_{\Omega} + u}{\sqrt{n}}) \fnorm{\Omegat - \Omegastar} + \tau_{\max} \nu_{\max} (\frac{\omega_{\Gamma} + u}{\sqrt{n}}) \fnorm{\Gammat - \Gammastar} \\
	&\qquad + \sqrt{\frac{\tau_{\max}}{\nu_{\min}}} \fnorm{\Omegat - \Omegastar} (\frac{\omega_{\Gamma} + \omega_{\Omega} + u}{\sqrt{n}}) + \tau_{\max}^{\frac{1}{2}} \nu_{\max}^{\frac{1}{2}} (\frac{\omega_{\Gamma} + u}{\sqrt{n}}))
\end{align*}
holds with probability at least $1 - 8\exp(- u^2)$, when $n \geq  (\omega_{\Gamma} + \omega_{\Gamma} + u)^2$. Here we use Assumption \ref{SingularValue_Omega}, \ref{SingularValue_Sigma} and $\max(\fnorm{\Gammat - \Gammastar}, \fnorm{\Omegat - \Omegastar}) \leq R$.

\subsection{Proof of Lemma \ref{Sample_Omega}}
We first rewrite $\nabla_{\Omega} f(\Gammat, \Omegat) - \nabla_{\Omega} f_n(\Gammat, \Omegat)$ as
\begin{equation}
	\nabla_{\Omega} f(\Gammat, \Omegat) - \nabla_{\Omega} f_n(\Gammat, \Omegat) = (\Gammat - \Gammastar)^T (\bmSigma_{\bmX} - \frac{\bmX^T \bmX}{n}) (\Gammat - \Gammastar) + \frac{2}{n} (\mGamma_{t} - \bmGamma_{\star})^T \bmX^T \bmE + (\Omegastar^{-1} - \frac{1}{n} \bmE^T\bmE).  \label{Sample_Difference_Omega}
\end{equation}

With the definition of $\calC_{2 s_{\Omega}}$, we could derive
\begin{equation}
	\fnorm{(\nabla_{\Omega} f(\Gammat, \Omegat) - \nabla_{\Omega} f_n(\Gammat, \Omegat))_{\calT}} \leq \usup{\bmV \in \calC_{2 s_{\Omega}} \cap \calS^{m^2 - 1}} \iprod{\bmV}{\nabla_{\Omega} f(\Gammat, \Omegat) - \nabla_{\Omega} f_n(\Gammat, \Omegat)},
\end{equation}
where we use the fact $\Card(\calT) \leq 2 s_{\Omega}$.

In this way, to bound $\fnorm{(\nabla_{\Omega} f(\Gammat, \Omegat) - \nabla_{\Omega} f_n(\Gammat, \Omegat))_{\calT}}$ and $\usup{\bmV \in \calC_{2s_{\Omega}} \cap \calS^{m^2 - 1}} \iprod{\bmV}{\nabla_{\Omega} f(\Gammat, \Omegat) - \nabla_{\Omega} f_n(\Gammat, \Omegat)}$, we need to deal with three suprema of random processes.

The random process associated with the first term of (\ref{Sample_Difference_Omega}) could be written as
\begin{equation}
	\usup{\bmV \in \calC_{2s_{\Omega}} \cap \calS^{m^2 - 1}} \iprod{\bmV}{(\Gammat - \Gammastar)^T (\bmSigma_{\bmX} - \frac{\bmX^T \bmX}{n}) (\Gammat - \Gammastar)} \leq \usup{\begin{subarray}{c}
			\bmU \in \calC_{2s_{\Gamma}} \cap \calS^{dm - 1} \\ \bmV \in \calC_{2s_{\Omega}} \cap \calS^{m^2 - 1}
	\end{subarray}} \iprod{\bmV}{\bmU^T (\bmSigma_{\bmX} - \frac{\bmX^T \bmX}{n}) \bmU} \fnorm{\Gammat - \Gammastar}^2.
\end{equation}
We could rearrange the random process $X_{\bmU, \bmV} - X_{\bmW, \bmZ}$ as
\begin{align}
	&X_{\bmU, \bmV} - X_{\bmW, \bmZ} \nonumber \\
	&= \EE[\frac{1}{n} \Vect{\bmX^T}^T (\bmI_n \otimes (\bmU \bmV^T \bmU^T - \bmW \bmZ^T \bmW^T)) \Vect{\bmX^T}] - \frac{1}{n} \Vect{\bmX^T}^T (\bmI_n \otimes (\bmU \bmV^T \bmU^T - \bmW \bmZ^T \bmW^T)) \Vect{\bmX^T}.
\end{align}
From the facts
\begin{equation}
	\bmU \bmV^T \bmU^T - \bmW \bmZ^T \bmW^T = \frac{1}{2} (\bmU - \bmW) \bmV^T (\bmU + \bmW)^T + \frac{1}{2} (\bmU + \bmW) \bmV^T (\bmU - \bmW)^T + \bmW (\bmV - \bmZ)^T \bmW^T
\end{equation}
and
\begin{equation}
	\fnorm{\bmU \bmV^T \bmU^T - \bmW \bmZ^T \bmW^T}^2 \leq 8 \fnorm{\bmU - \bmW}^2 + 2 \fnorm{\bmV - \bmZ}^2,
\end{equation}
%and
%\begin{align}
%	\norm{\bmU \bmV^T \bmU^T - \bmW \bmZ^T \bmW^T} &\leq 2 \norm{\bmU - \bmW} + \norm{\bmV - \bmZ}.
%\end{align}
we could derive the mixed tail according to Lemma \ref{Quadratic_form_XX}
\begin{align}
	&P(\abs{\iprod{\bmV}{\bmU^T (\bmSigma_{\bmX} - \frac{\bmX^T \bmX}{n}) \bmU} - \iprod{\bmZ}{\bmW^T (\bmSigma_{\bmX} - \frac{\bmX^T \bmX}{n}) \bmW}} > u) \nonumber \\
	&\leq 2 \exp(- c \min (\frac{u^2}{\frac{8}{n} \norm{\bmSigma_{\bmX}}^2 \fnorm{(\begin{smallmatrix}
				\bmU \\ \bmV
			\end{smallmatrix}) - (\begin{smallmatrix}
				\bmW \\ \bmZ
			\end{smallmatrix})}^2}, \frac{u}{\frac{2\sqrt{2}}{n} \norm{\bmSigma_{\bmX}} \fnorm{(\begin{smallmatrix}
				\bmU \\ \bmV
			\end{smallmatrix}) - (\begin{smallmatrix}
				\bmW \\ \bmZ
			\end{smallmatrix})}})).
\end{align}
Combining with Lemma \ref{Suprema}, we have the following lemma.
\begin{lemma} \label{SampleOmegaLemmaOne}
	When $n \geq C (\omega_{\Gamma} + \omega_{\Omega} + u)^2$, we could derive
	\begin{equation}
		P(\usup{\begin{subarray}{c}
				\bmU \in \calC_{2s_{\Gamma}} \cap \calS^{dm - 1} \\ \bmV \in \calC_{2s_{\Omega}} \cap \calS^{m^2 - 1}
		\end{subarray}} \abs{\iprod{\bmV}{\bmU^T (\bmSigma_{\bmX} - \frac{\bmX^T \bmX}{n}) \bmU}} > C\norm{\bmSigma_{\bmX}}(\frac{\omega_{\Gamma} + \omega_{\Omega} + u}{\sqrt{n}})) \leq 2 \exp(- u^2).
	\end{equation}
\end{lemma}

The random process associated with the second term of (\ref{Sample_Difference_Omega}) could be written as
\begin{equation}
	\usup{\bmV \in \calC_{2s_{\Omega}} \cap \calS^{m^2 - 1}} \iprod{\bmV}{\frac{2}{n} (\Gammat - \Gammastar)^T \mX ^T \mE} \leq \usup{\begin{subarray}{c}
			\bmU \in \calC_{2s_{\Gamma}} \cap \calS^{dm - 1} \\ \bmV \in \calC_{2s_{\Omega}} \cap \calS^{m^2 - 1}
	\end{subarray}} \iprod{\bmV}{\frac{2}{n} \mU^T \mX^T \mE} \fnorm{\Gammat - \Gammastar}.
\end{equation}
The random process $X_{\bmU, \bmV} - X_{\bmW, \bmZ}$ could be rearranged as
\begin{equation}
	X_{\bmU, \bmV} - X_{\bmW, \bmZ} = \Vect{\mE^T}^T (\bmI_n \otimes (\mV^T \mU^T - \mZ^T \mW^T)) \Vect{\mX^T}.
\end{equation}
From the fact
\begin{equation}
	\mV^T \mU^T - \mZ^T \mW^T = (\mV - \mZ)^T \mU^T + \mZ^T (\mU - \mW)^T,
\end{equation}
we could derive the mixed tail according to Lemma \ref{Quadratic_form_Independent}
\begin{align}
	&P(\abs{\iprod{\bmV}{\frac{2}{n} \mU^T \mX^T \mE} - \iprod{\bmZ}{\frac{2}{n} \mW^T \mX^T \mE}} > u) \nonumber \\
	&\leq 2 \exp(- c \min(\frac{u^2}{\frac{8}{n} \norm{\Omegastar^{-\frac{1}{2}}}^2 \norm{\bmSigma_{\bmX}^{\frac{1}{2}}}^2 \fnorm{(\begin{smallmatrix}
				\bmU \\ \bmV
			\end{smallmatrix}) - (\begin{smallmatrix}
				\bmW \\ \bmZ
			\end{smallmatrix})}^2}, \frac{u}{\frac{2\sqrt{2}}{n} \norm{\Omegastar^{-\frac{1}{2}}} \norm{\bmSigma_{\bmX}^{\frac{1}{2}}} \fnorm{(\begin{smallmatrix}
				\bmU \\ \bmV
			\end{smallmatrix}) - (\begin{smallmatrix}
				\bmW \\ \bmZ
			\end{smallmatrix})}})).
\end{align}
Combining with Lemma \ref{Suprema}, we have the following lemma.
\begin{lemma} \label{SampleOmegaLemmaTwo}
	When $n \geq  (\omega_{\Gamma} + \omega_{\Omega} + u)^2$, we could derive
	\begin{equation}
		P(\usup{\begin{subarray}{c}
				\bmU \in \calC_{2s_{\Gamma}} \cap \calS^{dm - 1} \\ \bmV \in \calC_{2s_{\Omega}} \cap \calS^{m^2 - 1}
		\end{subarray}} \abs{\iprod{\bmV}{\frac{2}{n} \bmE^T \bmX \bmU}} > C \norm{\Omegastar^{-\frac{1}{2}}} \norm{\bmSigma_{\bmX}^{\frac{1}{2}}} (\frac{\omega_{\Gamma} + \omega_{\Omega} + u}{\sqrt{n}})) \leq 2 \exp(- u^2).
	\end{equation}
\end{lemma}

The random process associated with the third term of (\ref{Sample_Difference_Omega}) could be written as
\begin{equation}
	\usup{\bmV \in \calC_{2s_{\Omega}} \cap \calS^{m^2 - 1}} \iprod{\bmV}{\Omegastar^{-1} - \frac{1}{n} \bmE^T\bmE} .
\end{equation}
The random process $X_{\bmV} - X_{\bmZ}$ could be rearranged as
\begin{equation}
	X_{\bmV} - X_{\bmZ} = \EE[\frac{1}{n} \Vect{\bmE^T}^T (\bmI_n \otimes(\bmV^T - \bmZ^T)) \Vect{\bmE^T}] - \frac{1}{n} \Vect{\bmE^T}^T (\bmI_n \otimes(\bmV^T - \bmZ^T)) \Vect{\bmE^T}.
\end{equation}
We could derive the mixed tail according to Lemma \ref{Quadratic_form_XX}
\begin{equation}
	P(\abs{\iprod{\bmV - \bmZ}{\Omegastar^{-1} - \frac{1}{n} \bmE^T\bmE}} > u) \leq 2 \exp(- c \min(\frac{u^2}{\frac{1}{n} \norm{\Omegastar^{-1}}^2\fnorm{\bmV - \bmZ}^2}, \frac{u}{\frac{1}{n} \norm{\Omegastar^{-1}} \fnorm{\bmV - \bmZ}})).
\end{equation}
Combining with Lemma \ref{Suprema}, we have the following lemma.
\begin{lemma} \label{SampleOmegaLemmaThree}
	When $n \geq (\omega_{\Omega} + u)^2$, we could derive
	\begin{equation}
		P(\usup{\bmV \in \calC_{2s_{\Omega}} \cap \calS^{m^2 - 1}} \abs{\iprod{\bmV}{\Omegastar^{-1} - \frac{1}{n} \bmE^T\bmE}} > C \norm{\Omegastar^{-1}}(\frac{\omega_{\Omega} + u}{\sqrt{n}})) \leq 2 \exp(- u^2).
	\end{equation}
\end{lemma}

Taking Lemma \ref{SampleOmegaLemmaOne}, \ref{SampleOmegaLemmaTwo} and \ref{SampleOmegaLemmaThree} into consideration, we could derive the event
\begin{align*}
	&\fnorm{(\nabla_{\Omega} f(\Gammat, \Omegat) - \nabla_{\Omega} f_n(\Gammat, \Omegat))_{\calT}} \\
	&\leq \usup{\bmV \in \calC_{2s_{\Omega}} \cap \calS^{m^2 - 1}} \iprod{\bmV}{\nabla_{\Omega} f(\Gammat, \Omegat) - \nabla_{\Omega} f_n(\Gammat, \Omegat)} \\
	&\leq C(\norm{\bmSigma_{\bmX}}(\frac{\omega_{\Gamma} + \omega_{\Omega} + u}{\sqrt{n}}) \fnorm{\Gammat - \Gammastar}^2 + \norm{\Omegastar^{-\frac{1}{2}}} \norm{\bmSigma_{\bmX}^{\frac{1}{2}}} (\frac{\omega_{\Gamma} + \omega_{\Omega} + u}{\sqrt{n}}) \fnorm{\Gammat - \Gammastar} + \norm{\Omegastar^{-1}}(\frac{\omega_{\Omega} + u}{\sqrt{n}})) \\
	&\leq C(\tau_{\max} R (\frac{\omega_{\Gamma} + \omega_{\Omega} + u}{\sqrt{n}})\fnorm{\Gammat - \Gammastar} + \sqrt{\frac{\tau_{\max}}{\nu_{\min}}} \fnorm{\Gammat - \Gammastar} (\frac{\omega_{\Gamma} + \omega_{\Omega} + u}{\sqrt{n}}) + \frac{1}{\nu_{\min}} (\frac{\omega_{\Omega} + u}{\sqrt{n}}))
\end{align*}
hold with probability at least $1 - 6\exp(- u^2)$, when $n \geq  (\omega_{\Gamma} + \omega_{\Gamma} + u)^2$. Here we use Assumption \ref{SingularValue_Omega}, \ref{SingularValue_Sigma} and $\fnorm{\Gammat - \Gammastar} \leq R$.

\subsection{Proof of Lemma \ref{Initialization_Gamma_IHT}}
From the optimality of $\bmGamma_{\text{ini}}$, we could derive
\begin{align}
	\frac{1}{2} \fnorm{\bmY - \bmX \bmGamma_{\text{ini}}}^2 &\leq \frac{1}{2} \fnorm{\bmY - \bmX \Gammastar}^2.
\end{align}

After rearrangement, we could get
\begin{align}
	\frac{1}{2n}\fnorm{\bmX(\bmGamma_{\text{ini}} - \Gammastar)}^2 &\leq \frac{1}{n} \iprod{\bmE}{\bmX(\bmGamma_{\text{ini}} - \Gammastar)}.  \label{InitialProcessGamma}
\end{align}

The left hand of (\ref{InitialProcessGamma}) could be rewritten as
\begin{equation}
	\frac{1}{2n} \fnorm{\bmX(\bmGamma_{\text{ini}} - \Gammastar)}^2 = \frac{1}{2n} \iprod{\bmU}{\bmX^T \bmX \bmU} \fnorm{\bmGamma_{\text{ini}} - \Gammastar}^2,
\end{equation}
where $\bmU \in \calC_{2s_{\Gamma}} \cap \calS^{dm - 1}$. Here we use the fact $\bmGamma_{\text{ini}} - \Gammastar \in \calC_{2s_{\Gamma}}$.

Then we illustrate the random process $X_{\bmU} = \iprod{\bmU}{(\bmSigma_{\bmX} - \frac{\bmX^T \bmX}{n}) \bmU}$ has a mixed tail.

We rearrange $X_{\bmU} - X_{\bmW}$ as
\begin{align*}
	&X_{\bmU} - X_{\bmW} \\
	&= \EE [\frac{1}{n} \Vect{\bmX^T}^T (\bmI_n \otimes (\bmU \bmU^T - \bmW \bmW^T)) \Vect{\bmX^T}] - \frac{1}{n} \Vect{\bmX^T}^T (\bmI_n \otimes (\bmU \bmU^T - \bmW \bmW^T)) \Vect{\bmX^T}.
\end{align*}
From the fact
\begin{equation}
	\bmU \bmU^T - \bmW \bmW^T = \frac{1}{2} (\bmU + \bmW)(\bmU - \bmW)^T + \frac{1}{2} (\bmU - \bmW)(\bmU + \bmW)^T,
\end{equation}
we could derive
\begin{align*}
	&P(\abs{\iprod{\bmU}{(\bmSigma_{\bmX} - \frac{\bmX^T \bmX}{n}) \bmU} - \iprod{\bmW}{(\bmSigma_{\bmX} - \frac{\bmX^T \bmX}{n}) \bmW}} > u) \\
	&\leq 2 \exp(- c \min(\frac{u^2}{\frac{4}{n}  \norm{\bmSigma_{\bmX}}^2 \fnorm{\bmU - \bmW}^2}, \frac{u}{\frac{2}{n} \norm{\bmSigma_{\bmX}} \fnorm{\bmU - \bmW}})),
\end{align*}
where we use Lemma \ref{Quadratic_form_XX}. Then we could derive the following statement by Lemma \ref{Suprema}.
\begin{lemma}
	When $n \geq (\omega_{\Gamma} + u)^2$, we could derive
	\begin{equation}
		P(\usup{\begin{subarray}{c}
				\bmU \in \calC_{2s_{\Gamma}} \cap \calS^{dm - 1}
		\end{subarray}} \abs{\iprod{\bmU}{(\bmSigma_{\bmX} - \frac{\bmX^T \bmX}{n}) \bmU}} > C\norm{\bmSigma_{\bmX}}(\frac{\omega_{\Gamma} + u}{\sqrt{n}})) \leq 2 \exp(- u^2).
	\end{equation}
\end{lemma}
From the above lemma we could derive
\begin{equation}
	\frac{1}{2n} \fnorm{\bmX(\bmGamma_{\text{ini}} - \Gammastar)}^2 \geq \frac{1}{2}(\lambda_{\min}(\bmSigma_{\bmX}) - C\lambda_{\max}(\bmSigma_{\bmX})\frac{\omega_{\Gamma} + u}{\sqrt{n}}) \fnorm{\bmGamma_{\text{ini}} - \Gammastar}^2,
\end{equation}
with probability at least $1 - 2 \exp(- u^2)$.

The right hand of (\ref{InitialProcessGamma}) could be rewritten as
\begin{equation}
	\frac{1}{n} \iprod{\bmE}{\bmX(\bmGamma_{\text{ini}} - \Gammastar)} = \frac{1}{n} \iprod{\bmV}{\bmX^T \bmE} \fnorm{\bmGamma_{\text{ini}} - \Gammastar},
\end{equation}
where $\bmV \in \calC_{2s_{\Gamma}} \cap \calS^{dm - 1}$.

Then we illustrate the random process $X_{\bmV} = \frac{1}{n}\iprod{\bmV}{\bmX^T \bmE}$ has a mixed tail.
\begin{equation}
	X_{\bmV} - X_{\bmZ} = \frac{1}{n} \iprod{\bmV}{\bmX^T \bmE} - \frac{1}{n} \iprod{\bmZ}{\bmX^T \bmE} = \frac{1}{n} \Vect{\bmE^T}^T (\bmI_n \otimes ( \bmV^T -  \bmZ^T)) \Vect{\bmX^T}.
\end{equation}
With Lemma \ref{Quadratic_form_Independent} and Lemma \ref{Suprema}, we could derive
\begin{align*}
	P(\abs{\frac{1}{n}\iprod{\bmV}{\bmX^T \bmE} - \frac{1}{n}\iprod{\bmZ}{\bmX^T \bmE}} > u) \leq 2 \exp(- c \min(\frac{u^2}{\frac{1}{n}  \norm{\Omegastar^{-\frac{1}{2}}}^2 \norm{\bmSigma_{\bmX}^{\frac{1}{2}}}^2 \fnorm{\bmU - \bmW}^2}, \frac{u}{\frac{1}{n} \norm{\Omegastar^{-\frac{1}{2}}} \norm{\bmSigma_{\bmX}^{\frac{1}{2}}} \fnorm{\bmU - \bmW}}))
\end{align*}
and the following lemma.
\begin{lemma}
	Under the condition of $n \geq  (\omega_{\Gamma} + u)^2$, we could derive
	\begin{equation}
		P(\usup{\bmV \in \calC_{2s_{\Gamma}} \cap \calS^{dm - 1}} \abs{\iprod{\bmV}{\frac{1}{n} \bmX^T \bmE}} > C \norm{\Omegastar^{-\frac{1}{2}}} \norm{\bmSigma_{\bmX}^{\frac{1}{2}}} (\frac{\omega_{\Gamma} + u}{\sqrt{n}})) \leq 2\exp(- u^2).
	\end{equation}
\end{lemma}
Taking the two processes into consideration, we could derive
\begin{align}
	\fnorm{\Gammaini - \Gammastar} &\leq C \lambda_{\max}(\bmSigma_{\bmX}^{\frac{1}{2}})\frac{\omega_{\Gamma} + u}{\sqrt{n} (\lambda_{\min}(\bmSigma_{\bmX}) - C \lambda_{\max}(\bmSigma_{\bmX}) \frac{\omega_{\Gamma} + u}{\sqrt{n}})} \norm{\Omegastar^{-\frac{1}{2}}} \nonumber \\
	&\leq 2C \frac{\lambda_{\max}(\bmSigma_{\bmX}^{\frac{1}{2}})}{\lambda_{\min}(\bmSigma_{\bmX})}\frac{\omega_{\Gamma} + u}{\sqrt{n}} \norm{\Omegastar^{-\frac{1}{2}}} \nonumber \\
	&\leq 2C \frac{\omega_{\Gamma} + u}{\sqrt{n}} \frac{\tau_{\max}^{\frac{1}{2}}}{\tau_{\min} \nu_{\min}^{\frac{1}{2}}} ,
\end{align}
with probability at least $1 - 4 \exp(- u^2)$, when $\sqrt{n} \geq 2 C \frac{\tau_{\max}}{\tau_{\min}} (\omega_{\Gamma} + u)$. Here, we use Assumption \ref{SingularValue_Omega} and \ref{SingularValue_Sigma}.
\subsection{Proof of Lemma \ref{Initialization_Omega_IHT}}
From the optimality of $\bmOmega_{\text{ini}}$, we could derive
\begin{equation}
	\frac{1}{2} \fnorm{\Omegaini - \Omegastar - (\bmS^{-1} - \Omegastar)}^2 \leq \frac{1}{2} \fnorm{\bmS^{-1} - \Omegastar}^2.
\end{equation}

After rearrangement, we could derive
\begin{align}
	\frac{1}{2} \fnorm{\Omegaini - \Omegastar}^2 &\leq \iprod{\Omegaini - \Omegastar}{\bmS^{-1} - \Omegastar} \nonumber \\
	&\leq \fnorm{\Omegaini - \Omegastar} \fnorm{\bmS^{-1} - \Omegastar} \nonumber \\
	&= \fnorm{\Omegaini - \Omegastar} \fnorm{\bmS^{-1}(\Omegastar^{-1} - \bmS)\Omegastar} \nonumber \\
	&\leq \fnorm{\Omegaini - \Omegastar} \norm{\bmS^{-1}} \norm{\Omegastar} \fnorm{\Omegastar^{-1} - \bmS},
\end{align}
where the second inequality is from the Cauchy–Schwarz inequality and we use $\fnorm{\bmA \bmB} \leq \norm{\bmA} \fnorm{\bmB}$ for two matrices $\bmA$ and $\bmB$ in the last inequality.

We still need to deal with two terms associated with random processes, $\fnorm{\Omegastar^{-1} - \bmS}$ and $\norm{\bmS^{-1}}$.

\begin{lemma} \label{Covariance_Deviation_IHT}
	The event
	\begin{equation}
		\fnorm{\Omegastar^{-1} - \bmS} \leq C \frac{\tau_{\max}^2}{\tau_{\min}^2 \nu_{\min}} \frac{m + \omega_{\Gamma} + u}{\sqrt{n}}
	\end{equation}
	holds with probability at least $1 - 12 \exp(-u^2)$, when $\sqrt{n} \geq 2 C \frac{\tau_{\max}}{\tau_{\min}} (m + \omega_{\Gamma} + u)$.
\end{lemma}

Our method to bound $\norm{\bmS^{-1}}$ is inspired by \cite{Banerjee2018ImprovedAlt}. To upper bound $\norm{\bmS^{-1}}$, we need to lower bound $\lambda_{\min}(\bmS)$.
\begin{lemma} \label{Covariance_IHT}
	The event
	\begin{equation}
		\lambda_{\min}(\bmS) \geq \frac{c}{\nu_{\max}}
	\end{equation}
	holds with probability $1 - 10 \exp(- u^2)$, when $\sqrt{n} > 2 C \frac{\tau_{\max} \nu_{\max}}{\tau_{\min} \nu_{\min}} (\sqrt{m} + \omega_{\Gamma} + u)$.
	%		, we could derive $\norm{\bmS^{-1}} \leq C \nu$.
\end{lemma}
Then we could derive $\norm{\bmS^{-1}} \leq \nu_{\max} / c$.	

Considering the two above lemmas, we derive the final result.

The event
\begin{align}
	\fnorm{\Omegaini - \Omegastar} &\leq 2 \norm{\bmS^{-1}} \norm{\Omegastar} \fnorm{\Omegastar^{-1} - \bmS} \\
	&\leq 2 \frac{\nu_{\max}}{c} \nu_{\max} C  \frac{\tau_{\max}^2}{\tau_{\min}^2 \nu_{\min}} \frac{m + \omega_{\Gamma} + u}{\sqrt{n}} \\
	&\leq C \frac{\tau_{\max}^2 \nu_{\max}^2 }{\tau_{\min}^2 \nu_{\min}} \frac{m + \omega_{\Gamma} + u}{\sqrt{n}}
\end{align}
holds with probability $1 - 18 \exp(- u^2)$, when $\sqrt{n} > 2 C \frac{\tau_{\max} \nu_{\max}}{\tau_{\min} \nu_{\min}} (m + \omega_{\Gamma} + u)$.

\subsection{Proof of Lemma \ref{Covariance_Deviation_IHT}}
The term $\fnorm{\Omegastar^{-1} - \bmS}$ could be rewritten as
\begin{equation*}
	\begin{split}
		&\fnorm{\Omegastar^{-1} - \bmS} \\
		&= \fnorm{(\Gammaini - \Gammastar)^T (\frac{\bmX^T \bmX}{n} - \bmSigma_{\bmX}) (\Gammaini - \Gammastar) - (\Gammaini - \Gammastar)^T \frac{\bmX^T \bmE}{n} - \frac{\bmE^T \bmX}{n} (\Gammaini - \Gammastar) + \frac{\bmE^T \bmE}{n} - \Omegastar^{-1} \\
			&\quad + (\Gammaini - \Gammastar)^T \bmSigma_{\bmX} (\Gammaini - \Gammastar)} \\
		&= \usup{\bmV \in \calS^{m^2 - 1}} \iprod{\bmV}{(\Gammaini - \Gammastar)^T (\frac{\bmX^T \bmX}{n} - \bmSigma_{\bmX}) (\Gammaini - \Gammastar) - (\Gammaini - \Gammastar)^T \frac{\bmX^T \bmE}{n} - \frac{\bmE^T \bmX}{n} (\Gammaini - \Gammastar) + \frac{\bmE^T \bmE}{n} - \Omegastar^{-1} \\
			&\quad + (\Gammaini - \Gammastar)^T \bmSigma_{\bmX} (\Gammaini - \Gammastar)}.
	\end{split}
\end{equation*}

We still bound these terms by Lemma \ref{Quadratic_form_XX}, \ref{Quadratic_form_Independent} and \ref{Suprema}.

From the facts
\begin{equation}
	\bmU \bmV^T \bmU^T - \bmW \bmZ^T \bmW^T = \frac{1}{2} (\bmU + \bmW) \bmV^T (\bmU - \bmW)^T + \frac{1}{2} (\bmU - \bmW) \bmV^T (\bmU + \bmW)^T + \bmW(\bmV - \bmZ)^T \bmW^T
\end{equation}
and
\begin{equation}
	\fnorm{\bmU \bmV^T \bmU^T - \bmW \bmZ^T \bmW^T}^2 \leq 8 \fnorm{\bmU - \bmW}^2 + 2 \fnorm{\bmV - \bmZ}^2,
\end{equation}
we could derive the mixed tail according to Lemma \ref{Quadratic_form_XX}
\begin{align}
	&P(\abs{\iprod{\bmV}{\bmU^T \frac{\bmX^T \bmX}{n} \bmU} - \EE \iprod{\bmV}{\bmU^T \frac{\bmX^T \bmX}{n} \bmU} - \iprod{\bmZ}{\bmW^T \frac{\bmX^T \bmX}{n} \bmW} + \EE \iprod{\bmZ}{\bmW^T \frac{\bmX^T \bmX}{n} \bmW}} > u) \nonumber \\
	&\leq 2 \exp(- c \min (\frac{u^2}{\frac{8}{n} \norm{\bmSigma_{\bmX}}^2 \fnorm{(\begin{smallmatrix}
				\bmU \\ \bmW
			\end{smallmatrix}) - (\begin{smallmatrix}
				\bmV \\ \bmZ
			\end{smallmatrix})}^2}, \frac{u}{\frac{2 \sqrt{2}}{n} \norm{\bmSigma_{\bmX}} \fnorm{(\begin{smallmatrix}
				\bmU \\ \bmW
			\end{smallmatrix}) - (\begin{smallmatrix}
				\bmV \\ \bmZ
			\end{smallmatrix})}})).
\end{align}
Then the supremum of the random process could be bounded as
\begin{equation}
	P(\usup{\begin{subarray}{c}
			\bmU \in \calC_{2s_{\Gamma}} \cap \calS^{dm - 1} \\ \bmV \in \calS^{m^2 - 1}
	\end{subarray}} \abs{\iprod{\bmV}{\bmU^T \frac{\bmX^T \bmX}{n} \bmU} - \EE \iprod{\bmV}{\bmU^T \frac{\bmX^T \bmX}{n} \bmU}} > C\norm{\bmSigma_{\bmX}}(\frac{m + \omega_{\Gamma} + u}{\sqrt{n}})) \leq 2 \exp(- u^2),
\end{equation}
when $n \geq (m + \omega_{\Gamma} + u)^2$, according to Lemma \ref{Suprema}.

Following the procedure of Lemma \ref{SampleOmegaLemmaTwo}, the second and third terms could be bounded as
\begin{equation}
	P(\usup{\begin{subarray}{c}
			\bmU \in \calC_{2s_{\Gamma}} \cap \calS^{dm - 1} \\ \bmV \in \calS^{m^2 - 1}
	\end{subarray}} \abs{\iprod{\bmV}{\bmU^T \frac{\bmX^T \bmE}{n}}} > C \norm{\Omegastar^{-\frac{1}{2}}} \norm{\bmSigma_{\bmX}^{\frac{1}{2}}} \frac{m + \omega_{\Gamma} + u}{\sqrt{n}}) \leq 2 \exp(- u^2),
\end{equation}
when $n \geq (m + \omega_{\Gamma} + u)^2$.

Following the procedure of Lemma \ref{SampleOmegaLemmaThree}, the fourth term could be bounded as
\begin{equation}
	P(\usup{\bmV \in \calS^{m^2 - 1}} \abs{\iprod{\bmV}{\frac{\bmE^T \bmE}{n} - \Omegastar^{-1}}} > C \norm{\Omegastar^{-1}} \frac{m + u}{\sqrt{n}}) \leq 2 \exp(- u^2),
\end{equation}
when $n \geq  (m + u)^2$.

The last determined term could be bounded as
\begin{equation*}
	\fnorm{(\Gammaini - \Gammastar)^T \bmSigma_{\bmX} (\Gammaini - \Gammastar)} \leq \norm{\bmSigma_{\bmX}} \fnorm{\Gammaini - \Gammastar}^2.
\end{equation*}

Taking all terms into consideration, we could derive the event
\begin{align*}
	&\fnorm{\Omegastar^{-1} - \bmS} \\
	&\leq C(\norm{\bmSigma_{\bmX}}(1 + \frac{m + \omega_{\Gamma} + u}{\sqrt{n}}) \fnorm{\bmGamma_{\text{ini}} - \Gammastar}^2 + \norm{\Omegastar^{-\frac{1}{2}}} \norm{\bmSigma_{\bmX}^{\frac{1}{2}}} \frac{m + \omega_{\Gamma} + u}{\sqrt{n}} \fnorm{\bmGamma_{\text{ini}} - \Gammastar} + \norm{\Omegastar^{-1}} \frac{m + u}{\sqrt{n}}) \\
	&\leq C(\frac{\tau_{\max}^2}{\tau_{\min}^2 \nu_{\min}} (1 + \frac{m + \omega_{\Gamma} + u}{\sqrt{n}}) \frac{(\omega_{\Gamma} + u)^2}{n} + \frac{\tau_{\max}}{\tau_{\min} \nu_{\min}} \frac{m + \omega_{\Gamma} + u}{\sqrt{n}} \frac{\omega_{\Gamma} + u}{\sqrt{n}} + \frac{1}{\nu_{\min}} \frac{m + u}{\sqrt{n}}) \\
	&\leq C \frac{\tau_{\max}^2}{\tau_{\min}^2 \nu_{\min}} \frac{m + \omega_{\Gamma} + u}{\sqrt{n}}
\end{align*}
holds with probability at least $1 - 12 \exp(- u^2)$, when $\sqrt{n} \geq 2 C \frac{\tau_{\max}}{\tau_{\min}} (m + \omega_{\Gamma} + u)$.
\subsection{Proof of Lemma \ref{Covariance_IHT}}
We could rewrite $\bmv^T \bmS \bmv$ as
\begin{align*}
	&\bmv^T \bmS \bmv \nonumber \\
	&=\bmv^T ((\Gammaini - \Gammastar)^T \frac{\bmX^T \bmX}{n} (\Gammaini - \Gammastar) - (\Gammaini - \Gammastar)^T \frac{\bmX^T \bmE}{n} - \frac{\bmE^T \bmX}{n} (\Gammaini - \Gammastar) + \frac{\bmE^T \bmE}{n}) \bmv \nonumber \\
	&=\bmv^T ((\Gammaini - \Gammastar)^T (\frac{\bmX^T \bmX}{n} - \bmSigma_{\bmX}) (\Gammaini - \Gammastar) - 2 (\Gammaini - \Gammastar)^T \frac{\bmX^T \bmE}{n} + \frac{\bmE^T \bmE}{n} - \Omegastar^{-1}\\
	&\quad + (\Gammaini - \Gammastar)^T \bmSigma_{\bmX} (\Gammaini - \Gammastar) + \Omegastar^{-1}) \bmv \nonumber \\
	&\geq \bmv^T ((\Gammaini - \Gammastar)^T (\frac{\bmX^T \bmX}{n} - \bmSigma_{\bmX}) (\Gammaini - \Gammastar) - 2 (\Gammaini - \Gammastar)^T \frac{\bmX^T \bmE}{n} + \frac{\bmE^T \bmE}{n} - \Omegastar^{-1} + \Omegastar^{-1}) \bmv \nonumber ,
\end{align*}
where we use the fact that $(\Gammaini - \Gammastar)^T \bmSigma_{\bmX} (\Gammaini - \Gammastar)$ is positive semidefinite.

We need to deal with three random processes. The first term is bounded by the following lemma.
\begin{lemma} \label{Covariance_IHT_first}
	The event
	\begin{equation}
		\uinf{\begin{subarray}{c}
				\bmU \in \calC_{2s_{\Gamma}} \cap \calS^{dm - 1} \\ \bmv \in \calS^{m - 1}
		\end{subarray}} \bmv^T \bmU^T (\frac{\bmX^T \bmX}{n} - \bmSigma_{\bmX}) \bmU \bmv \fnorm{\Gammaini - \Gammastar}^2 \geq -C\norm{\bmSigma_{\bmX}} \frac{\sqrt{m} + \omega_{\Gamma} + u}{\sqrt{n}} \fnorm{\Gammaini - \Gammastar}^2
	\end{equation}
	holds with probability $1 - 2 \exp(- u^2)$, when  $n >  (\sqrt{m} + \omega_{\Gamma} + u)^2$.
\end{lemma}

The second term could be rewritten as
\begin{equation}
	\bmv^T (\Gammaini - \Gammastar)^T \frac{\bmX^T \bmE}{n} \bmv = \bmv^T \bmU^T \frac{\bmX^T \bmE}{n} \bmv \fnorm{\Gammaini - \Gammastar},
\end{equation}
where $\bmU \in \calC_{2s_{\Gamma}} \cap \calS^{dm - 1}$.

We could rearrange $X_{\bmU, \bmv} - X_{\bmW, \bmz}$ as
\begin{align*}
	X_{\bmU, \bmv} - X_{\bmW, \bmz} &= \bmv^T \bmU^T \frac{\bmX^T \bmE}{n} \bmv - \bmz^T \bmW^T \frac{\bmX^T \bmE}{n} \bmz = \frac{1}{n} \Vect{\bmE^T} (\bmI_n \otimes (\bmv \bmv^T \bmU^T - \bmz \bmz^T \bmW^T)) \Vect{\bmX^T}.
\end{align*}
From the facts
\begin{equation}
	\bmv \bmv^T \bmU^T - \bmz \bmz^T \bmW^T = \frac{1}{2} (\bmv + \bmz)(\bmv - \bmz)^T \bmU^T + \frac{1}{2} (\bmv - \bmz)(\bmv + \bmz)^T \bmU^T + \bmz \bmz^T (\bmU - \bmW)^T
\end{equation}
and
\begin{equation}
	\fnorm{\bmv \bmv^T \bmU^T - \bmz \bmz^T \bmW^T}^2 \leq 8 \ltwonorm{\bmv - \bmz}^2 + 2 \fnorm{\bmU - \bmW}^2 \leq 8 \fnorm{(\begin{smallmatrix}
			\bmU \\ \bmv^T
		\end{smallmatrix}) - (\begin{smallmatrix}
			\bmW \\ \bmz^T
		\end{smallmatrix})}^2,
\end{equation}
we could derive the mixed tail according to Lemma \ref{Quadratic_form_Independent}
\begin{equation}
	\begin{split}
		&P(\abs{\frac{1}{n}\Vect{\bmE^T} (\bmI_n \otimes (\bmv \bmv^T \bmU^T - \bmz \bmz^T \bmW^T)) \Vect{\bmX^T}} > u) \\
		&\leq 2 \exp(-\min (\frac{u^2}{\frac{8}{n} \norm{\bmSigma_{\bmX}^{\frac{1}{2}}}^2 \norm{\Omegastar^{-\frac{1}{2}}}^2 \fnorm{(\begin{smallmatrix}
					\bmU \\ \bmv^T
				\end{smallmatrix}) - (\begin{smallmatrix}
					\bmW \\ \bmz^T
				\end{smallmatrix})}^2}, \frac{u}{\frac{2\sqrt{2}}{n} \norm{\bmSigma_{\bmX}^{\frac{1}{2}}} \norm{\Omegastar^{-\frac{1}{2}}} \fnorm{(\begin{smallmatrix}
					\bmU \\ \bmv^T
				\end{smallmatrix}) - (\begin{smallmatrix}
					\bmW \\ \bmz^T
				\end{smallmatrix})}})).
	\end{split}
\end{equation}
Then we could derive from Lemma \ref{Suprema}
\begin{equation}
	P(\usup{\begin{subarray}{c}
			\bmU \in \calC_{2s_{\Gamma}} \cap \calS^{dm - 1} \\ \bmv \in \calS^{m - 1}
	\end{subarray}} \abs{\bmv^T \bmU^T \frac{\bmX^T \bmE}{n} \bmv} > C \norm{\bmSigma_{\bmX}^{\frac{1}{2}}} \norm{\Omegastar^{-\frac{1}{2}}} \frac{\sqrt{m} + \omega_{\Gamma} + u}{\sqrt{n}}) \leq 2 \exp(- u^2),
\end{equation}
when $n >  (\sqrt{m} + \omega_{\Gamma} + u)^2$.

Now we deal with the third term. From the facts
\begin{align}
	\bmv \bmv^T - \bmz \bmz^T = \frac{1}{2} (\bmv + \bmz)(\bmv - \bmz)^T + \frac{1}{2} (\bmv - \bmz)(\bmv + \bmz)^T
\end{align}
and
\begin{align}
	\fnorm{\bmv \bmv^T - \bmz \bmz^T}^2 \leq 4 \ltwonorm{\bmv - \bmz}^2,
\end{align}
we could get the mixed tail according to Lemma \ref{Quadratic_form_XX}
\begin{equation}
	\begin{split}
		&P(\abs{\bmv^T(\frac{1}{n} \bmE^T \bmE - \Omegastar^{-1})\bmv - \bmz^T(\frac{1}{n} \bmE^T \bmE - \Omegastar^{-1})\bmz} > u) \leq 2 \exp(-\min (\frac{u^2}{\frac{4}{n} \norm{\Omegastar^{-1}}^2 \ltwonorm{\bmv - \bmz}^2}, \frac{u}{\frac{2}{n} \norm{\Omegastar^{-1}} \ltwonorm{\bmv - \bmz}})).
	\end{split}
\end{equation}
Then we could derive
\begin{equation}
	P(\usup{\bmv \in \calS^{m - 1}} \abs{\bmv^T (\frac{1}{n} \bmE^T \bmE - \Omegastar^{-1}) \bmv} > C \norm{\Omegastar^{-1}} \frac{\sqrt{m} + u}{\sqrt{n}}) \leq 2 \exp(- u^2),
\end{equation}
when $n >  (\sqrt{m} + u)^2$, according to Lemma \ref{Suprema}.

%The third item could be bounded by the following lemma.
%\begin{lemma}[Theorem 6 \cite{Banerjee2014Estimation}]
%	Suppose $\bmX$ is a matrix with $n$ independent sub-Gaussian rows with $\EE[\bmX_i^T \bmX_i] = \bmSigma_{\bmX}$ and $\subgnorm{\bmX_i \bmSigma_{\bmX}^{-\frac{1}{2}}} \leq K$. Then the event
%	\begin{equation}
%		\lambda_{\min}(\bmSigma_{\bmX} \mid \calA) (1 - C \frac{\gwidth{\calA}}{\sqrt{n}}) \leq \uinf{\bmu \in \calA} \frac{1}{n} \ltwonorm{\bmX \bmu}^2 \leq \usup{\bmu \in \calA} \frac{1}{n} \ltwonorm{\bmX \bmu}^2 \leq \lambda_{\max}(\bmSigma_{\bmX} \mid \calA) (1 - C \frac{\gwidth{\calA}}{\sqrt{n}})
%	\end{equation}
%hold with probability $1 - 2 \exp(-c\gwidth{\calA}^2)$, where $\lambda_{\min}(\bmSigma_{\bmX} \mid \calA) = \uinf{\bmu \in \calA} \bmu^T \bmSigma_{\bmX} \bmu$ and $\lambda_{\max}(\bmSigma_{\bmX} \mid \calA) = \usup{\bmu \in \calA} \bmu^T \bmSigma_{\bmX} \bmu$.
%\end{lemma}
%We could derive
%\begin{align*}
%	\frac{1}{n} \bmv^T \bmE^T \bmE \bmv \geq \lambda_{\min}(\Omegastar^{-1})(1 - C \frac{\sqrt{m} + u}{\sqrt{n}}),
%\end{align*}
%with probability $1 - 2 \exp(-u^2)$, where we set $c \omega_{\Gamma}^2 = u^2$.

Taking all parts into consideration, we could derive
\begin{align*}
	&\quad \bmv^T \bmS \bmv \\
	&\geq -C\norm{\bmSigma_{\bmX}} \frac{\sqrt{m} + \omega_{\Gamma} + u}{\sqrt{n}} \fnorm{\Gammaini - \Gammastar}^2 - 2 C \frac{\sqrt{m} + \omega_{\Gamma} + u}{\sqrt{n}} \norm{\bmSigma_{\bmX}^{\frac{1}{2}}} \norm{\Omegastar^{-\frac{1}{2}}} \fnorm{\Gammaini - \Gammastar} - C \norm{\Omegastar^{-1}} \frac{\sqrt{m} + u}{\sqrt{n}} \\
	&\quad + \lambda_{\min}(\Omegastar^{-1}) \nonumber \\
	&\geq - C \frac{\tau_{\max}^2}{\tau_{\min}^2 \nu_{\min}} \frac{(\sqrt{m} + \omega_{\Gamma} + u)^3}{n^{\frac{3}{2}}} - C \frac{\tau_{\max}}{\tau_{\min} \nu_{\min}} \frac{(\sqrt{m} + \omega_{\Gamma} + u)^2}{n} - C \frac{1}{\nu_{\min}} \frac{\sqrt{m} + u}{\sqrt{n}} + \frac{1}{\nu_{\max}}\\
	&\geq \frac{c}{\nu_{\max}},
\end{align*}
with probability $1 - 10 \exp(- u^2)$, when $\sqrt{n} > 2 C \frac{\tau_{\max} \nu_{\max}}{\tau_{\min} \nu_{\min}} (\sqrt{m} + \omega_{\Gamma} + u)$, where we use Lemma \ref{Initialization_Gamma_IHT}.
\subsection{Proof of Lemma \ref{Covariance_IHT_first}}
We could rewrite the term as
\begin{equation}
	\lambda_{\min}((\Gammaini - \Gammastar)^T (\frac{\bmX^T \bmX}{n} - \bmSigma_{\bmX}) (\Gammaini - \Gammastar)) \geq \uinf{\begin{subarray}{c}
			\bmU \in \calC_{2s_{\Gamma}} \cap \calS^{dm - 1} \\ \bmv \in \calS^{m - 1}
	\end{subarray}} \bmv^T \bmU^T (\frac{\bmX^T \bmX}{n} - \bmSigma_{\bmX}) \bmU \bmv \fnorm{\Gammaini - \Gammastar}^2.
\end{equation}
From the facts
\begin{equation}
	\bmU \bmv \bmv^T \bmU^T - \bmW \bmz \bmz^T \bmW^T = (\bmU - \bmW) \bmv \bmv^T \bmU^T + \frac{\bmW(\bmv + \bmz)(\bmv - \bmz)^T\bmU^T}{2} + \frac{\bmW(\bmv - \bmz)(\bmv + \bmz)^T\bmU^T}{2} + \bmW \bmz \bmz^T (\bmU - \bmW)^T
\end{equation}
and
\begin{equation}
	\fnorm{\bmU \bmv \bmv^T \bmU^T - \bmW \bmz \bmz^T \bmW^T}^2 \leq 6 \fnorm{\bmU - \bmW}^2 + 16 \ltwonorm{\bmv - \bmz}^2,
\end{equation}
we could derive the mixed tail according to Lemma \ref{Quadratic_form_XX}
\begin{equation}
	\begin{split}
		&P(\abs{\Vect{\bmX^T}^T(\bmU \bmv \bmv^T \bmU^T - \bmW \bmz \bmz^T \bmW^T)\Vect{\bmX} - \EE[\Vect{\bmX^T}^T(\bmU \bmv \bmv^T \bmU^T - \bmW \bmz \bmz^T \bmW^T)\Vect{\bmX}]} > u) \\
		&\leq 2 \exp(-\min (\frac{u^2}{\frac{16}{n} \norm{\bmSigma_{\bmX}}^2 \fnorm{(\begin{smallmatrix}
					\bmU \\ \bmv^T
				\end{smallmatrix}) - (\begin{smallmatrix}
					\bmW \\ \bmz^T
				\end{smallmatrix})}^2}, \frac{u}{\frac{4}{n} \norm{\bmSigma_{\bmX}} \fnorm{(\begin{smallmatrix}
					\bmU \\ \bmv^T
				\end{smallmatrix}) - (\begin{smallmatrix}
					\bmW \\ \bmz^T
				\end{smallmatrix})}})).
	\end{split}
\end{equation}
Then we could derive
\begin{equation}
	P(\usup{\begin{subarray}{c}
			\bmU \in \calC_{2s_{\Gamma}} \cap \calS^{dm - 1} \\ \bmv \in \calS^{m - 1}
	\end{subarray}} \abs{\bmv^T \bmU^T \frac{\bmX^T \bmX}{n} \bmU \bmv - \EE [\bmv^T \bmU^T \frac{\bmX^T \bmX}{n} \bmU \bmv]} > C \norm{\bmSigma_{\bmX}} \frac{\sqrt{m} + \omega_{\Gamma} + u}{\sqrt{n}}) \leq 2 \exp(- u^2),
\end{equation}
when $n >  (\sqrt{m} + \omega_{\Gamma} + u)^2$, according to Lemma \ref{Suprema}.

\section{Additional experimental materials}
\subsection{Structured matrices estimation}
In this part, we present the sparse patterns of the estimated matrices produced by Algorithm \ref{Alg_AltIHT} and our initialization (Algorithm \ref{Alg_AltIHT_Initialization}).

We set $d = m = 100$, $s_{\Gamma}^{\star} = 100$. The rows of the predictor matrix $\bmX$ are generated independently from the distribution $\calN(\vzero, \mI_{d})$. The precision matrix follows a block diagonal graph. Every block is a $5 \times 5$ matrix, whose diagonal entries are $1$ and the other entries are $0.3$. The number of measurements is set as $3000$.

\begin{figure}[ht]
	%	\centering
	\vskip -0.1in
	%\captionsetup{aboveskip=0pt}
	\centerline{\subfigure[Original regression matrix $\Gammastar$]{\includegraphics[width=0.5\linewidth]{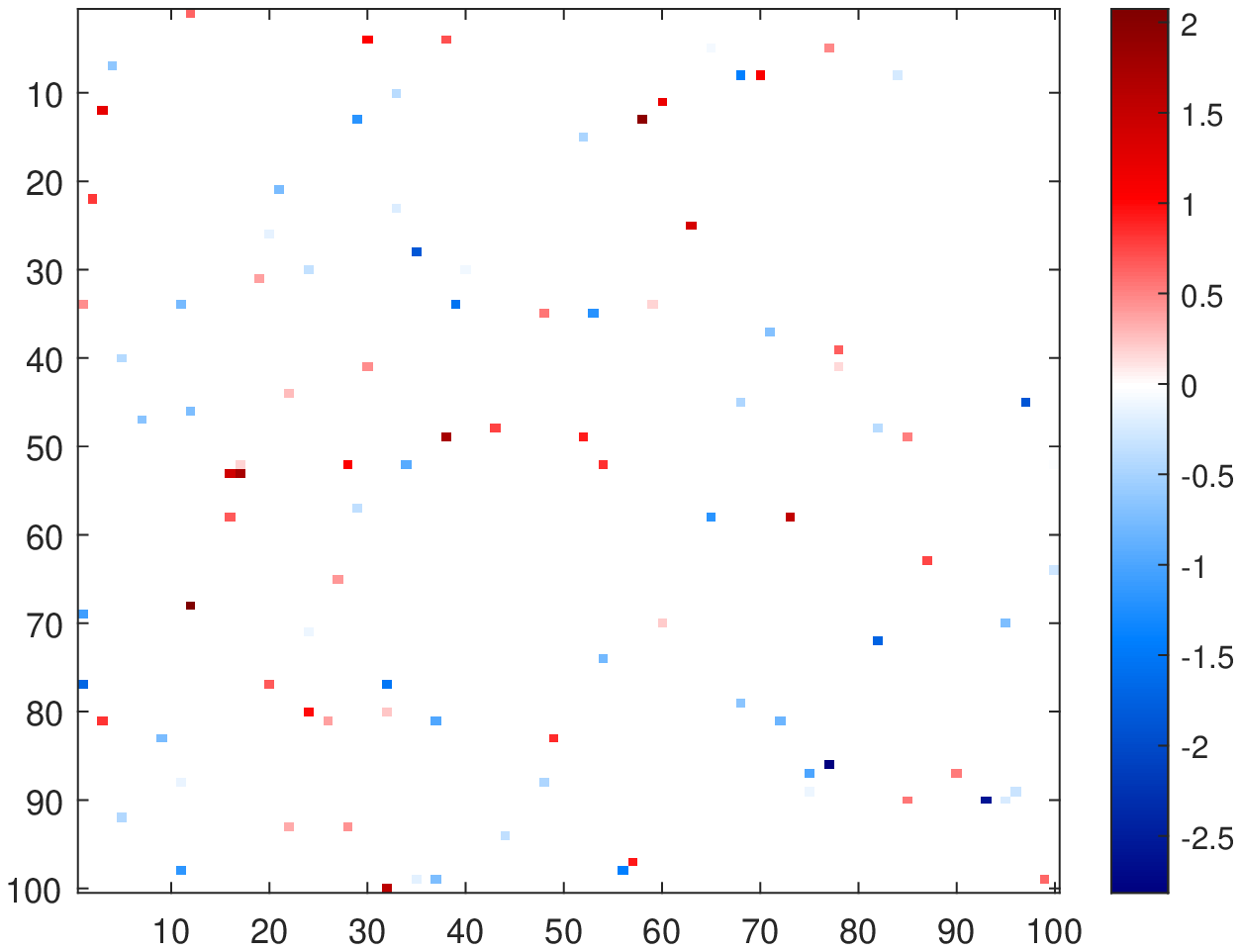}
			% where an .eps filename suffix will be assumed under latex,
			% and a .pdf suffix will be assumed for pdflatex
			\label{OriginalGamma}}
		%		\hspace{0.1em}%
		\hfill
		\subfigure[Estimated regression matrix $\hat{\mGamma}$]{\includegraphics[width=0.5\linewidth]{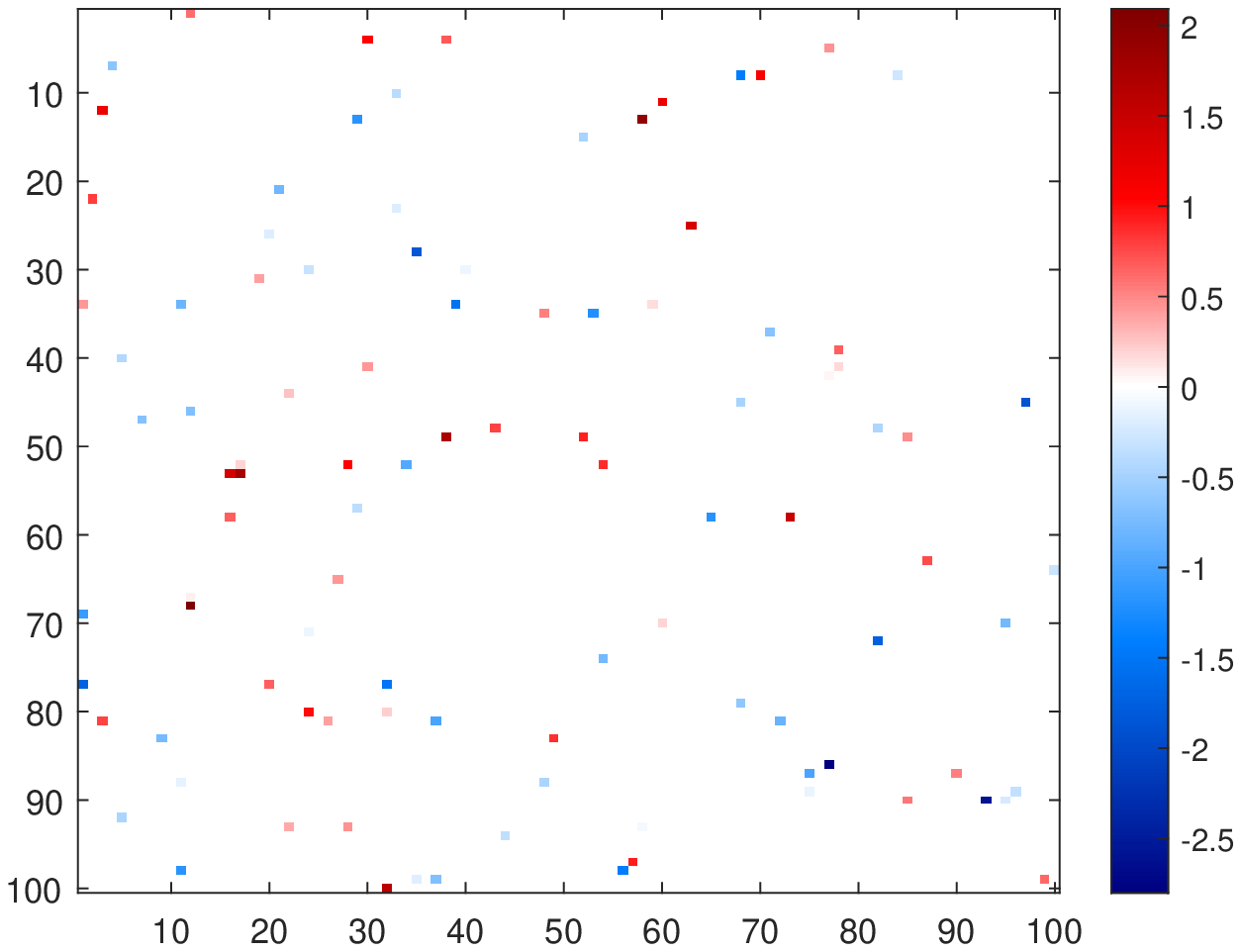}
			% where an .eps filename suffix will be assumed under latex,
			% and a .pdf suffix will be assumed for pdflatex
			\label{RecoveredGamma}}}
	\vskip -0.1in
	\caption{Comparison between the original regression coefficient matrix $\Gammastar$ and its estimation $\hat{\mGamma}$.}
	\label{Fig_EstimationGamma}
\end{figure}

\begin{figure}[ht]
	%	\centering
	\vskip -0.1in
	%\captionsetup{aboveskip=0pt}
	\centerline{\subfigure[Original precision matrix $\Omegastar$]{\includegraphics[width=0.5\linewidth]{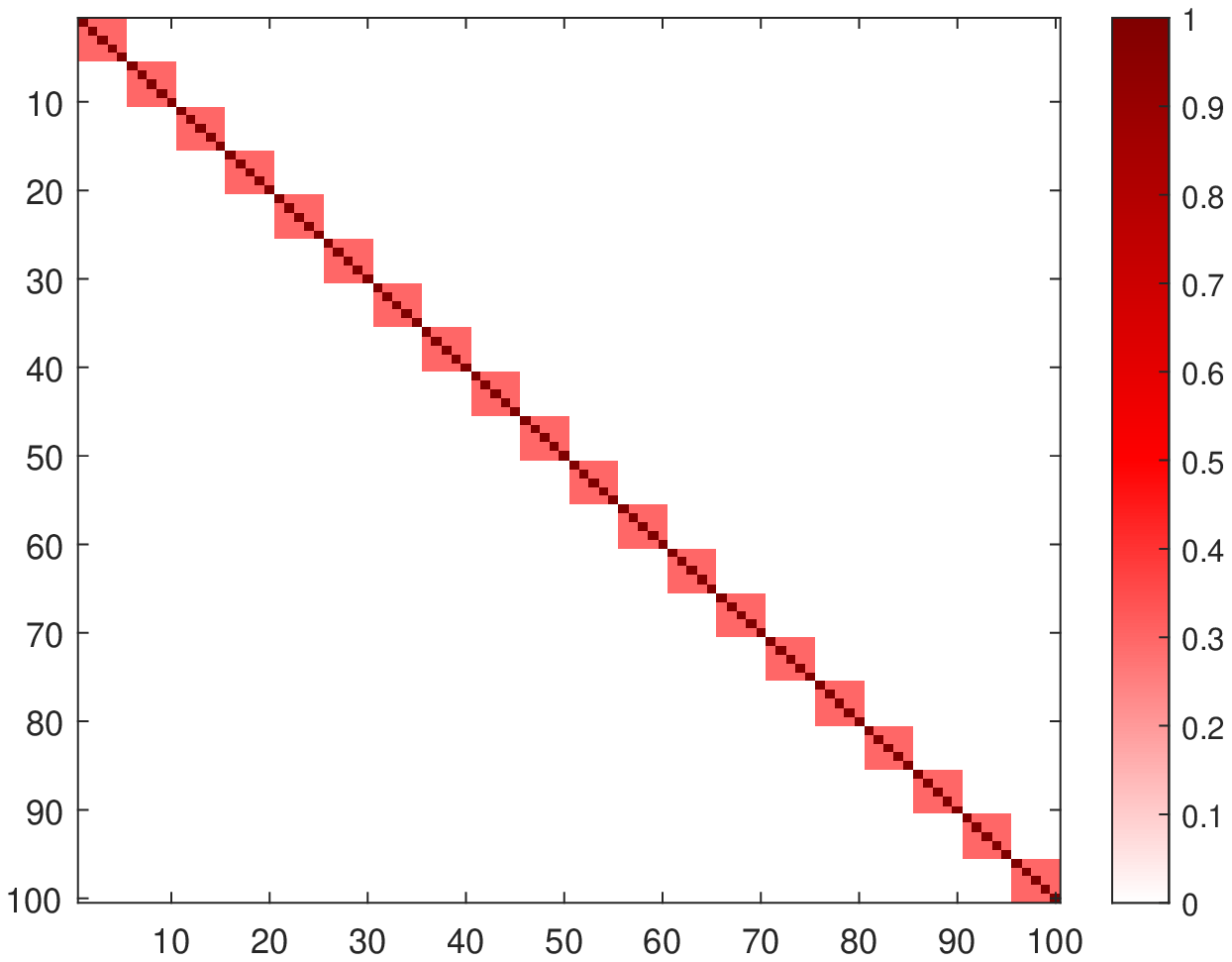}
			% where an .eps filename suffix will be assumed under latex,
			% and a .pdf suffix will be assumed for pdflatex
			\label{OriginalOmega}}
		%		\hspace{0.1em}%
		\hfill
		\subfigure[Estimated precision matrix $\hat{\mOmega}$]{\includegraphics[width=0.5\linewidth]{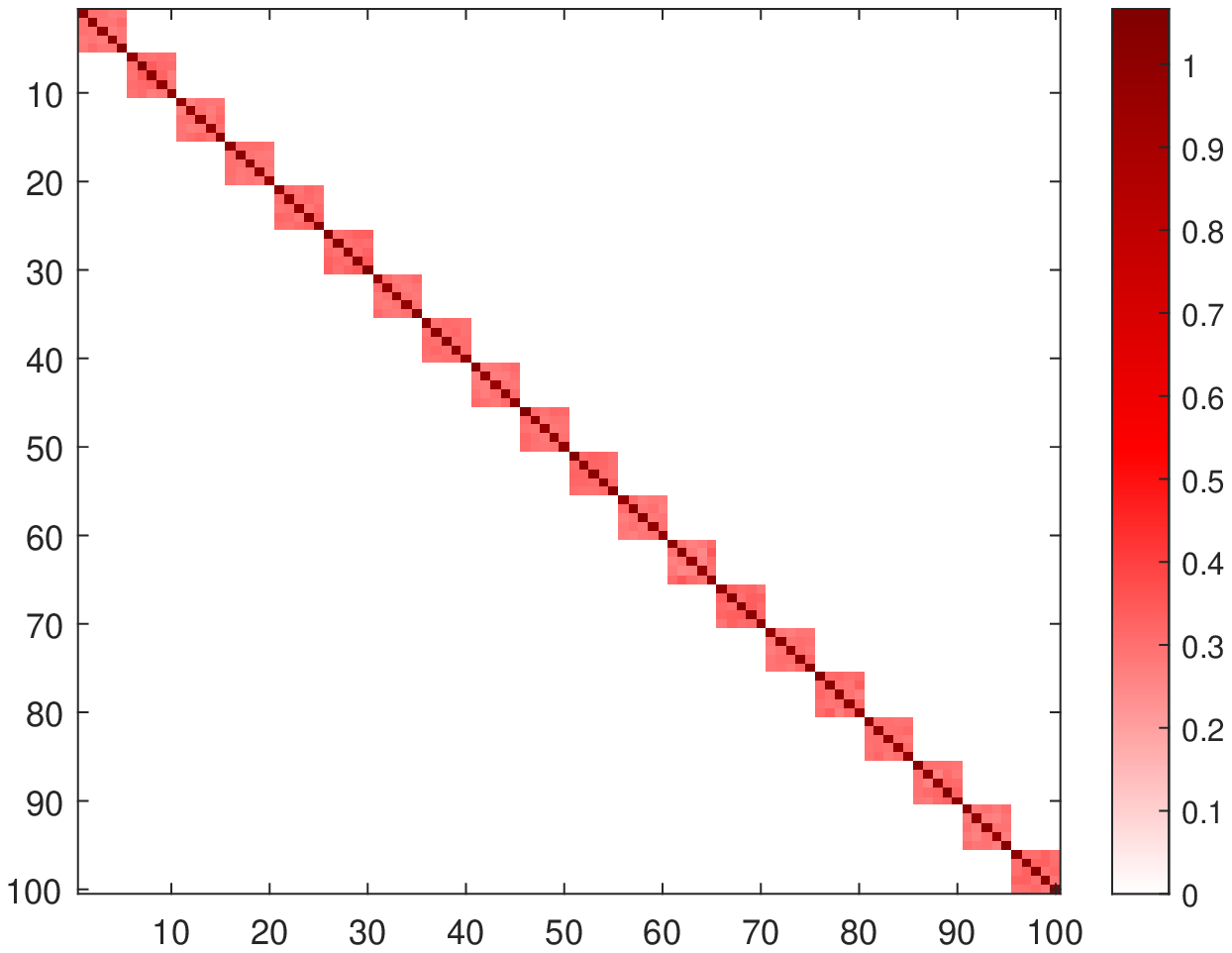}
			% where an .eps filename suffix will be assumed under latex,
			% and a .pdf suffix will be assumed for pdflatex
			\label{RecoveredOmega}}}
	\vskip -0.1in
	\caption{Comparison between the original precision matrix $\Omegastar$ and its estimation $\hat{\mOmega}$.}
	\label{Fig_EstimationOmega}
\end{figure}

In Figure \ref{Fig_EstimationGamma} and \ref{Fig_EstimationOmega}, we compare the original regression coefficient matrix $\Gammastar$ and the precision matrix $\Omegastar$ with their estimations $\hat{\mGamma}$ and $\hat{\mOmega}$ respectively. These figures illustrate that Algorithm \ref{Alg_AltIHT} and our initialization (Algorithm \ref{Alg_AltIHT_Initialization}) could recover the sparse structures of $\Gammastar$ and $\Omegastar$, and verify our theoretical results. For Algorithm \ref{General_AltPGD} and \ref{General_Initialization} with the $l_1$-norm, the results are similar and we do not include them in this manuscript.

\bibliographystyle{IEEEtran}
%\bibliography{definitions,bibliofile}
%%
%% where we here have assume the existence of the files
%% definitions.bib and bibliofile.bib.
%% BibTeX documentation can be obtained at:
%% http://www.ctan.org/tex-archive/biblio/bibtex/contrib/doc/
%%%%%%

\bibliography{reference}
%\bibliography{C:/Users/201308/Documents/private/writing/reference}

%% Or you use manual references (pay attention to consistency and the
%% formatting style!):
%\begin{thebibliography}{9}
%
%\bibitem{Laport:LaTeX}
%L.~Lamport,
%  \emph{\LaTeX: A Document Preparation System,}
%  Addison-Wesley, Reading, Massachusetts, USA, 2nd~ed., 1994.
%
%\bibitem{GMS:LaTeXComp}
%F.~Mittelbach, M,~Goossens, J.~Braams, D.~Carlisle, and
%C.~Rowley, \emph{The {\LaTeX} Companion,} Addison-Wesley,
%Reading, Massachusetts, USA, 2nd~ed., 2004.
%
%\bibitem{oetiker_latex}
%T.~Oetiker, H.~Partl, I.~Hyna, and E.~Schlegl, \emph{The Not So Short
%  Introduction to {\LaTeX2e}}, version 5.06, Jun.~20, 2016. [Online].
%  Available: \url{https://tobi.oetiker.ch/lshort/}
%
%\bibitem{typesetmoser}
%S.~M. Moser, \emph{How to Typeset Equations in {\LaTeX}}, version 4.6,
%  Sep. 29, 2017. [Online]. Available:
%  \url{http://moser-isi.ethz.ch/manuals.html#eqlatex}
%
%\bibitem{IEEE:pdfsettings}
%IEEE, \emph{Preparing Conference Content for the IEEE Xplore Digital
%  Library.} [Online]. Available:
%  \url{http://www.ieee.org/conferences_events/conferences/organizers/pubs/preparing_content.html}
%
%\bibitem{IEEE:AuthorToolbox}
%IEEE, \emph{Author Digital Toolbox.} [Online.] Available:
%  \url{http://www.ieee.org/publications_standards/publications/authors/authors_journals.html}
%
%\end{thebibliography}

\end{document}